\documentclass[english]{iopart}

\usepackage{graphicx}
\usepackage{url}
 \usepackage{amssymb}

\newcounter{fig}
\begin{document}

\title[Replicable functions]
{\Large Modular correspondences and replicable functions (unabridged version)}

\vskip .3cm 

\author{J.-M. Maillard$^\pounds$}
\address{$^\pounds$ LPTMC, UMR 7600 CNRS, 
Sorbonne Universit\'e, Tour 23,
 5\`eme \'etage, case 121, 
 4 Place Jussieu, 75252 Paris Cedex 05, France} 

\ead{maillard@lptmc.jussieu.fr}

\begin{abstract}

Landen transformation, and more generally modular correspondences,
can be seen to be exact symmetries of some integrable lattice models, like the square Ising model, or the Baxter
model. They are solutions of remarkable Schwarzian equations and have some compositional properties.
Most of the known examples correspond, in an elliptic curves framework, to an automorphy
property of pullbacked $\, _2F_1$ hypergeometric functions, associated with modular forms.
It is, however, important to underline that these Schwarzian equations go beyond an 
elliptic curves, and hypergeometric functions framework.
The question of a modular correspondence interpretation
of the solutions of these ``Schwarzian'' equations was clearly an open question. This paper
tries to shed some light on this open question. We first shed some light on the very nature of a
one-parameter series solution of the Schwarzian equation.  This one-parameter series is not generically
a modular correspondence series, but it actually reduces to an
 infinite set of modular correspondence series for an infinite set
of ($N$-th root of unity) values of the parameter. We also  provide
an example of two-parameter series, with a compositional property,
solution of a Schwarzian equation. We  finally provide simple pedagogical
examples that are very similar to modular correspondence series, but are far beyond the
elliptic curves framework. These last examples show that the modular correspondence-like
series, or the nome-like series, are not necessarily globally bounded. 
The results of that paper can be seen as an incentive to study  differentially algebraic 
series with integer coefficients, in physics or enumeratice combinatorics. 

\end{abstract}


\noindent {\bf AMS Classification scheme numbers}: 34M55, 47Exx, 32Hxx, 32Nxx, 34Lxx, 34Mxx, 14Kxx, 14H52 

\vskip .3cm


\vskip .2cm

 \today

 
\vskip .1cm
 
\section{Introduction: infinite order symmetries.}
\label{int}

There is no need to underline the crucial role played by the concept of symmetry in physics,
theoretical physics, mathematical physics. We will not consider here continuous symmetry
groups (Lie groups) but rather {\em discrete} symmetries, not necessarily corresponding to
geometrical symmetries (Coxeter groups, Weyl groups of infinite-dimensional Kac–Moody algebras),
or finite groups. At first sight we do not expect any representation
of these {\em  discrete} symmetries as {\em linear} transformations of vector spaces
(no representation theory). Examples of such {\em discrete symmetries}, without
representation as  {\em linear} transformations, are, for instance,
{\em birational transformations}~\cite{Noetherian,FactoBirat},
which are known to be (infinite order) discrete  symmetries of
integrable models~\cite{YBE,BeMaVi92}. 
Such discrete symmetries  can be studied, {\em per se}, 
in a discrete dynamical perspective\footnote[2]{One can recall that the theory of iteration 
of rational functions was seen, in the pioneering work of Julia, Fatou and Ritt,
as a method for investigating  functional equations~\cite{Fatou,Ritt,Fatou2,Eremenko}}.

The simplest example of such  discrete symmetries corresponds to a (univariate) transformation 
$\, x \, \rightarrow \, \, y(x)$ preserving some structures\footnote[1]{These structures can be
linear (or non-linear) differential equations, systems of partial differential equations~\cite{Barry},
functional equations, etc ...}.  These structures must be {\em invariant}, or  {\em covariant},
under the previous transformations $\, x \, \rightarrow \, \, y(x)$. The simplest example of
``structure'' is certainely just a function.  Let us consider a function $\, \Phi(x)$,  let us
discard the (too simple) invariance situation, where we have a functional equation
$\,  \Phi(y(x)) \, = \, \,  \Phi(x)$, and let us consider the following ``covariance'' property
for a function $\, \Phi(x)$
\begin{eqnarray}
\label{automorph}
\hspace{-0.95in}&& \quad \quad  \quad  \quad  \quad  \quad  \quad \quad  \quad 
  \Phi\Bigl(y(x)\Bigr) \,   \, = \, \, \, \,\, 
 {\cal A}(x) \cdot \,  \Phi(x), 
\end{eqnarray}
where the ``automorphy'' cofactor $\, {\cal A}(x)$ can be described in terms of the symmetry
transformation $\, y(x)$. Along this line the function $\, \Phi(x)$ can be seen as
an ``automorphic'' function~\cite{Ford} with respect to the transformation
$\, x \, \rightarrow \, \, y(x)$: the composition of the  transformation $\, y(x)$
with itself, clearly yields another ``covariance'' or ``automorphy'' property
\begin{eqnarray}
\label{automorph2}
\hspace{-0.95in}&& \,   \quad  \quad 
\Phi\Bigl(y(y(x))\Bigr)
\,   \, = \, \, \, \,\, 
 {\cal A}\Bigl(y(x)\Bigr) \cdot \,  \Phi\Bigl(y(x)\Bigr)
 \,\, = \, \, \, \,\,
 \Bigl( {\cal A}\Bigl(y(x)\Bigr) \cdot \, {\cal A}(x) \Bigr) \cdot \, \Phi(x), 
\end{eqnarray}
and so on, for every $\, n$-th iteration of $\, y(x)$ with itself.
From a mathematical view-point such an ``automorphy property'' (\ref{automorph})
is reminiscent of the theory of
{\em automorphic forms}~\cite{Ford,Stiller,Schwarzian2,Schwarzian}
(which can be generalized to
Hilbert modular forms for two, or more, variables), which generalizes the theory of
{\em modular forms}~\cite{Koblitz,1-2-3,Berndt,ModularForms,Hilbert,SuperMaier}.  
In the case where $\, y(x)$  is not only a rational function, but a linear fractional transformation,
the ``covariance'' property (\ref{automorph})
can be illustrated by the Poincar\'e series~\cite{McMullen,Fricke,Janos},
and other Theta-Fuchsian functions or series~\cite{Schwarzian,Dalzell,Farkas,Schoeneberg}. 
From a physics view-point such an ``automorphy property'' (\ref{automorph})
is reminiscent of the  renormalization group theory, revisited
by Wilson~\cite{Migdal,Fisher}, seen as a fundamental symmetry in lattice statistical
mechanics or field theory.  The graduate student example of exact renormalization
calculation of the partition function of the one-dimensional Ising model, displayed
in~\cite{Renorm}, relies on  an ``automorphy relation'' (\ref{automorph}), where $\, \Phi(x)$
is the partition function per site, and $\, y(x)$ corresponds  to the renormalization
transformation symmetry $\, \tanh(K) \, \, \rightarrow \, \,  \tanh(K)^2$.

In the following we will not restrict the transformation symmetry $\, y(x)$
to be a linear fractional transformation: the function $\, y(x)$ is a series, analytic
at $\, x= \, 0$, it can be a rational function, an algebraic function, a D-finite function,
a D-D-finite function\footnote[5]{A D-D-finite function is a function solution of a
  linear differentiable operator with $\, D$-finite function coefficients~\cite{DDFinite}.},
a {\em differentially algebraic} function\footnote[9]{A differentially algebraic function
is a function solution of a non-linear differential equation of the form 
$ \, P(x, \, y, \, y', \, y'', \, \cdots \, y^{(n)}) \, = \, \, 0$,
where $\, P$ is a polynomial.}, ...

To be more specific,  let us give a simple, but highly pedagogical,
illustration of a ``covariance'' property (\ref{automorph}), which corresponds to
$\, \Phi(x)$ being a selected $\, _2F_1$ hypergeometric function~\cite{Schwarz1,Schwarz2} 
\begin{eqnarray}
\label{modularform2explici}
\hspace{-0.95in}&& \quad \quad  \quad  \quad  \quad 
 _2F_1\Bigl([{{1} \over {12}}, \, {{5} \over {12}}], \, [1], \, y(x)  \Bigr)
\,\, \, = \, \, \, \,\, 
 {\cal A}(x) \cdot \,
 _2F_1\Bigl([{{1} \over {12}}, \, {{5} \over {12}}], \, [1], \,  x  \Bigr),
\end{eqnarray}
where the ``automorphic prefactor'' $\, {\cal A}(x)\, $  reads
\begin{eqnarray} 
\label{modularform3}
\hspace{-0.95in}&& \quad   \quad \quad  \quad \quad  \quad \quad\quad
{\cal A}(x)  \, \, = \, \,  \,  
 \lambda \cdot \,  \Bigl( {{u(x)} \over { u(y(x)) }}
 \cdot \,  y'(x) \Bigr)^{1/2},
\end{eqnarray}
and where $\, u(x)$ is related~\cite{Schwarz1,Schwarz2}
to the wronskian of the order-two linear differential
operator annihilating $\, \Phi(x)$, namely the  $\, _2F_1$ hypergeometric function
$\, _2F_1([{{1} \over {12}}, \, {{5} \over {12}}], \, [1], \,  x)$.

\vskip .1cm 

\subsection{Modular forms, correspondences and physics.}
\label{modularforms}

The simplest example of a transformation $\, x \,\,  \rightarrow \,\,  y \, =\,  \, y(x)$
occurring in the ``automorphy'' relation  (\ref{modularform2explici}), or occurring as
an {\em exact generator} of the {\em renormalization group} of the square Ising model,
or even of the Baxter model~\cite{Heegner}, 
corresponds to the {\em Landen transformation}~\cite{Heegner,bo-ha-ma-ze-07b}
\begin{eqnarray} 
\label{Landen}
\hspace{-0.95in}&& \quad \quad \quad \quad \quad \quad \quad \quad \quad \quad \quad
k \, \, \quad  \longrightarrow \,\,  \quad k_L \, = \, \, 
{{2 \sqrt{k}} \over {1+k}},
\end{eqnarray}
or to its compositional inverse, the inverse Landen transformation.
As it should, the {\em critical point} of the 
square Ising model (resp. Baxter model) is a {\em fixed point}~\cite{Heegner}
of the Landen transformation: $\, k\,= \, 1$.

Let us introduce the $\, j$-invariant\footnote[8]{The $j$-invariant~\cite{Heegner,Canada} 
(see also Klein's modular invariant) regarded as a function of a complex 
  variable $\, \tau$ (the ratio of periods), is a modular function
  of weight zero for $\, SL(2,\, \mathbb{Z})$.} 
of an elliptic curve of modulus $\, k$, and its 
transform by the Landen transformation (\ref{Landen})
\begin{eqnarray} 
\label{jjprime}
\hspace{-0.95in}&& \quad \quad \quad
 j(k) \, = \, \, \, \, 256
\cdot {{(1-k^2+k^4)^3} \over {k^4 \cdot (1-k^2)^2}},
\quad \, \,  \quad  j(k_L) \, = \, \, \, \, 
16 \cdot {\frac { (1+14\,{k}^{2}+{k}^{4})^3}{
 (1-{k}^{2})^{4} \cdot {k}^{2} }},
\end{eqnarray}
and let us also introduce the two corresponding 
{\em Hauptmoduls}~\cite{Heegner}:
\begin{eqnarray} 
\label{Haupt}
\hspace{-0.95in}&& \quad \quad \quad \quad \quad \quad \quad \quad \quad \quad
x \, \, = \, \, \, {{1728} \over {j(k)}}, 
\quad \quad \quad \, 
y \, \, = \,\, \,  {{1728} \over {j(k_L)}}. 
\end{eqnarray}
These two Hauptmoduls (\ref{Haupt}) are related by the 
{\em modular equation}~\cite{Andrews,Atkin,Hermite,Hanna,Morain,Weisstein}:
\begin{eqnarray}
\hspace{-0.95in}&& \quad 
1953125\, \, {x}^{3}{y}^{3} \, \, -187500\,\, {x}^{2}{y}^{2} \cdot \, (x+y) \, \, 
+375\, \, xy \cdot \, (16\,{x}^{2}-4027\,xy+16\,{y}^{2})
\nonumber \\
\label{modularcurvetau2}
\hspace{-0.95in}&& \quad \quad  \quad 
\quad  
 \,  -64\, \, (x+y)  \cdot \, ({x}^{2}+1487\,xy+{y}^{2}) 
\,\,  +110592\,xy
 \, \,  \,= \, \,\,  \, \, 0. 
\end{eqnarray}
The algebraic function $\, y \, = \, \, y(x)$, defined from the
modular curve (\ref{modularcurvetau2}),  is a
{\em multivalued function}, but we can single out the series 
expansion\footnote[9]{This series (\ref{seriesmodularcurve})
has a radius of convergence $\, 1$, even if the discriminant 
of the modular equation (\ref{modularcurvetau2}) which vanishes
at $\, x \, = \, 1$, vanishes for values inside the 
unit radius of convergence, for instance at 
$\, x \, = \, -64/125$.}:
\begin{eqnarray}
\label{seriesmodularcurve}
\hspace{-0.95in}&& \quad  \quad  \quad  \quad
 y \, \, \, = \, \, \,  \,  \,
{\frac {1}{1728}} \cdot \, {x}^{2} \, \,  \, \,
+{\frac {31}{62208}} \cdot \, {x}^{3} \,\, 
\, +{\frac {1337}{3359232}} \cdot \,{x}^{4} \, \,   \,\,
+{\frac {349115}{1088391168}} \cdot \,{x}^{5}
\nonumber \\ 
\hspace{-0.95in}&& \quad \quad  \quad  \quad \quad  \quad 
+{\frac {20662501}{78364164096}} \cdot \, {x}^{6} \,\, \,  \, 
+{\frac {1870139801}{8463329722368}} \cdot \,{x}^{7} \, \,\,\, 
 \, + \, \, \cdots 
\end{eqnarray}
The transformation 
$\, x \, \rightarrow  \, \, y(x) \, = \, y$, where $\, y$ 
is given by the modular equation (\ref{modularcurvetau2}), 
is thus an {\em algebraic} transformation,  corresponding to the
Landen transformation (\ref{Landen}), or to the inverse Landen transformation.
The emergence of a {\em modular form}~\cite{IsingCalabi,IsingCalabi2,Christol}
corresponds to the remarkable automorphy identity (\ref{modularform2explici})
on the {\em same} hypergeometric function, but where the pullback $\, x$
is changed  $\, x \, \rightarrow  \, \, y(x) \, = \, y$,  according to the 
modular equation (\ref{modularcurvetau2}). 

Let us consider another important modular equation. 
The modular equation of order three corresponding to 
$\, \, \tau \, \rightarrow \, 3 \cdot \, \tau$,
or $\, \tau \, \rightarrow \,  \, \tau/3$, 
reads\footnote[5]{Legendre already knew (1824) this 
order three modular equation in the form 
$\, (k \lambda)^{1/2} \, +   (k'  \lambda')^{1/2} 
 =  1$, where $\, k$ and $\, k'$, and $\,\lambda$, $\,\lambda' \, $
are pairs of complementary moduli $\, k^2\,+k'^2\,=\,1$, 
$\, \lambda^2\, +\lambda'^2 = \, 1$, and Jacobi derived that 
modular equation~\cite{Jacobi,Nova}.}: 
\begin{eqnarray}
\label{orderthree}
\hspace{-0.95in}&&  \quad  \quad  \quad  \quad 
{k}^{4} \,\, +12\,{k}^{3}\lambda \, \, \, +6\,{k}^{2}{\lambda}^{2} \,\, 
+12\,k{\lambda}^{3}\,\, +{\lambda}^{4} \,  \,  \,  \,
  -16\,{k}^{3}{\lambda}^{3} \,  -16\,k\lambda
\,\, \,  \,  = \, \,\,  \,   0.
\end{eqnarray}
Recalling that 
\begin{eqnarray}
\label{orderthreexy}
\hspace{-0.95in}&&  \,  
x \,\, = \, \,\, {{27} \over {4}} \cdot \, 
{\frac { {k}^{4} \cdot \, (1 \, -{k}^{2})^{2}}{ ({k}^{4}-{k}^{2}+1)^{3}}}
\,\, = \, \,\, {{1728} \over {j(k)}}, \quad \, 
y \, \,  = \, \,\,{{27} \over {4}} \cdot \,  {\frac {{\lambda}^{4} \cdot \,
 \left( 1-{\lambda}^{2} \right)^{2}}{({\lambda}^{4}-{\lambda}^{2}+1)^{3}}}
\,\, = \, \,\, {{1728} \over {j(\lambda)}}, 
\end{eqnarray}
gives a modular equation $\, P(x, \, y) \, = \, \, 0$, 
yielding the series expansion:
\begin{eqnarray}
\label{orderthreey}
\hspace{-0.95in}&& \quad 
 y \,\, = \, \,\,\,  {\frac {{x}^{3}}{2985984}}
\,  \,\, +{\frac {31 \,x^4 }{71663616}} \,  \,
+{\frac {36221\, x^5  }{82556485632}} \, 
\,\,  +{\frac {29537101 \, {x}^{6}}{71328803586048}} 
\,\, \,   \, + \, \,\, \cdots 
\end{eqnarray}
Note that these two series  (\ref{seriesmodularcurve}) and (\ref{orderthreey}) {\em commute}.
An alternative rational parametrization of this last modular equation
$\, P(x, \, y) \, = \, \, 0$ can be found
in~\cite{SuperMaier}:
\begin{eqnarray}
\label{modularequa3ratio}
  \hspace{-0.98in}&&   \, \,  \quad \quad
 x \, \, = \, \, \, {{ 1728 \, t } \over {(t+27) \cdot \, (t+3)^3 }}
      \quad \quad  \quad \hbox{ and:} \quad \quad \quad \, \,  
 y \, \, = \, \, \, {{ 1728 \, t^3 } \over {(t+27) \cdot \, (t+243)^3 }}. 
\end{eqnarray}
Again we have an automorphy relation (\ref{modularform2explici}) where
$\, y(x)$ is given by  (\ref{orderthreey}) with an algebraic
``automorphic prefactor'' $\, {\cal A}(x)$.

\vskip .1cm

\subsection{Schwarzian condition}
\label{Schwarzian}

More generally, the Gauss hypergeometric function 
$\,\, _2F_1([\alpha, \, \beta], \, [\gamma], \, x) \, $
is solution of the second order linear 
differential operator\footnote[1]{Note that $\, A(x)$ is 
the log-derivative of 
$\, \, u(x) \, \,= \,\, \, x^{\gamma} \cdot \, (1 \, -x)^{\alpha+\beta+1-\gamma}$.}: 
\begin{eqnarray}
\label{Gaussdiff}
\hspace{-0.95in}&& \quad \quad  \quad \quad 
 \Omega \,\,  = \, \, \, \,  
D_x^2    \,  \, + \, A(x) \cdot \,  D_x  \, \, + \, B(x), 
\quad \quad \quad  \quad  \quad  \quad \quad  \hbox{where:} 
 \\
\hspace{-0.95in}&& \quad    
A(x) \,\,  = \, \, \, 
{{ (\alpha +\beta+1) \cdot  \, x \,  \, -\gamma} \over { x \cdot \, (x\, -1)}}
  \, \, = \, \,  \,  {{u'(x)} \over { u(x)}}, 
\quad  \quad \quad \quad 
B(x) \,\,  = \, \, \,  {{\alpha  \, \beta } \over {x \cdot \, (x\, -1) }}.
\nonumber 
\end{eqnarray}
An automorphy relation, like (\ref{modularform2explici}) but
on $\,\, _2F_1([\alpha, \, \beta], \, [\gamma], \, x)$, amounts to
saying that the  second order linear differential operator (\ref{Gaussdiff}),
pullbacked by $\, x \, \rightarrow \, \, y(x)$, reduces to the conjugate of
the linear differential operator (\ref{Gaussdiff}). Let us assume that
the pullback $\, y(x)$ is an algebraic series like in
(\ref{seriesmodularcurve}) and (\ref{orderthreey}).
A  straightforward calculation~\cite{Schwarz2} 
allows to find the algebraic cofactor $\, {\cal A}(x)$
in terms of the algebraic function pullback $\, y(x)$:
\begin{eqnarray}
\label{modularform3bis}
\hspace{-0.95in}&& \quad   \quad \quad \quad \quad  \quad \quad  \quad \quad
{\cal A}(x)  \, \, = \, \,  \,  
 \lambda \cdot \, \Bigl( {{u(x)} \over { u(y(x)) }}
 \cdot \,  y'(x) \Bigr)^{-1/2}. 
\end{eqnarray}
Expression (\ref{modularform3bis}) for  $\, {\cal A}(x)$       
is such that the two order-two linear differential operators 
(of a similar form as (\ref{Gaussdiff}))
have the same $\, D_x$ coefficient. The identification
of these two  linear differential operators thus corresponds 
(beyond (\ref{modularform3bis})) to {\em just one (non-linear) condition} that can 
be rewritten (after some algebra ...) in the following {\em Schwarzian} form:
\begin{eqnarray}
\label{condition1}
\hspace{-0.95in}&& \quad   \quad \quad \quad  \quad    \quad \quad  \quad 
 W(x)  \, \, \,  \,-W(y(x)) \cdot  \, y'(x)^2
\, \, \,  \,+ \,  \{ y(x), \, x\}  \, \,\, \,  = \,\, \, \,  \, 0, 
\end{eqnarray}
where  
\begin{eqnarray}
\label{wherecond}
\hspace{-0.95in}&& \quad   \quad \quad  \quad \quad  \quad   \quad   \quad 
W(x)  \, \, = \, \,  \, \, \,  A'(x) \, \, \,  + \, \, 
   {{A(x)^2} \over {2 }} \, \,  \, \,  -2 \cdot \, B(x),
\end{eqnarray}
and where $\,\{ y(x), \, x\}$ denotes the 
{\em Schwarzian derivative}~\cite{What}:
\begin{eqnarray}
\label{Schwa}
\hspace{-0.95in}&&  \, \,  \,    \, 
\{ y(x), \, x\}    \, \, = \, \,  \,  \,  \,
{{y'''(x) } \over{ y'(x)}} 
 \,  \,  - \, \, {{3} \over {2}}
 \cdot \, \Bigl({{y''(x)} \over{y'(x)}}\Bigr)^2
 \, \, = \, \,  \,  \,  \,
{{ d } \over { dx  }} \Bigl( {{y''(x) } \over{ y'(x)}}   \Bigr) 
\, \, - {{1} \over {2}} \cdot \, \Bigl( {{y''(x) } \over{ y'(x)}}   \Bigr)^2. 
\end{eqnarray}
For $\, _2F_1\Bigl([{{1} \over {12}}, \, {{5} \over {12}}], \, [1], \,  x)$, the
``automorphy'' condition (\ref{modularform2explici}) yields 
the Schwarzian  condition (\ref{condition1})  with: 
\begin{eqnarray}
\label{whereconda}
\hspace{-0.95in}&&  \quad  \quad  \quad \quad \quad  \quad \quad \quad  \quad  \quad \, \,  
W(x)  \, \,  = \, \,  \,  
-{{32\,{x}^{2}-41\,x+36}  \over {72 \cdot \, {x}^{2} \cdot \, (x\, -1)^2 }}. 
\end{eqnarray}
The algebraic series (\ref{seriesmodularcurve}) and (\ref{orderthreey}), associated with 
different modular equations (like (\ref{modularcurvetau2})),  are both solutions of the {\em same}
Schwarzian  condition (\ref{condition1}) with $\, W(x)$ given by  (\ref{whereconda}). 
These two {\em modular correspondences} series (\ref{seriesmodularcurve}) and (\ref{orderthreey}),
associated with {\em modular curves}, are thus {\em algebraic series}. Consequently, the prefactor 
 (\ref{modularform3bis}) is an {\em algebraic function}.  

\vskip .1cm

\subsection{One-parameter solution series of the Schwarzian condition (\ref{condition1})}
\label{reversibility}

Trying to generalize the modular equation (\ref{modularcurvetau2}), and
its associated algebraic series  (\ref{seriesmodularcurve}), let us try to find the series
of the form $\, a \, \cdot x^2 \, + \, \, \cdots$, solutions of the Schwarzian
equation (\ref{condition1}) with $\, W(x)$ given by (\ref{whereconda}). It
is straightforward to find that such series is,
in fact, the following {\em one-parameter} series:
\begin{eqnarray}
\label{seriesmodcurvea}
\hspace{-0.96in}&&  
 y_2 \, \, = \, \, \,
 a \cdot  \,{x}^{2} \, 
+{\frac {31 \cdot \, a{x}^{3}}{36}}
 \,-{\frac {a \cdot \,
 \left( 5952\,a-9511 \right) }{13824}}\cdot \, {x}^{4}
\,-{\frac {a \cdot \, 
\left( 14945472\,a-11180329 \right) }{20155392}} \cdot \, {x}^{5}
\nonumber \\ 
\hspace{-0.96in}&& \quad \quad  \,  
 \,+{\frac {a \cdot \, 
\left( 88746430464\,{a}^{2}-677409785856\,a+338926406215 \right)
}{743008370688}} \cdot \,  {x}^{6} 
\, \, \,\, + \, \, \cdots 
\end{eqnarray}
which actually reduces to (\ref{seriesmodularcurve}) for
$\, a \, = \, \, 1/1728$.
Similarly, one also finds a one-parameter 
family of solution-series of the Schwarzian condition
(\ref{condition1}) of the form 
$\, \, b \cdot \, x^3 \, + \, \cdots\, \, $, namely
\begin{eqnarray}
\label{seriesmodcurve3a}
\hspace{-0.95in}&& \quad  \, \,   \,
y_3 \, \, \, = \, \, \,  \, \, \, 
b \cdot \, {x}^{3} \, \,  \, 
+{\frac {31\,b }{24}} \cdot \, {x}^{4} \, \,  \, 
+{\frac {36221\,b }{27648}} \cdot \, {x}^{5}  \,  \, \, \, 
-{\frac {b  \cdot \, \left( 23141376\,b-66458485 \right)
}{53747712}} \cdot \,  {x}^{6}
\nonumber \\ 
\hspace{-0.95in}&& \quad  \quad \quad  \quad  \quad   \, 
\, -{\frac {b   \cdot \, (183649959936\,b-187769367601) 
 }{165112971264}}  \cdot \, {x}^{7}
\, \, \,\,  \, + \, \, \cdots 
\end{eqnarray}
which reduces to  (\ref{orderthreey}) for 
$\, b \, = \, \, 1/2985984 \, = \, 1/1728^2$.

\vskip .1cm 

{\bf Remark \ref{int}.1:} Generically the two series
(\ref{seriesmodcurvea}) and (\ref{seriesmodcurve3a})
are differentially {\em algebraic series} (being solution of a Schwarzian condition
(\ref{condition1}), with $\, W(x)$ given by (\ref{whereconda})). 
For {\em selected} values of the parameter, like $\, a \, = \, \, 1/1728$ and $\, b  \, = \, 1/1728^2$,
these series become {\em algebraic series} ({\em correspondences} associated with {\em modular curves}).
Are there other selected values of the parameters for which the series becomes an  algebraic series ?
Are there selected values of the parameters for which the series become (non algebraic)
$\, D$-finite series ?  Are there selected values of the parameters for which the series
become $\, D$-$D$-finite\footnote[1]{$D$-finite functions are solutions
of linear differential operators with polynomial coefficients,  $\, D$-$D$-finite
series are solutions  of linear differential operators with $D$-finite function
coefficients, etc ...} series~\cite{DD,DD1} ? 

\vskip .1cm

\subsection{The nome and mirror maps \\} 
\label{Schwarzmirror}

Let us recall the concept of 
{\em mirror map}~\cite{IsingCalabi,IsingCalabi2,Candelas,Doran,Doran2,LianYau,Kratten} 
relating the reciprocal of the $\,j$-function and the nome, 
with the well-known series with {\em integer}
coefficients\footnote[2]{In Maple the series (\ref{mirror}) 
can be obtained substituting $\, L= \, EllipticModulus(q^{1/2})^2$,
 in $ 1/j \,  = \,  \,$
$ \,L^2 \cdot \,(L-1)^2/(L^2-L+1)^3/256$. See https://oeis.org/A066395 
for the series (\ref{mirror})
and https://oeis.org/A091406 for the series (\ref{mirror2}).} :
\begin{eqnarray}
\label{mirror}
\hspace{-0.95in}&& \quad  \quad 
\tilde{X}(q) \, \, = \, \, \, \, q \,\,\, \, -744\,{q}^{2} 
\,\, +356652\,{q}^{3} \,\, \, -140361152\,{q}^{4} \,\,\, +49336682190\,{q}^{5}
 \nonumber \\ 
\hspace{-0.95in}&& \quad  \quad \quad \, 
-16114625669088\,{q}^{6}  \,\,
\, +4999042477430456\,{q}^{7} \, \, \,\, \, + \, \, \cdots
\end{eqnarray}
and the {\em nome} which is
its compositional  inverse:
\begin{eqnarray}
\label{mirror2}
\hspace{-0.95in}&& \quad 
\tilde{Q}(x) \, \, = \, \, \,  \,
x \,  \, \, +744\,{x}^{2} \, \, +750420\,{x}^{3} 
\, \, +872769632\,{x}^{4} \,  \,
+1102652742882\,{x}^{5} 
\nonumber \\ 
\hspace{-0.95in}&& \quad \quad \,  \, +1470561136292880\,{x}^{6}
 \, +2037518752496883080\,{x}^{7} \, \,\, \, + \, \, \cdots
\end{eqnarray}
The series (\ref{mirror}) corresponds to $\, x$ being the reciprocal of the
$\, j$-function: $\, 1/j$ . As a consequence of the (modular 
form) hypergeometric identities (\ref{modularform2explici}) 
(see  (\ref{Haupt})), 
we need $\, x$ to be identified 
with the {\em Hauptmodul} $\, 1728/j$.

The series $\, X(q) \, = \, \,  1728 \cdot \, \tilde{X}(q)$
(with $\tilde{X}(q)$ given by (\ref{mirror})) is solution of 
the Schwarzian equation:
\begin{eqnarray}
\label{Harnad111}
\hspace{-0.95in}&& \quad  \quad  \quad \quad  \quad 
\{X(q), \, q \} \, \, \,\,   -{{1} \over { 2 \, q ^2}} \, \, \,    \,
 - \, W\Bigl(X(q)\Bigr)
\cdot \,  \Bigl( {{ d X(q)} \over {d q}}  \Bigr)^2
\, \, \, = \,\,\, \, 0. 
\end{eqnarray}
The series  $\, Q(x)  \, = \, \tilde{Q}(x/1728)$
(with $\,\tilde{Q}(x)$ given by (\ref{mirror2})) is solution 
of the Schwarzian equation:
\begin{eqnarray}
\label{Harnad214}
\hspace{-0.95in}&& \quad  \quad  \quad  \quad  \quad 
 \, \{Q(x), \, x \} \, \, \, \, 
 + {{1} \over {2  \cdot \, Q(x)^2 }} 
 \cdot  \Bigl({{ d Q(x)} \over {d x}} \Bigr)^2
\,  \, +  \, W(x) \, \,  \, =  \, \, \, \, \, 0.
\end{eqnarray}
The two mirror map series (\ref{mirror}), (\ref{mirror2}) 
thus correspond to {\em differentially algebraic}~\cite{Selected,IsTheFull} series:
they are solutions of simple (non-linear) Schwarzian equations like in 
(\ref{condition1}).

The two one-parameter series (\ref{seriesmodcurvea}) and (\ref{seriesmodcurve3a})
correspond respectively to:
\begin{eqnarray}
\label{XQ}
\hspace{-0.95in}&& \quad  \quad  \quad  \quad  \quad  \quad \quad
X\Bigl( a \cdot Q(x)^2  \Bigr)     \quad  \quad  \quad \quad \hbox{and:} \quad
\quad  \quad  \quad   X\Bigl( b \cdot Q(x)^3  \Bigr).
\end{eqnarray}
More generally, all the series 
\begin{eqnarray}
\label{XQn}
\hspace{-0.95in}&& \quad  \quad  \quad \quad \quad  \quad \quad \quad \quad \quad \quad
y_n(a, \, x) \,\,  \, = \, \, \, \, X\Bigl( a \cdot Q(x)^n   \Bigr),  
\end{eqnarray}
are solutions of the Schwarzian  condition (\ref{condition1}). For the selected values
$\, a = \, 1/1728^{n-1}$ these series (\ref{XQn})   turn out to be
{\em algebraic series}: they are  series actually associated with  {\em correspondences},
{\em modular curves}.  The composition of two such series is also solution
of the Schwarzian  condition (\ref{condition1}).
One easily finds that
\begin{eqnarray}
\label{compo}
\hspace{-0.98in}&& \, \quad  \quad \quad  \quad  \quad 
y_n\Bigl(a, \, y_m(b, \, x)\Bigr) 
\, \, = \, \, \,  \, y_{m\, n}(a \cdot \, b^n, \, x)
\, \, = \, \, \,  \,  \, a \cdot \, b^n \cdot \, x^{m\, n} \, \,  \,\,  + \, \, \cdots 
\nonumber \\
\hspace{-0.98in}&& \, \quad  \quad \quad  \quad  \quad 
y_m\Bigl(b, \, y_n(a, \, x)\Bigr)
\, \, = \, \, \, y_{m\, n}( b\cdot \, a^m, \, x)
\, \, = \, \, \,  \,  \,  b \cdot \, a^m \cdot \, x^{m\, n} \, \, \, \,  + \, \, \cdots 
\end{eqnarray}
Generically the two series $\, y_n$ and $\, y_m$ {\em do not commute}.

{\bf Remark \ref{int}.2:} Do note that these two series {\em do commute} for the selected
values $\,a=\, 1/1728^{n-1}$ and $\, b= \, 1/1728^{m-1}$, of the parameters $\, a$ and $\, b$,
actually associated with modular
correspondences (algebraic series). In that case, one has the identity:
\begin{eqnarray}
\label{compo}
\hspace{-0.95in}&& \quad  \quad  \quad\quad
a \cdot \, b^n \,\, = \, \, \, b \cdot \, a^m \,\, = \, \, \,
{{1} \over {1728^{n-1}}} \cdot \, \Bigl({{1} \over {1728^{m-1}}}\Bigr)^{n}
\,\, = \, \, \,{{1} \over {1728^{\, m\, n\, -1}}}.
\end{eqnarray}
Also note that if one assumes that the parameters $\, a$ (resp. $\, b$) are of the form
$\, \rho^{n-1}$ (resp. $\, \rho^{m-1}$) with $\, \rho$ different from $\, 1/1728$ or  $\, 1$,
the series $\, y_n(a, \, x)$ and  $\, y_m(b, \, x)$ {\em still commute}\footnote[5]{In terms
  of the nome, this amounts to noticing that transformations
  $\, q \rightarrow \, \alpha^{n-1} \cdot q^n \, $
  and $\, q \rightarrow \, \alpha^{m-1} \cdot q^m \, $  commute.},
even if they are {\em not} algebraic series but {\em only differentially algebraic series}.
The compositional identities (\ref{compo}) are inherited from
the fact that the composition of two algebraic series is an algebraic series, and that 
the  {\em composition of two solutions of  the Schwarzian
  condition} (\ref{condition1})  {\em must}\footnote[1]{This is also a clear
consequence of the automorphy property (\ref{automorph}).}
{\em   also be a solution of the Schwarzian
condition} (\ref{condition1}).  Such properties are reminiscent of the
concept of {\em replicable}
functions~\cite{Replicable,Replicable2,Replicable3,Replicable4,Replicable5,Replicable6,Replicable7,Replicable8}. 

This set of solution series (\ref{XQn}) of the Schwarzian
condition (\ref{condition1}), can also be obtained by the composition
of algebraic series associated with modular {\em correspondences}
(which have no parameter, see (\ref{seriesmodularcurve}), (\ref{orderthreey}), ...),
together with the following  one-parameter series $\, X\Bigl( e \cdot Q(x) \Bigr) \, $
{\em also solution} of the Schwarzian  condition (\ref{condition1}). This series reads:
\begin{eqnarray}
\label{seriesmodcurve1a}
\hspace{-0.95in}&& \, \, \quad \,       \quad   \quad       
y(e, \, x) \, \, \, = \, \, \,  \, 
e \cdot \, x   \, \,\, \, + \, e  \cdot \, (e-1) \cdot \, S_e(x),
 \quad \quad \quad  \, \,    \quad  \quad \hbox{where:} 
\\
\hspace{-0.95in}&& \, \,  \quad   \quad \quad   \quad   \quad \,  \,       
 S_e(x)\, \, = \, \, \, \, 
-{\frac {31}{72}}  \cdot \,  {x}^{2} \, 
\, \, \, +{\frac { (9907\,e -20845) }{82944}} \cdot \,  {x}^{3}
\nonumber \\
\hspace{-0.95in}&& \, \,  \,   \,   \, \, 
\quad \quad    \quad   \quad  \quad    \quad      \quad  
\,\,  -{\frac {
 (4386286\,{e}^{2}-20490191\,e +27274051)
 }{161243136}}\cdot \,  {x}^{4}
\,\,  \,  \, \,+ \, \,\, \cdots 
\end{eqnarray}

\vskip .1cm

{\bf Remark \ref{int}.3:} It is straightforward to see that the series (\ref{seriesmodcurve1a})
is an order-$N$ transformation when the parameter $\, e$ is a $\, N$-th root of unity:
$\, e^N \, = \, \, 1$. These $\, N$-th root of unity are, thus, clearly {\em selected values}
of the parameters.  Are all these $\, N$-th root of unity series
algebraic series, or just D-finite series, or simply  differentially algebraic series ? 

\vskip .2cm 

 \subsection{Multivaled functions and reversibility}
\label{reversibility}

The Landen algebraic transformation (\ref{Landen}) amounts 
to multiplying 
({\em or dividing} because of the 
modular group symmetry $\tau \,\leftrightarrow \, 1/\tau$)
 the ratio $\tau$ of the two periods of the elliptic curves:
 $ \, \,\tau \, \, \longleftrightarrow \, \, 2\, \tau$.  
The other (isogeny) transformations\footnote[9]{See 
for instance (2.18) in~\cite{Canada}.} 
correspond to 
$\tau \,\leftrightarrow \, N \cdot \tau$,
for various integers $\, N$.

We, thus, see that a modular equation, like (\ref{modularcurvetau2}), 
yields {\em multivalued}
functions corresponding to the different series solutions of the
modular equation (for instance (\ref{seriesmodularcurve})
and its compositional inverse). More generally,
for $\tau \,\leftrightarrow \, N \cdot \tau$, we will have series
like $\, \, 1/1728^{N-1} \cdot \, x^N \, \, + \, \, \cdots\, $ and also
(their compositional inverse Puiseux series)
$\, 1728^{(N-1)/N} \cdot \, x^{1/N} \, \, + \, \, \cdots$

In the textbooks the renormalization group is often presented
as a semi-direct group\footnote[8]{In most of the graduate text book on
renormalization group, the critical fixed point is an attractive fixed point.
There is an ``arrow of time''. The renormalization group is seen as an irreversible process. }.
In fact the renormalization group generators have no reason to be such
irreversible transformations. They are, at first sight, {\em reversible transformations}.
The modular equation (\ref{modularcurvetau2}) has a $\, x \leftrightarrow \, y$
symmetric polynomial, corresponding to 
the Landen transformation, {\em  as well as its compositional inverse},  
the inverse Landen transformation. These two transformations are both {\em exact generators} of the
renormalization group of the square Ising model, or of the Baxter model~\cite{Heegner}. With this    
exact renormalization group representation
we see that the modular equation restores, {\em as a consequence of
its  $\, x \leftrightarrow \, y$ symmetry}, the {\em reversible character of the 
 renormalization group}, the price to pay being that the function $\, y(x)$ is actually 
{\em multivalued}.

The Schwarzian condition (\ref{condition1}) encapsulates~\cite{Schwarz1,Schwarz2} an
{\em infinite number of modular correspondences}
associated with their modular curves and modular forms~\cite{Koblitz,1-2-3,Berndt}. In these cases the
automorphy relation (\ref{modularform2explici}) corresponds to
{\em algebraic function} prefactors $\, {\cal A}(x)$. However, for series
with one-parameter, like  (\ref{seriesmodcurvea}) and (\ref{seriesmodcurve3a}),
which are {\em generically} differentially algebraic, we still have
an ``automorphy'' relation (\ref{modularform2explici}), but with
differentially algebraic ``automorphy'' prefactors $\, {\cal A}(x)$. 
We cannot expect a modular equation, but is there a way to still see such transformations 
(\ref{seriesmodcurvea}) and (\ref{seriesmodcurve3a}), as ``correspondences''
with some ``appropriate''
generalization of the concept of correspondences ?  

\vskip .2cm 

 \subsection{Correspondences, Schwarzian conditions and replicable functions}
\label{Questions}

The Schwarzian condition (\ref{condition1})  coincides exactly with 
one of the conditions G. Casale 
obtained~\cite{Casale,Casale2,Casale3,Casale4,Casale5,Casale6,Casale7}
 in a classification of 
Malgrange's $\, {\cal D}$-envelopes and $\, {\cal D}$-groupoids~\cite{Malgrange}  on 
$\mathbb{P}_1$. Denoting $\, y'(x)$, $\, y''(x)$ and $\, y'''(x)$ the 
first, second and third derivative of $\, y(x)$ with respect to $\, x$, 
these conditions\footnote[2]{Casale's condition (\ref{cas2}) is 
{\em exactly the same condition as} the one we already 
found in~\cite{Hindawi}, and this is not a coincidence.} read
respectively\footnote[5]{More generally see 
the concept of differential algebraic invariant of 
isogenies in~\cite{buium}.} 
\begin{eqnarray}
\label{cas2}
\hspace{-0.95in}&& \quad \quad  \quad \quad 
\mu(y) \cdot \, y'(x) \,\,  -\mu(x) \,\,  + \, {{y''(x)} \over{ y'(x)}}
 \, \,\,  = \, \, \, \, 0, 
\\
\label{Casale}
\hspace{-0.95in}&& \quad \quad  \quad \quad 
\nu(y) \cdot \, y''(x)^2 \,  \, -\nu(x) \, \, \, 
+ \, {{y'''(x)} \over{ y'(x)}}  \,\, 
 -{{3} \over {2}} \cdot \,  \Bigl({{y''(x)} \over{ y'(x)}}\Bigr)^2 
 \, \, = \, \,\,  \, 0, 
\end{eqnarray}
together with $\, \gamma(y) \cdot \,  y'(x)^n \, - \, \gamma(x)
\, \, = \, \, \, 0\, \, $ 
and $\, h(y) \, \, = \, \, \, h(x)$, 
corresponding respectively to rank two, rank three, 
together with rank one and rank nul groupo\"ids, 
where $\, \nu(x)$, $\, \mu(x)$, $\, \gamma(x)$ 
are {\em meromorphic} functions ($h(x)$ is holomorph).

\vskip .1cm 

The previous examples of Schwarzian condition (\ref{condition1})
correspond to {\em elliptic curves}
(modular curves, modular forms and modular correspondences), through pullbacked
$\, _2F_1$ hypergeometric functions~\cite{SuperMaier}. In subsection 3.2 of~\cite{Schwarz1} 
we have seen that the Schwarzian condition (\ref{condition1}) can actually occur with
{\em Heun} functions {\em which cannot be reduced to pullbacked}
$\, _2F_1$  {\em hypergeometric functions}\footnote[9]{See for instance
the two Heun functions given by (164) in~\cite{Shimura_Una}.},
and which {\em do not} correspond to globally bounded~\cite{Christol,ChristolUna} 
series. Similarly, we have seen
Schwarzian conditions (\ref{condition1}) corresponding to  (non globally bounded)
pullbacked $\, _2F_1$  hypergeometric functions, associated
with {\em Shimura curves}~\cite{Shimura,Goro}. 
The Malgrange-Casale approach for Schwarzian conditions (\ref{Casale}) suggests
that one should be able to find examples of such Schwarzian conditions
{\em far beyond modular curves, or even Shimura curves}
(and their associated modular forms~\cite{Koblitz,1-2-3,Berndt} and automorphic forms~\cite{Ford}).
If such generalizations exist,  are they also associated with {\em one-parameter} series ?  How
to describe them ? Can they necessarly be seen, eventually,
as generalization of correspondences ? 

\vskip .1cm 

In the next section we will first revisit the previous ``classical'' modular correspondence
results {\em with a different normalization}
of the pullback (see (\ref{modularform2explicit}) below) which makes the occurrence of series
with {\em integer} coefficients crystal clear. Revisiting these calculations with a key role
played by a function $\, F(x)$ defined below by (\ref{whereW}), we will be able to find some
new partial differential equations (see (\ref{wherecondn3}), or (\ref{wherecondn3also}) below),
in the parameter of the series\footnote[1]{See also (\ref{wherecondn3b})
below for more parameters.}. These new equations will help  us to find many examples of
replicable-like~\cite{Replicable2,Replicable3,Replicable4,Replicable5,Replicable6}
functions {\em far beyond modular curves or Shimura curves}~\cite{Shimura,Goro}.

\vskip .1cm 

\section{Recalls}
\label{more}

Some part of this section will be reminiscent of the results explained in~\cite{Schwarz1},
with the difference that we have {\em another normalization} of the pullback,
corresponding to change
$\, x \, \rightarrow \, 1728 \, x$, the ``automorphy'' relation (\ref{modularform2explici}) thus becoming
\begin{eqnarray}
\label{modularform2explicit}
\hspace{-0.95in}&& \quad  \quad \, \, 
 _2F_1\Bigl([{{1} \over {12}}, \, {{5} \over {12}}], \, [1], \, 1728 \cdot \, y  \Bigr)
\,\,  = \, \, \, \,\, 
 {\cal A}(x) \cdot \,
 _2F_1\Bigl([{{1} \over {12}}, \, {{5} \over {12}}], \, [1], \,  1728 \cdot \,  x  \Bigr), 
\end{eqnarray}
where\footnote[5]{Note a typo in (92) in~\cite{Schwarz1}. the exponent $-1/2$ in (92)
must be changed into  $\, 1/2$.}: 
\begin{eqnarray}
\label{modularform3}
\hspace{-0.95in}&& \quad   \quad  \quad  \quad  \quad \quad  \quad \quad  \quad \quad
{\cal A}(x)  \, \, = \, \,  \,  \lambda \cdot \, 
 \Bigl( {{u(x)} \over { u(y(x)) }}
 \cdot \,  y'(x) \Bigr)^{1/2}. 
\end{eqnarray}
As a consequence the (pullback) algebraic series $\, y = \, y(x)$,
corresponding to isogenies like (\ref{seriesmodularcurve}), (\ref{orderthreey}),  ...  
are normalized as $\, \, x \, \,  \longrightarrow \, \,  x^N \,\, + \, \, \cdots$,
and are series with {\em integer} coefficients. 

In our case, taking into account the exact expression of the wronskian, one has
$\, u(x)= \,  x \cdot \, (1 \, -1728\, x)^{1/2}$, and, thus, we get:
\begin{eqnarray}
\label{modularform3ter}
\hspace{-0.95in}&& \quad   \quad \quad  \quad \quad  \quad \quad
{\cal A}(x)  \, \, = \, \,  \,   \lambda \cdot \, 
 \Bigl( {{ x \cdot \, (1 \, -1728\, x)^{1/2} } \over {y \cdot \, (1 \, -1728\, y)^{1/2}  }}
 \cdot \,  y'(x) \Bigr)^{1/2}. 
\end{eqnarray}
Taking the square of (\ref{modularform2explicit}) we can thus rewrite
the ``automorphic'' relation (\ref{modularform2explicit}) as
\begin{eqnarray}
\label{modularform2explicitbis}
\hspace{-0.95in}&& \quad \quad  
\lambda \cdot \,  y \cdot \, (1 \, -1728 \cdot \, y)^{1/2} \cdot \,
 _2F_1\Bigl([{{1} \over {12}}, \, {{5} \over {12}}], \, [1], \, 1728 \cdot \, y  \Bigr)^2
\nonumber \\
 \hspace{-0.95in}&& \quad \quad  \quad  \quad   \quad  
\, = \, \, \, \,\, 
 x \cdot \, (1 \, -1728 \cdot \, x)^{1/2} \cdot \,  
 _2F_1\Bigl([{{1} \over {12}}, \, {{5} \over {12}}], \, [1],
   \,  1728 \cdot \,  x  \Bigr)^2 \cdot \,  {{d y } \over {d x}}, 
\end{eqnarray}
which is, in fact, nothing but
\begin{eqnarray}
\label{modularform2explicitnothingbut}
\hspace{-0.95in}&& \quad \quad \quad \quad \quad \quad \quad \quad \quad \quad \quad \quad \quad
    \lambda \cdot \, {{dx} \over {F(x)}} \, \, = \, \, \,  \,   {{dy} \over {F(y)}}. 
\end{eqnarray}
where $\, F(x)$ reads:
\begin{eqnarray}
\label{Fx}
\hspace{-0.95in}&& \quad \quad \quad 
F(x) \, \, = \, \, \, \, x \cdot \, (1 \, -1728 \cdot \, x)^{1/2} \cdot \,  
_2F_1\Bigl([{{1} \over {12}}, \, {{5} \over {12}}], \, [1], \,  1728 \cdot \,  x  \Bigr)^2
\\
\hspace{-0.95in}&& \quad \quad \quad \quad
\, \, = \, \, \, \, x \,\, -744\,x^2 \, -393768\, x^3
\,\, -357444672\, x^4 \,\, -394896727080\, x^5
\, \,\, + \, \, \, \cdots
\nonumber 
\end{eqnarray}
The elimination of the ``automorphic'' cofactor $\, {\cal A}(x)$
gives the Schwarzian equation on $\, y(x)$ 
\begin{eqnarray}
\label{condition1nbis}
\hspace{-0.95in}&& \quad  \quad  \quad \quad   \quad \quad \quad  
 W(x) 
\, \, \,  \,-W(y(x)) \cdot  \, y'(x)^2
\, \, \,  \,+ \,  \{ y(x), \, x\} 
\, \,\, \, = \,\, \, \,  \, 0, 
\end{eqnarray}
where now
\begin{eqnarray}
  \label{W}
  \,   \quad  \quad 
   W(x)   \, \, \,  = \, \, \,
   - {{1} \over {2}} \cdot \,
   {{ 1 \, -1968\, x  \, +2654208\, x^2} \over { x^2 \cdot \, (1 \, - 1728\, x)^2 }},    
\end{eqnarray}
namely:
\begin{eqnarray}
\label{EQSchwar}
\hspace{-0.95in}&& \quad   \quad  
 - {{1} \over {2}} \cdot \,  {{ 1 \, -1968\, x  \, +2654208\, x^2} \over { x^2 \cdot \, (1 \, - 1728\, x)^2 }}
  \\
\hspace{-0.98in}&& \quad   \quad  \quad \quad \quad
  \, +{{1} \over {2}} \cdot \,
 {{ 1 \, -1968\, y(x)  \, +2654208\, y(x)^2} \over { y(x)^2 \cdot \, (1 \, - 1728\, y(x))^2 }} \cdot  \, y'(x)^2
\, \, \,  \,  \,  +  \{ y(x), \, x\} 
\, \,\, \, = \,\, \, \,  \, 0. \nonumber 
\end{eqnarray}

\vskip .2cm 

\section{Modular equation, modular correspondence}
\label{modularcurve}

\vskip .1cm

\subsection{ $\, q \, \longrightarrow \, \, q^2$ }
\label{modularcurve2}

Let us consider the {\em modular equation}\footnote[1]{Which is nothing but
  (\ref{modularcurvetau2}) with the change of variables 
  $\, x \, \rightarrow  \, \, x/1728$, $\, y \, \rightarrow  \, \, y/1728$. }:
\begin{eqnarray}
\label{modularequa}
\hspace{-0.98in}&& \quad  \quad \quad \quad 
   \Gamma_2(x, \, y) \, \, = \, \, \,
     \, x\, y  \, \, \, - (x+y) \cdot \, (x^2+1487\, x\, y+y^2)  
 \nonumber \\
 \hspace{-0.98in}&& \quad   \quad  \quad \quad \quad \quad \,\,
 +10125\cdot \, x\, y \cdot \, (16\, x^2-4027\, x\, y+16\, y^2)
               \\
 \hspace{-0.98in}&&  \quad   \quad    \quad \quad 
     \, -8748000000 \cdot \, x^2\, y^2  \cdot \, (x+y) \, 
+ 157464000000000 \cdot \, x^3\, y^3
\, \, = \,\, \, \, 0,
\nonumber 
\end{eqnarray}
which has the following rational parametrization~\cite{SuperMaier}:
\begin{eqnarray}
\label{modularequa2ratioo}
  \hspace{-0.98in}&&   \, \,  \quad \quad \quad \quad \quad \quad
 x \, \, = \, \, \, {{ t } \over { (t+16)^3 }}
      \quad \quad  \quad \hbox{and:} \quad \quad \quad \, \,  
 y \, \, = \, \, \, {{ t^2 } \over { (t+256)^3 }}. 
\end{eqnarray}
It has the following {\em algebraic series} solutions with {\em integer} coefficients
\begin{eqnarray}
\label{modularequasol1}
\hspace{-0.95in}&& \quad 
y_2 \,\, = \,  \, \,  \, \,
x^2  \,\, \, +1488\, x^3 \, \, +2053632\, x^4 \,\,
+2859950080\, x^5  \,\, +4062412996608\, x^6
\nonumber \\
\hspace{-0.98in}&& \quad   \quad   \quad  \quad   
\, +5882951135920128\, x^7 \, \,  +8664340079503736832\, x^8 \, 
 \, \,  \, +  \,  \,  \, \cdots 
\end{eqnarray}
and
\begin{eqnarray}
\label{modularequasol2}
\hspace{-0.95in}&& \quad 
  y_{1/2} \,  \, = \, \, \, \, \,
  \omega \cdot \, x^{1/2}  \, \, \, \, -744 \cdot \, x^{2/2}
  \, \,  \, +357024 \cdot \, \omega \cdot \,x^{3/2}
     \,\,   \,  \, -140914688 \cdot \, x^{4/2}
 \nonumber \\
\hspace{-0.98in}&& \quad   \quad    \quad    \quad
 \, \,  \,+49735011840  \cdot \, \omega \cdot \, x^{5/2}
  \,\,  \,\,  -16324041375744 \cdot \, x^{6/2}
 \,  \, \,  \, \, +  \,  \,  \, \cdots 
\end{eqnarray}
where  $\, \, \omega^2 \, = \, 1$ (i.e. $\, \omega \, = \, \pm \, 1$).
These two  algebraic series can be written respectively: 
\begin{eqnarray}
\label{XQrewrit}
  \hspace{-0.95in}&& \,  \, \quad  \quad \quad \quad \quad \quad \, \, \,  
\tilde{X}\Bigl(  \tilde{Q}(x)^2 \Bigr)
\quad \quad \, \quad \, \, \,  \hbox{and:} \quad \, \quad \quad \, \, \, \,
\tilde{X}\Bigl(  \omega \cdot \,  \tilde{Q}(x)^{1/2} \Bigr).
\end{eqnarray}
They amount, respectively, to changing the nome as follows:  $\, q \, \longrightarrow \, \, q^2$, 
{\em together with its compositional inverse} $\, q \, \longrightarrow \, \, \omega \cdot \,  q^{1/2}$,
where  $\, \omega^2 \, = \, 1$.
These two series, (\ref{modularequasol1}) and (\ref{modularequasol2}),
  are actually solutions of the Schwarzian equation (\ref{condition1nbis})
  with $\, W(x)$, now, given by (\ref{W}).
Note that we have the following relation:
\begin{eqnarray}
\label{modularform2explicitfollow}
\hspace{-0.95in}&& \quad \quad   
2 \cdot \,  y_2 \cdot \, (1 \, -1728 \cdot \, y_2)^{1/2} \cdot \,
   _2F_1\Bigl([{{1} \over {12}}, \, {{5} \over {12}}], \, [1], \, 1728 \cdot \, y_2  \Bigr)^2
\nonumber \\
 \hspace{-0.95in}&& \quad \quad  \quad  \quad   \quad  
\, = \, \, \, \,\, 
 x \cdot \, (1 \, -1728 \cdot \, x)^{1/2} \cdot \,  
 _2F_1\Bigl([{{1} \over {12}}, \, {{5} \over {12}}], \, [1], \,
 1728 \cdot \,  x  \Bigr)^2 \cdot \,  {{d y_2 } \over {d x}}. 
\end{eqnarray}
We have a similar relation for $\, y_{1/2}$. Relation (\ref{modularform2explicitfollow}),
and the corresponding one for  $\, y_{1/2}$, are nothing but:
\begin{eqnarray}
\label{modularform2explicitnothingbut}
  \hspace{-0.95in}&& \quad \quad \quad \quad \, \, \, \, \, \,
  2 \cdot \,   {{dx} \over {F(x)}} \, \, = \, \, \,   {{dy_2} \over {F(y_2)}}
     \quad \quad \quad \hbox{and:} \quad \quad  \quad  
    {{1} \over {2}}  \cdot \,   {{dx} \over {F(x)}} \, \, = \, \, \,   {{dy_{1/2}} \over {F(y_{1/2})}}. 
\end{eqnarray}

\vskip .1cm

\subsection{ Linear ODE for $\, q \, \longrightarrow \, \, q^2$ }
\label{modularcurve2sec}

The previous {\em algebraic series}  (\ref{modularequasol1}), (\ref{modularequasol2})
are solutions of an order-three linear differential operator
$\, M_3 \, = \, \, M_1 \oplus \, M_2$
which is the {\em direct sum} (LCLM) of an order-two linear differential operator $\, M_2$,
and
an order-one linear differential operator $\, M_1$ with the following rational function solution:
\begin{eqnarray}
  \label{solM1}
 \hspace{-0.95in}&& \,     
S_1^{(1)}  \, \, = \, \, \, \,
 {{x } \over {496}} \cdot \,
{{(496 \, +13591125 \, x \, +2916000000\, x^2 ) } \over {(1 \, -54000\, x)^3}}
             \\
 \hspace{-0.95in}&& \quad \,  \quad   
  = \, \, \,\,
  x \, \, \, \, +{{93943125} \over {496}} \, x^2 \,\,   \,+{{680168390625} \over {31}} \, x^3
  \,  \, +{{63705259687500000} \over {31}} \, x^4
 \,\, \, \,\, + \, \, \, \cdots  \nonumber
\end{eqnarray}
Let us introduce the two formal solution series of order-two linear differential
operator $\, M_2$, namely:
\begin{eqnarray}
 \label{solM2}
 \hspace{-0.95in}&& \quad \, \, \,    
 S_1^{(2)}  \, \, = \, \, \,\,
   x \,\,  \, +{{46971563} \over {248}} \, x^2 \,  \, +{{680168390718} \over {31}} \, x^3
 \,  \, +{{63705259687628352} \over {31}} \, x^4 \,  \,\, + \, \, \, \cdots 
\nonumber \\
  \hspace{-0.95in}&& \quad \, \, \,
S_2^{(2)}  \, \, = \, \, \,   x^{1/2} \cdot \,
\Bigl( 1 \,  \, \, +357024\, \, x \, \, \, +49735011840 \,\,  x^2 \,  \,+5091284519436288 \,\,  x^3
\nonumber \\
  \hspace{-0.95in}&& \quad \quad \quad \quad  \quad  \quad \quad  \quad \quad \, \, \,  \,
 +445924637193878765568 \, \, x^4 \,\, \, \,+ \,\, \,\cdots   \Bigr)                
 \nonumber          
\end{eqnarray}                  
One has the following relation
\begin{eqnarray}
 \label{relationy2S}
 \hspace{-0.95in}&& \quad \quad  \quad \quad  \quad  \quad  \quad \, \, \,    
 y_2   \, \, = \, \, \,
 496 \cdot \, ( S_1^{(2)} \,  -S_1^{(1)} ),  
\end{eqnarray}     
\begin{eqnarray}
 \label{relationy2S2}
 \hspace{-0.95in}&& \quad  \quad \quad \quad  \quad   \quad  \quad \, \, \,    
 y_{1/2}(\omega, \, x)   \, \, = \, \, \, \, \,  \omega \cdot \, S_2^{(2)} 
\, \, \, \,  -248  \cdot \, ( 2 \, S_1^{(1)} \,  +S_1^{(2)} ), 
\end{eqnarray}     
where $\, \omega^2 \, = \, 1$.

\vskip .1cm

\subsection{ $\, q \, \longrightarrow \, \, q^3$ }
\label{modularcurve3}
Let us consider the {\em modular equation}:
\begin{eqnarray}
\label{modularequa3}
  \hspace{-0.98in}&&   \quad \quad \quad \, \, \, 
1855425871872000000000 \cdot \, x^3\, y^3 \cdot \, (y+x)
\nonumber \\
 \hspace{-0.95in}&&    \quad    \quad   \quad   \quad \quad               \, 
  +16777216000000 \cdot \, y^2\, x^2 \cdot \, (27\, x^2-45946\, x\, y+27\, y^2)
 \nonumber \\
 \hspace{-0.95in}&&   \quad  \quad    \quad  \quad   \quad \quad     \, 
  +36864000 \cdot \, x\, y \cdot \, (y+x) \cdot \, (x^2+241433\, x\, y+y^2)
 \nonumber \\
 \hspace{-0.95in}&& \quad  \quad  \quad   \quad  \quad \quad  \quad           \,  \, 
 +(x^4-1069956\, x^3\, y+2587918086\, x^2\, y^2-1069956\, x\, y^3+y^4) \,
\nonumber \\
 \hspace{-0.95in}&&   \quad  \quad  \quad  \quad  \quad  \quad \quad \quad   \,  \,
+2232 \cdot \, x\, y \cdot \, (y+x) \,  \,  \, \,
-x\, y    \, \, \, = \, \, \, \, \, 0, 
\end{eqnarray}
which has the following rational parametrization~\cite{SuperMaier}:
\begin{eqnarray}
\label{modularequa3ratio}
  \hspace{-0.98in}&&   \, \,  \quad \quad
 x \, \, = \, \, \, {{ t } \over {(t+27) \cdot \, (t+3)^3 }}
      \quad \quad  \quad \hbox{and:} \quad \quad \quad \, \,  
 y \, \, = \, \, \, {{ t^3 } \over {(t+27) \cdot \, (t+243)^3 }}. 
\end{eqnarray}
This  {\em modular equation} (\ref{modularequa3})  has the following {\em algebraic} series solutions
\begin{eqnarray}
\label{modularequasol13}
\hspace{-0.95in}&& \quad 
 y_3 \,  \, = \, \,  \, \, \,
 x^3 \,  \, \,+2232\, x^4 \,   \,\,+3911868\, x^5 \,  \, \,+6380013816\, x^6
 \,  \,\,  +10139542529238\, x^7
 \nonumber \\
\hspace{-0.95in}&& \quad     \quad   \quad   \quad   \quad
 \, +15969813236020944\, x^8 \,   \, \, +25104342383076998772\, x^9
\, \, \,\, \, + \, \, \, \cdots              
\end{eqnarray}
and its compositional inverse
\begin{eqnarray}
\label{modularequasol13b}
\hspace{-0.95in}&& \, \, 
 y_{1/3}(\omega, \, x) \,  \, = \, \,  \,\,\,
\omega \cdot \, x^{1/3} \,\, \, \,-744 \cdot \, \omega^2 \cdot \, x^{2/3} \,\,\, \,
  +356652 \cdot \, x^{3/3}   \,\,\, -140360904 \cdot \, \omega \cdot \, x^{4/3}
\nonumber \\
 \hspace{-0.95in}&& \quad \quad   \quad  \quad  \quad
 +49336313166 \cdot \, \omega^2\, x^{5/3}  \,\, \, \, -16114360320000 \cdot \, x^{6/3}
 \, \, \, \, +  \, \, \, \cdots 
\end{eqnarray}              
where  $\, \omega^3 \, = \, 1$.
The radius of convergence of the series (\ref{modularequasol13}) is
$\, R \, = \, 1/1728$, corresponding to the vanishing of the discriminant of
the modular equation (\ref{modularequa3}).
These two series can be written respectively 
\begin{eqnarray}
\label{XQrewrit3}
\hspace{-0.95in}&& \,  \, \quad  \quad \quad \quad \quad  \quad \quad 
\tilde{X}\Bigl(  \tilde{Q}(x)^3 \Bigr) \quad \quad \quad  \quad
\hbox{and:} \quad \quad \quad \quad   
 \tilde{X}\Bigl(  \omega \cdot \,  \tilde{Q}(x)^{1/3} \Bigr),
\end{eqnarray}
where  $\, \omega^3 \, = \, 1$.
They amount, respectively, to changing the nome as follows:
$\, q \, \longrightarrow \, \, q^3$, together with
its compositional inverse
$\, q \, \longrightarrow \, \, \omega \cdot \,  q^{1/3}$
where  $\, \omega^3 \, = \, 1$.
These two algebraic series, (\ref{modularequasol13}) and (\ref{modularequasol13b}),
are actually solutions of the Schwarzian equation (\ref{condition1nbis}),
with $\, W(x)$ given by (\ref{W}). Note that we have the following relation:
\begin{eqnarray}
\label{modularform2explicitfollow3}
\hspace{-0.95in}&& \quad \,\, \,  
 3 \cdot \,  y_3 \cdot \, (1 \, -1728 \cdot \, y_3)^{1/2} \cdot \,
      _2F_1\Bigl([{{1} \over {12}}, \, {{5} \over {12}}], \, [1], \, 1728 \cdot \, y_3  \Bigr)^2
\nonumber \\
 \hspace{-0.95in}&& \quad \quad  \quad  \quad   \quad  
\, = \, \, \, \,\, 
 x \cdot \, (1 \, -1728 \cdot \, x)^{1/2} \cdot \,  
 _2F_1\Bigl([{{1} \over {12}}, \, {{5} \over {12}}], \, [1], \,
 1728 \cdot \,  x  \Bigr)^2 \cdot \,  {{d y_3 } \over {d x}}. 
\end{eqnarray}
We have a similar relation for $\, y_{1/3}$. Relation (\ref{modularform2explicitfollow3}),
and the corresponding one for  $\, y_{1/3}$,
are nothing but:
\begin{eqnarray}
\label{modularform3explicitnothingbut}
\hspace{-0.95in}&&  \quad \quad \quad \quad \, \, 
3 \cdot \,   {{dx} \over {F(x)}} \, \, = \, \, \,   {{dy_3} \over {F(y_3)}}
\quad  \quad \quad \hbox{and:} \quad \quad \quad \, \, 
    {{1} \over {3}}  \cdot \,   {{dx} \over {F(x)}} \, \, = \, \, \,   {{dy_{1/3}} \over {F(y_{1/3})}}. 
\end{eqnarray}

\vskip .1cm

\subsection{ Linear ODE for $\, q \, \longrightarrow \, \, q^3$ }
\label{modularcurve2sec3}

The previous algebraic series  (\ref{modularequasol13}), (\ref{modularequasol13b})
are solutions of an order-four linear differential operator
$\, M_3 \, = \, \, M_1 \oplus \, M_3$, 
which is the direct sum (LCLM) of an order-three linear differential operator $\, M_3$,
and an order-one linear differential operator $\, M_1$ with the rational function solution
\begin{eqnarray}
\label{modular311}
\hspace{-0.95in}&& \quad \quad \, \, \quad
S_1^{ (1)} \, \, = \, \, \, {{x } \over { 267489}} \cdot \,
{{ p_3(x)} \over { (1 \,  +12288000\, x)^3}}
  \\
\hspace{-0.95in}&&  \quad \quad \quad \, \,  \quad \quad
 \, \, = \, \,\, \,\,\,
 x \,\, \,\,\, -{{447621120000} \over {9907}} \, x^2 \, \,\,\,
 +{{324554085892096000000} \over {267489}} \, x^3
 \, \,\,\, \,\, + \, \, \, \cdots
\nonumber 
\end{eqnarray}
where:
\begin{eqnarray}
\label{modular311p}
\hspace{-0.95in}&& \quad \quad  \quad \quad 
p_3(x) \, \, = \, \, \,\,
 267489 \, \,\, -2225055744000 \,\, x \, \,\, +192711491584000000\, x^2
         \nonumber \\
 \hspace{-0.95in}&& \quad \quad \quad  \quad \quad \quad \quad \quad
 \, -463856467968000000000\, x^3.
\end{eqnarray}
The solutions of the  order-three linear differential operator $\, M_3$ read:
\begin{eqnarray}
\label{modular312}
  \hspace{-0.95in}&& \quad \quad \quad \quad
S_1^{ (2)} \, \, = \, \, \,\,\,
x \,\, \, \,  -{{447621120000} \over {9907}} \,x^2  \,  \,\,
   +{{108184695297365333333} \over {89163}} \, x^3
   \,\, \, \, + \, \, \, \cdots,  
 \nonumber
\end{eqnarray}
\begin{eqnarray}
\label{modular311p2}
\hspace{-0.95in}&& \quad \quad \quad \quad
S_2^{ (2)} \, \, = \, \, \,
x^{2/3} \cdot \,
\Bigl( 1  \,  \,  \, \, -{{8222718861} \over {124}} \, x
\,\,\, +{{62192008621897866} \over {31}} \, x^2
 \nonumber \\ 
\hspace{-0.95in}&& \quad \quad \quad  \quad \quad \quad \quad \quad \quad \quad 
  -{{2837950236255383813660913} \over {62}} \, x^3
 \,    \,\, \, + \, \, \, \cdots \Bigr),              
\end{eqnarray}
\begin{eqnarray}
\label{modular311p2}
\hspace{-0.95in}&& \quad \quad \quad \quad
S_3^{ (2)} \, \, = \, \, \,
x^{1/3} \cdot \,
 \Bigl( 1 \, \, \,  -140360904\, x \, \, \, +4998903239356308\, x^2
 \nonumber   \\
 \hspace{-0.95in}&& \quad \quad \quad  \quad  \quad \quad \quad \quad  \quad 
\,  -122558022956400494032656\, x^3 \, 
 \,  \, \,  + \, \, \, \cdots \Bigr). 
\end{eqnarray}                   
The solutions (\ref{modularequasol13}) and   (\ref{modularequasol13b})
of the modular equation (\ref{modularequa3}) can be expressed in terms of the
 solutions of the previous linear differential operators $\, M_1$ and $\, M_3$
\begin{eqnarray}
\label{modular312y3}
  \hspace{-0.95in}&& \quad \quad \quad \quad \quad \quad \quad
 y_3 \, \, = \, \,  \, \, 267489 \cdot \, (S_1^{(1)} \,  -S_1^{(2)}). 
\end{eqnarray}    
and:
\begin{eqnarray}
\label{modular312y3}
  \hspace{-0.95in}&& \quad  \,  \,  \,
y_{1/3}(\omega, \, x)    \, \, = \, \,  \,  \,\,
\omega \cdot \, S_3^{ (2)}  \,  \, \, \,  -744 \cdot \,   \omega^2 \cdot \,   S_2^{ (2)}
\, \,   \,  \,  +89163 \cdot \, \Bigl(3 \cdot \,   S_1^{(1)}  \,\, +S_1^{(2)}\Bigr).  
\end{eqnarray}   

\vskip .1cm

\subsection{ $\, q \, \longrightarrow \, \, q^5$ }
\label{modularcurve5}

We are not going to give explicitely  the modular equation corresponding to
$\, q \, \longrightarrow \, \, q^5$
because it starts becoming a bit too large. Let us just say that it can (easily) be
obtained by the elimination of $\, t$ in its rational parameterization~\cite{SuperMaier}:
\begin{eqnarray}
\label{modular5}
  \hspace{-0.95in}&& \quad \quad \,  \quad 
  x \, \, = \, \, {{ t} \over { (t^2 +10\,t +5)^3}} \quad \quad \quad
  \hbox{and:} \quad  \quad \quad \, 
 y \, \, = \, \, {{ t^5} \over { (t^2 +250\, t +3125)^3 }}. 
\end{eqnarray}
This modular curve $\, \Gamma_5(x, \, y) = \Gamma_5(y, \, x) = \,  0$,
has the following {\em algebraic series} solutions
\begin{eqnarray}
\label{modularequasol15}
\hspace{-0.95in}&& \quad 
y_5 \,  \, = \, \,  \,\,
x^5 \,\,\, +3720\, x^6 \, \, \,+9287460\, x^7  \,\, \,
 +19648405600\, x^8 \,\, \, +38124922672650\, x^9
\nonumber \\
 \hspace{-0.95in}&& \quad \quad  \quad  \quad 
 \, +70330386411705000\, x^{10} \,\, \, +125698841122545005000\, x^{11}
 \, \, \, \, + \, \, \, \cdots 
\end{eqnarray}
and
\begin{eqnarray}
\label{modularequasol15b}
\hspace{-0.95in}&& \quad 
 y_{1/5} \,  \, = \, \,  \, \,\,\,
 \omega \cdot \, x^{1/5} \,\,  \,  -744 \cdot \, \omega^2 \cdot \, x^{2/5} \,\, \,
+356652\, \omega^3 \cdot \, x^{3/5} \,\, \, - 140361152 \cdot \, \omega^4 \cdot \, x^{4/5}
 \nonumber \\
 \hspace{-0.95in}&& \quad \quad \quad   \quad  \, 
 +49336682190\, \cdot \,   x^{5/5} \, \,\,  -{{80573128344696} \over {5 }} \cdot \, \omega \cdot \, x^{6/5}
 \,\,\, \, \, + \, \, \cdots 
\end{eqnarray}
where  $\, \omega^5 \, = \, 1$.
The series (\ref{modularequasol15}) and (\ref{modularequasol15b}) are (algebraic) solutions of
an order-six linear differential operator $\, L_6 \, = \, \, L_1 \oplus \, L_5$, which is
the direct sum of an order-one linear differential operator with a rational function solution ($p_5(x)$
is a polynomial with integer coefficients)
\begin{eqnarray}
\label{ratio5}
  \hspace{-0.98in}&& \quad\,  \quad \, 
  r(x) \, = \, \, \,
  {{x } \over {41113901825}} \cdot \,
  {{ p_5(x) } \over { 1 \, +654403829760\, x \, +5209253090426880\, x^2 }} 
\\
 \hspace{-0.98in}&& \quad \quad \quad \quad \quad  \quad  \quad 
 \, = \, \, \, \, \,
 x  \,\,\,  \,  \,   \,  -{{4085556703324323840000} \over {1644556073}} \, x^2 \, \,\,  \, 
\,  + \, \, \cdots
 \nonumber 
\end{eqnarray}                 
and an irreducible order-five  linear differential operator operator $\, L_5$.
The solutions of $ \, L_5$ read
\begin{eqnarray}
\label{solL5}
  \hspace{-0.95in}&& \quad \quad    \quad  \quad  \quad  \quad 
  S_0\, = \, \, \, \,  \, x  \, \,\,   \, -{{4085556703324323840000} \over {1644556073}} \, x^2 \,\, 
    \, \,   \, \, + \, \, \, \cdots
\end{eqnarray}
and: 
\begin{eqnarray}
\label{solL5and}
  \hspace{-0.98in}&&
S_1  \, = \, \, x^{1/5} \cdot \,
\Bigl( 1 \, -{{80573128344696} \over {5}} \, x \, \,
 +{{851459104996461085786368168} \over {25}} \, x^2 \, \, + \, \, \, \cdots   \Bigr),
\nonumber \\
\hspace{-0.98in}&& 
S_2  \, = \, \, x^{2/5} \cdot \,
\Bigl( 1 \, -{{3124401548255651} \over {465}} \, x \, \,
 +{{9703780710544581292971588992} \over {775}}  \, x^2 \, \, + \, \, \, \cdots   \Bigr),
\nonumber \\
\hspace{-0.98in}&& 
S_3  \, = \, \, x^{3/5} \cdot \,
 \Bigl( 1 \, -{{621945576635752328} \over {148605}} \, x \,
 +{{31428560280309440232822493239667} \over {4458150}} \, x^2  \Bigr),
\nonumber \\
\hspace{-0.98in}&& 
S_4  \, = \, \, x^{4/5} \cdot \,
\Bigl(  1 \, -{{2163813797006375923833} \over  {701805760}} \, x \,
\\
\hspace{-0.98in}&& \quad \quad \quad \quad \quad  \quad \quad \quad \quad  \quad  \quad 
+{{2096632093647521705592575109262587} \over {438628600}} \, x^2
\, \, + \, \, \, \cdots   \Bigr).
\nonumber                      
\end{eqnarray}  
The series (\ref{modularequasol15}) can be written as a linear combination
of (\ref{ratio5}) and (\ref{solL5}) :
\begin{eqnarray}
\label{solL5combi}
  \hspace{-0.95in}&& \quad \, \quad \quad \quad  \quad  \quad  \quad  \quad 
  y_5    \, \, = \, \, \,   41113901825 \cdot \, \Bigl(r(x) \,  \, - S_0  \Bigr).           
\end{eqnarray}  
The series  (\ref{modularequasol15b}) can be written as a linear combination
of the solutions of  (\ref{solL5and}):
\begin{eqnarray}
\label{solL5andcombi}
\hspace{-0.98in}&& \quad  \quad \, \, 
y_{1/5}    \, \, = \, \, \,\, \,  \,
\omega \cdot \, S_1 \,\, \,\,  -744 \cdot \, \omega^2 \cdot \, S_2 \,\, \, \,
+356652 \cdot \, \omega^3 \cdot \, S_3  \,\,  \, -140361152 \cdot \, \omega^4 \cdot \, S_4
\nonumber \\
\hspace{-0.98in}&& \quad \quad   \quad   \quad  \quad   \quad   \quad  
+8222780365     \cdot \, S_0   \,\, \,\,    +41113901825 \cdot r(x),  
\end{eqnarray} 
where  $\, \omega^5 \, = \, 1$.
The series (\ref{modularequasol15}) and (\ref{modularequasol15b}) can be written respectively 
\begin{eqnarray}
\label{XQrewrit5}
\hspace{-0.95in}&& \,  \, \quad  \quad \,   \, \quad 
y_5 \, = \, \, \tilde{X}\Bigl(  \tilde{Q}(x)^5 \Bigr)
\quad \quad  \quad \, \,  \hbox{and:} \quad \quad \quad    \, \,  \,\,\,
y_{1/5} \, = \, \, \tilde{X}\Bigl(  \omega \cdot \,  \tilde{Q}(x)^{1/5} \Bigr),
\end{eqnarray}
where  $\, \omega^5 \, = \, 1$.
They amount, respectively, to changing the nome as follows:  $\, q \, \longrightarrow \, \, q^5$,
and  its compositional inverse
$\, q \, \longrightarrow \, \, \omega \cdot \,  q^{1/5}$
where  $\, \omega^5 \, = \, 1$.
These two series, (\ref{modularequasol15}) and (\ref{modularequasol15b}),
are actually solutions of the Schwarzian equation (\ref{condition1nbis}),
 with $\, W(x)$ given by (\ref{W}).
Note that we have the following relation:
\begin{eqnarray}
\label{modularform2explicitfollow5}
\hspace{-0.95in}&& \quad \quad \, 
 5 \cdot \,  y_5 \cdot \, (1 \, -1728 \cdot \, y_5)^{1/2} \cdot \,
  _2F_1\Bigl([{{1} \over {12}}, \, {{5} \over {12}}], \, [1], \, 1728 \cdot \, y_5  \Bigr)^2
\nonumber \\
 \hspace{-0.95in}&& \quad \,  \quad  \quad   \quad   \,\, 
\, = \, \, \, \,\, 
 x \cdot \, (1 \, -1728 \cdot \, x)^{1/2} \cdot \,  
 _2F_1\Bigl([{{1} \over {12}}, \, {{5} \over {12}}], \, [1], \,  1728 \cdot \,  x  \Bigr)^2
 \cdot \,  {{d y_5 } \over {d x}}, 
\end{eqnarray}
i.e.
\begin{eqnarray}
\label{modularform2explicitfollow5onF}
  \hspace{-0.95in}&&  \quad \quad \quad \quad \quad \quad \quad \quad \quad \quad
5 \cdot \, F(y_5) \, \, = \, \, \, F(x) \cdot \, {{ d y_5} \over { dx}},
\end{eqnarray}
 and:                    
\begin{eqnarray}
\label{modularform5explicitnothingbut}
  \hspace{-0.95in}&& \quad \quad \quad \quad 
5 \cdot \,   {{dx} \over {F(x)}} \, \, = \, \, \,   {{dy_5} \over {F(y_5)}}
\quad  \quad \quad \, \hbox{and:} \quad \quad \quad \,\,\, \, 
{{1} \over {5}}  \cdot \,   {{dx} \over {F(x)}} \, \, = \, \, \,   {{dy_{1/5}} \over {F(y_{1/5})}}. 
\end{eqnarray}
    {\bf Remark \ref{modularcurve}.1:} The series
    (\ref{modularequasol2}), (\ref{modularequasol13b}),  (\ref{modularequasol15b})
(and also (\ref{modularequasol14b}) below) {\em can be seen to be functions of}
$\,\,\,  \omega \cdot \, x^{1/N}\,\,$  with $\, \, \omega^N \, = \, \, 1$. 

\vskip .2cm

\subsection{ $\, q \, \longrightarrow \, \, q^4$ }
\label{modularcurve4}

We are not going to give explicitely  the modular equation corresponding to
$\, q \, \longrightarrow \, \, q^4$
because it  becomes a bit too large. Let us just say that it can (easily) be obtained
by the elimination of $\, t \, $ in its rational parameterization~\cite{SuperMaier}:  
\begin{eqnarray}
\label{modular4}
  \hspace{-0.95in}&& \quad \quad \,\,
  x \, \, = \, \, {{ t \cdot (t+16) } \over { (t^2 \, +16\,t \, +16)^3}}
      \quad \quad \quad \,\hbox{and:} \quad  \quad \quad \,\,
  y \, \, = \, \, {{ t^4 \cdot \, (t +16) }\over { (t^2 \, +256\, t \, +4096)^3 }}. 
\end{eqnarray}
This {\em modular curve} $\, \Gamma_4(x, \, y) = \Gamma_4(y, \, x) = \,  0\, \, $ can also be obtained
from the elimination of the variable $\, z$ between the  (fundamental) modular equation 
$\, \Gamma_2(x, \, z)  \,  = \,  0$, given by (\ref{modularequa}), and the same modular equation
$\, \Gamma_2(z, \, y)  \,  = \,  0$. The calculation of the resultant, in $\, z$,
between $\, \Gamma_2(x, \, z)$
and $\, \Gamma_2(z, \, y)$ factorizes, and gives $\, (x\, -y)^2 \cdot \, \Gamma_4(x, \, y)$.
This modular curve $\, \Gamma_4(x, \, y) = \Gamma_4(y, \, x) = \,  0$,  has
the following algebraic series solutions
\begin{eqnarray}
\label{modularequasol14}
\hspace{-0.95in}&& \quad \quad 
  y_4 \,  \, = \, \,  \, \,  \, 
  x^4 \,  \,  \, \, +2976 \, x^5 \, \,  \,  +6322896 \, x^6  \, \,  \, +11838151424 \, x^7
  \, \,  \,  +20872495228416\, x^8 
\nonumber \\
 \hspace{-0.95in}&& \quad \quad \quad  \quad  
  +35647177050980352 \, x^9 \, \,  \, +59796357134115627008\, x^{10}
\nonumber \\
 \hspace{-0.95in}&& \quad \quad  \quad  \quad \quad \quad 
\,  +99264875397039869263872 \, x^{11}
 \,  \,  \, \, + \, \, \, \cdots 
\end{eqnarray}
\begin{eqnarray}
\label{modularequasol14b}
\hspace{-0.95in}&& 
 y_{1/4}(\omega, \, x) \,  \, = \, \,   \, \, 
 \omega \cdot \, x^{1/4} \, \,   \,  \, -744 \cdot \, \omega^2 \cdot \, x^{1/2} \, \,  \,
 +356652 \cdot \, \omega^3 \cdot \, x^{3/4}\, \,\,   \,   -140361152 \cdot \, x^{4/4} 
 \nonumber \\
  \hspace{-0.95in}&& \quad \quad \quad \, 
 \, +49336682376 \cdot \, \omega \cdot \, x^{5/4} \, \,  \,
 -16114625945856 \cdot \, \omega^2 \cdot \, x^{6/4} \, \, 
  \nonumber \\
  \hspace{-0.95in}&& \quad \quad  \, \quad \quad \quad   \, 
  +4999042676442272 \cdot \, \omega^3 \cdot \,  x^{7/4} \,\,  \,
  -1492669488513712128 \cdot \, x^{8/4}
 \nonumber \\
 \hspace{-0.95in}&& \quad \quad  \quad \quad \quad \quad \quad \,    \quad   
\, +432762805367932714848 \cdot \, \omega \cdot \, x^{9/4}
          \, \,   \, \, + \, \, \, \cdots 
\end{eqnarray}
where $\, \omega^4 = \, 1$, {\em together} with the ({\em involutive}) series:
\begin{eqnarray}
\label{modularequasol14c}
\hspace{-0.95in}&& \quad 
 y_{1} \,  \, = \, \,  \,  \,
-x  \,  \,   \,\,   \, -1488\, x^2 \,\,   \,  \,  -2214144\, x^3 \, \, \,\, 
-3337633792\, x^4 \, \, \, \,  -5094329942016\, x^5
\nonumber \\
  \hspace{-0.95in}&& \quad \, \,  \quad                 \,
-7859077093785600\, x^6 \, \,\,   \,  -12234039128005541888\, x^7
  \\
  \hspace{-0.95in}&& \quad \quad \, \,  \quad   \,
-19190712499154486034432\, x^8 \, \,  \,  -30301349938167862039412736\, x^9
\, \,  \, + \, \, \, \cdots
\nonumber   
\end{eqnarray}
The radius of convergence of the series (\ref{modularequasol14}), or (\ref{modularequasol14c}),
is $\, R \, = \, 1/1728$, corresponding to the vanishing of the discriminant of
the modular equation  $\, \Gamma_4(x, \, y) = \Gamma_4(y, \, x) = \,  0$.
These three series (\ref{modularequasol14}), (\ref{modularequasol14b}) and (\ref{modularequasol14c}),
can be written respectively 
\begin{eqnarray}
\label{XQrewrit5}
  \hspace{-0.95in}&& \,  \, \quad  \quad
  \tilde{X}\Bigl(  \tilde{Q}(x)^4 \Bigr) \quad \quad \,  \,
  \hbox{and:} \quad \quad  \,  \,
\tilde{X}\Bigl(  \omega \cdot \,  \tilde{Q}(x)^{1/4} \Bigr)
   \quad  \, \,  \quad \hbox{and:} \ \,  \quad \quad    \tilde{X}\Bigl(  -\tilde{Q}(x) \Bigr),  
\end{eqnarray}
where  $\, \omega^4 \, = \, 1$.
These series can be obtain from the series (\ref{modularequasol1}) and
(\ref{modularequasol2}) of subsection (\ref{modularcurve2}).
It is straightforward to see\footnote[1]{The composition/iteration
of {\em multivalued} functions, like algebraic functions, is a bit tricky,
we have, however, no problem to compose {\em algebraic series},
for instance $\, x \, \rightarrow \, y_2(x)\,  \rightarrow \, y_4(x) \, = \, \, y_2( y_2(x))$.}
 that $\, y_4(x) \, = \, y_2(y_2(x))$, and that $\, y_{1/4}(x) \, = \, y_{1/2}(y_{1/2}(x))$, 
which amounts, {\em on the nome}, to performing
$\, q \, \longrightarrow \, q^2 \, \longrightarrow \, (q^2)^2 \, = \, \, q^4$
and similarly
$\, q \, \longrightarrow \,  \pm \, q^{1/2} \, \longrightarrow \, \pm \, (\pm \, q^{1/2})^{1/2} \, = \, \, \omega \cdot \, q^{1/4}$,
where $\, \omega^4 = \, 1$.
However, the composition of $\, y_2$ and $\, y_{1/2}\, $  also corresponds, on the nome, to
\begin{eqnarray}
\label{nome}
\hspace{-0.95in}&& \,  \,  \,  \,   q \, \rightarrow \,  \,
\pm \, q^{1/2} \, \rightarrow \,  \,  (\pm \, q^{1/2})^2 \, = \, \,  q
\quad \quad  \, \hbox{or:} \quad \quad  \,  \, 
q \, \rightarrow \, \,  q^2 \, \rightarrow \, \,   \pm \, (q^2)^{1/2} \, = \, \,  \pm \,  q.  
\end{eqnarray}
Getting rid of the identity transformation, we get $\, q \, \rightarrow \, -q$, which
precisely corresponds to the {\em involutive} series (\ref{modularequasol14c}).
These three series
(\ref{modularequasol14}), (\ref{modularequasol14b}) and (\ref{modularequasol14c}) 
are actually solutions of the Schwarzian equation (\ref{condition1nbis}),
with $\, W(x)$ given by (\ref{W}).
Note that we have the following relation:
\begin{eqnarray}
\label{modularform4explicitfollow4}
\hspace{-0.95in}&& \quad \quad  
 4 \cdot \,  y_4 \cdot \, (1 \, -1728 \cdot \, y_4)^{1/2} \cdot \,
  _2F_1\Bigl([{{1} \over {12}}, \, {{5} \over {12}}], \, [1], \, 1728 \cdot \, y_4  \Bigr)^2
\nonumber \\
 \hspace{-0.95in}&& \quad \quad  \quad  \quad   \quad  
\, = \, \, \, \,\, 
 x \cdot \, (1 \, -1728 \cdot \, x)^{1/2} \cdot \,  
 _2F_1\Bigl([{{1} \over {12}}, \, {{5} \over {12}}], \, [1], \,  1728 \cdot \,  x  \Bigr)^2
 \cdot \,  {{d y_4 } \over {d x}}, 
\end{eqnarray}
\begin{eqnarray}
\label{modularform4explicitnothingbut}
  \hspace{-0.95in}&&  \quad  \quad  \quad   \quad 
4 \cdot \,   {{dx} \over {F(x)}} \, \, = \, \, \,   {{dy_4} \over {F(y_4)}}
   \quad \quad  \quad \quad \hbox{and:} \quad  \quad  \quad  \quad 
{{1} \over {4}}  \cdot \,   {{dx} \over {F(x)}} \, \, = \, \, \,   {{dy_{1/4}} \over {F(y_{1/4})}}
\nonumber \\
 \hspace{-0.95in}&&  \quad \quad  \quad \quad   \quad \,  \quad  \quad  \quad 
  \quad \hbox{and:} \quad  \quad    \quad  \quad 
    {{dx} \over {F(x)}} \, \, = \, \, \,   {{dy_{1}} \over {F(y_{1})}}.
\end{eqnarray}

\vskip .1cm

\subsection{Linear differential operators for $\, q \, \rightarrow \, q^4$}
\label{linearoper}

The previous algebraic series  (\ref{modularequasol14}), (\ref{modularequasol14b})
and  (\ref{modularequasol14c}) 
are solutions of an order-six linear differential operator
$\, M_6 \, = \, \, M_1 \oplus \, M_2 \oplus \, M_3$, 
which is the {\em direct sum} (LCLM) of an order-three linear differential
operator $\, M_3$, an order-three linear differential operator $\, M_2$,
and an order-one linear differential operator $\, M_1$.
Let us introduce $\, S_i^{(n)}$ ($i \, = 1, \cdots \, n$)
the (normalized) solution-series of the linear differential 
operators $\, M_n$ ($n \, = \, 1, 2, 3$), namely the (normalized)
solution of the order-one
linear differential operator $\, M_1$
\begin{eqnarray}
  \label{sol}
\hspace{-0.98in}&&  
 S_1^{(1)} = \, 
  x \, +{{1990225984684950000} \over {187148203}}  \cdot  x^2
  +{{12420842277895932711852000000} \over {187148203}} \cdot  x^3
   + \,  \cdots,  
\nonumber
\end{eqnarray}
together with the two (normalized) solutions of the order-two
linear differential operator $\, M_2$
\begin{eqnarray}
  \label{solmore}               
  \hspace{-0.98in}&& 
S_1^{(2)}  = \, 
x \,  +{{331704330780824752} \over {31191367}} \cdot  x^2
\, \,  +{{2070140379649322118641630976} \over {31191367}} \cdot x^3
 \, \,\,   + \,  \,  \cdots, 
\nonumber \\
  \hspace{-0.98in}&& 
S_2^{(2)} \, = \, \,
x^{1/2} \cdot \,
\Bigl( 1 \,  \, \,+{{671442747744} \over {31}} \cdot \, x \, \,\, \, 
+{{5106946630014945047040} \over {31}} \cdot \, x^2
 \, \,\,  \,  + \, \, \,\cdots  \Bigr),    
 \nonumber
\end{eqnarray}
and  the (normalized) solutions of the order-three
linear differential operator $\, M_3$:
\begin{eqnarray}
  \label{solmore2}
  \hspace{-0.98in}&& 
S_1^{(3)}  = \,  \,  \, 
x \, \,  \,  +1488\, x^2 \, \,  \,  +2214144 \, x^3 \, \,   \,  +3337633793\, x^4
  \, \,   \,  +5094329944992\, x^5
  \,  +  \, \,\cdots, 
\nonumber \\
  \hspace{-0.98in}&& 
S_2^{(3)} \, = \, \,  \,
x^{1/4} \cdot \, \Bigl( 1 \, \,  \, +49336682376\, x \,\,  +432762805367932714848 \, x^2
  \,  \,   \, + \, \, \cdots  \Bigr), 
 \nonumber \\
    \hspace{-0.98in}&& 
 S_3^{(3)} \, = \, \,   x^{3/4} \cdot \, \Bigl( 1 \, \, + {{1249760669110568} \over {89163}} \, x
\,\,  +{{2838197560249922422013408} \over {29721}} \, x^2  \, \, + \, \, \cdots  \Bigr).
\nonumber        
\end{eqnarray}
Note that $\, S_1^{(1)}$, the solution of $\, M_1$, is a rational function:
\begin{eqnarray}
  \label{ratio}
  \hspace{-0.95in}&&  \quad  \, \,   \quad
    S_1^{(1)} \, \, = \, \, \, \,
x \cdot \, {{ p_5(x) } \over { 187148203 \cdot \,
 (1 \,\,  -2835810000\, x \,\,  +6549518250000 \, x^2)^3}},  
\end{eqnarray}
where:
\begin{eqnarray}
  \label{ratiowhere}
  \hspace{-0.95in}&&  \,  
p_5(x) \, \, = \, \, \,
   187148203 \,\,  +398075748036660000 \,\, x \,  \,+4173268788948807866250000 \,\, x^2
\nonumber \\
\hspace{-0.95in}&&  \quad                
 \, +62885546488332818428095703125 \, x^3 \, \,  -31422194354407801042441406250000 \,x^4
\nonumber \\
  \hspace{-0.95in}&&  \quad    \quad    \quad    \quad   \,
 -121645442598919219468125000000000000\, \,x^5. 
\end{eqnarray}
 
{\bf Remark \ref{modularcurve}.2:} The order of the linear operator $\, M_6$, corresponds to the
four series of the form  $\, y_{1/4}$,  together with the series $\, y_4$, and
the series $\, y_1$, namely $\, 6 \, = \, \, 4 \, +1\, +1$. The series $\, y_4$
(given by (\ref{modularequasol14}))
can be seen to be an (algebraic) {\em analytic continuation of the involutive series}
$\, y_1$ (given by (\ref{modularequasol14c})).

{\bf Remark \ref{modularcurve}.3:}  Taking into account
$\, 1\, +\omega\, +\omega^2\, +\omega^3 \, = \, \, 0$, 
let us consider the sum of the four  algebraic series $\, y_{1/4}$. This sum reads:
\begin{eqnarray}
  \label{soly1sur4sumser}
  \hspace{-0.95in}&&    \,  \quad  \, \,     \,
  y_{1/4}(1, \, x) \,  \,  \, + y_{1/4}(\omega, \, x)
  \,  \, + y_{1/4}(\omega^2, \, x) \,\,  \, + y_{1/4}(\omega^3, \, x)
 \nonumber \\                   
 \hspace{-0.95in}&&  \quad   \quad  \, \,    \,
\, \, = \, \, \,  \,
 - 561444608 \, x \, \, \,   \, - 5970677954054848512\, x^2 \, \,
\nonumber \\
\hspace{-0.95in}&&  \quad  \quad  \quad  \quad \,  \, 
  -37262526833687798135553785856\, x^3 \, \\
  \hspace{-0.95in}&&  \quad  \quad \quad  \quad  \quad  \,  \, 
 - 185766744391994261104411840078449475584  \, x^4
 \nonumber \\
\hspace{-0.95in}&&  \quad  \quad  \quad  \quad \quad \quad \,  \, 
  \, - 817583724079763955212555161997757454304107560960 \, x^5 \, 
\,\,  \, + \, \, \, \cdots
 \nonumber       
\end{eqnarray}
We  have the following relations:
\begin{eqnarray}
  \label{soly1sur4sum20}
  \hspace{-0.95in}&&    \,  \quad  \quad \quad  \quad    \,
  y_{1/4}(1, \, x) \,\,\,  + y_{1/4}(\omega, \, x) \,\,\,
  + y_{1/4}(\omega^2, \, x) \,\,\, + y_{1/4}(\omega^3, \, x)
                     \nonumber \\
  \hspace{-0.95in}&&  \quad   \quad  \quad\quad   \quad  \quad  \quad  \quad
 \, \, = \, \, \,  \,
 -187148202 \cdot \, S_1^{(2)} \, \,  \, -374296406  \cdot \, S_1^{(1)}            
\end{eqnarray}
\begin{eqnarray}
  \label{soly1sur4sum21}
  \hspace{-0.95in}&&    \,  \quad  \quad \quad   \quad   \, 
 y_{1/4}(1, \, x) \, \,  \, +\omega \cdot \,  y_{1/4}(\omega, \, x) \, \, \, 
 +\omega^2 \cdot y_{1/4}(\omega^2, \, x) \,\, \, \,  +\omega^3 \cdot y_{1/4}(\omega^3, \, x)
 \nonumber \\
\hspace{-0.95in}&&  \quad   \quad  \quad \quad  \quad  \quad  \quad  \quad 
\, \, = \, \, \,  \,   1426608 \cdot \, S_3^{(3)}, 
\end{eqnarray}
\begin{eqnarray}
  \label{soly1sur4sum22}
  \hspace{-0.95in}&&    \,  \quad  \quad \quad  \quad    \, 
y_{1/4}(1, \, x) \, \, \, +\omega^2 \cdot \, y_{1/4}(\omega, \, x) \, \, \,
    +\omega^4 \cdot \, y_{1/4}(\omega^2, \, x) \,\,  \,\,  +  \omega^6 \cdot \, y_{1/4}(\omega^3, \, x)
 \nonumber \\
  \hspace{-0.95in}&&  \quad   \quad   \quad \quad  \quad  \quad
\, \, = \, \, \,  \,
y_{1/4}(1, \, x) \, \,  - y_{1/4}(\omega, \, x)
\,  \, + y_{1/4}(\omega^2, \, x) \,\, \,  - y_{1/4}(\omega^3, \, x)
\nonumber \\
  \hspace{-0.95in}&&  \quad   \quad  \quad  \quad  \quad    \quad   \quad  \quad  \quad 
\, \, = \, \, \,  \,  -2976 \cdot \, S_2^{(2)}, 
\end{eqnarray}
\begin{eqnarray}
  \label{soly1sur4sum23}
  \hspace{-0.95in}&&   \quad  \,  \quad  \quad  \quad    \,  \, 
y_{1/4}(1, \, x)  \,\, \,\, +\omega^3 \cdot \,  y_{1/4}(\omega, \, x) \,\,\,\,
    +\omega^6 \cdot y_{1/4}(\omega^2, \, x) \,\,\, \,  +\omega^9 \cdot y_{1/4}(\omega^3, \, x)
 \nonumber \\
\hspace{-0.95in}&&  \quad  \quad  \quad  \quad  \quad  \quad  \quad \quad  \quad 
\, \, = \, \, \,  \,    4 \cdot \, S_2^{(3)}. 
\end{eqnarray}
Furthermore, we have the two relations:
\begin{eqnarray}
  \label{soly1sur4furthermore}
  \hspace{-0.95in}&&  \quad  \quad \quad \quad \quad  \quad \quad
187148202 \cdot \, \Bigl(S_1^{(2)} \, - S_1^{(1)}  \Bigr)   \, \, -S_1^{(1)}
  \, \, \,  \, = \, \,  \, \, y_1 \, + y_4,         
\end{eqnarray}
\begin{eqnarray}
  \label{soly1sur4furthermore3}
  \hspace{-0.95in}&&  \quad  \quad \quad \quad \quad  \quad \quad \quad  \quad \quad
S_1^{(3)}  \, \,  \, = \, \,  \, \,   -y_1 \, \, +y_4.
\end{eqnarray}
From (\ref{soly1sur4sum20}) and (\ref{soly1sur4furthermore}) we get
\begin{eqnarray}
  \label{S111}
  \hspace{-0.95in}&&  \quad  \quad \quad \quad \quad \quad \quad \quad
S_1^{(1)}  \, \,  \, = \, \,  \, \,
   -{{1} \over {561444609}} \cdot \, \Bigl( Q_0 \,  \,  \, +y_1\, \, +y_4) \Bigr),                   
\end{eqnarray}
\begin{eqnarray}
  \label{S12}
  \hspace{-0.95in}&&  \quad  \quad \quad \quad \quad \quad \quad \quad
S_1^{(2)}  \, \,  \, = \, \,  \, \,
    -{{1} \over {561444606}} \cdot \, \Bigl( Q_0 \,  \,  \, -2 \, y_1\, \, -2 \, y_4) \Bigr),   
\end{eqnarray}
where:
\begin{eqnarray}
  \label{Q0}
  \hspace{-0.95in}&&   \quad \quad \quad
 Q_0       \, \,  \, = \, \,  \, \,\, \, 
 y_{1/4}(1, \, x) \, \, \,  + y_{1/4}(\omega, \, x)
   \,\,  \,  + y_{1/4}(\omega^2, \, x) \,\, \,  \, + y_{1/4}(\omega^3, \, x).
\end{eqnarray}
We can, thus, express all the six series $\, S_i^{(n)}$, solutions of $\, M_6$,
in terms of the algebraic series
$\, y_1$, $\, y_4$ and $\,  y_{1/4}(\omega^n, \, x)$,
solutions of the modular equation $\, \Gamma_4(x, \, y)  = \,  0$.
Conversely the {\em algebraic series} $\, y_1$, $\, y_4$ and $\,  y_{1/4}(\omega^n, \, x)$
can be expressed in terms of the  six series $\, S_i^{(n)}$.
We have from (\ref{soly1sur4furthermore}) and  (\ref{soly1sur4furthermore3}) 
\begin{eqnarray}
  \label{converse1}
  \hspace{-0.95in}&&  \, \,   \quad \quad  \quad \quad  \quad
y_1 \, \, = \, \, \,\,
93574101 \cdot \, ( S_1^{(2)}  \, -S_1^{(1)}) \, \, \,  \, -{{1} \over {2}} \cdot \, (S_1^{(3)} \, +S_1^{(1)}), 
\end{eqnarray}
\begin{eqnarray}
  \label{converse2}
  \hspace{-0.95in}&&  \quad  \quad \quad  \quad \quad  \quad
y_4 \, \, = \, \, \,\,
93574101 \cdot \, ( S_1^{(2)}\,  -S_1^{(1)}) \, \, \, \,  +{{1} \over {2}} \cdot \, (S_1^{(3)} \, -S_1^{(1)}).
\end{eqnarray}
and from (\ref{soly1sur4sum20}), (\ref{soly1sur4sum21}), (\ref{soly1sur4sum22}), (\ref{soly1sur4sum23}):
\begin{eqnarray}
  \label{soly1sur4}
\hspace{-0.95in}&&    \quad  \quad     \quad \quad
  y_{1/4}(\omega, \, x)  \, \, = \, \, \, \,\,\,
\omega \cdot \, S_2^{(3)} \, \, \, \, -744  \cdot \, \omega^2  \cdot \, S_2^{(2)}
 \, \, \, + \, 356652  \cdot \,\omega^3 \cdot \, S_3^{(3)} \, \,
\nonumber \\
\hspace{-0.95in}&&  \quad  \quad  \quad \quad  \quad  \quad \quad  \quad \quad  
 - {{93574101} \over { 2}} \cdot \, \Bigl( S_1^{(2)}  \, +2 \,  S_1^{(1)}\Bigr) \, \,\,\,\,
 -{{1} \over {2}} \cdot \, S_1^{(1)}.
\end{eqnarray}
The identification of the LHS of the modular equation $\, \Gamma_4(x, \, y)  = \,  0\,$ with the polynomial 
\begin{eqnarray}
 \label{poly}
 \hspace{-0.95in}&&  \quad 
P(y) \, \, =\, \, \,   \Bigl(y \, -y_1\Bigr) \cdot  \,    \Bigl(y \, -y_4\Bigr) 
 \\
\hspace{-0.95in}&&  \quad \,\,
 \times  \,    \Bigl(y \, -y_{1/4}(1, \, x)  \Bigr) \cdot  \,
 \Bigl(y \, -y_{1/4}(\omega, \, x)  \Bigr) \cdot  \, \Bigl(y \, -y_{1/4}(\omega^2, \, x) \Bigr) \cdot  \,
   \Bigl(y \, -y_{1/4}(\omega^3, \, x) \Bigr),    \nonumber              
\end{eqnarray}
gives, straightforwardly, relation (\ref{S111}) together with  relation (\ref{ratio})
and also:
\begin{eqnarray}
 \label{polyprod}
  \hspace{-0.95in}&&  \quad  \quad \quad   \quad   \quad  \,\,
  y_1 \cdot \,   y_4 \cdot \,  y_{1/4}(1, \, x) \cdot \, y_{1/4}(\omega, \, x)
  \cdot \, y_{1/4}(\omega^2, \, x)  \cdot \, y_{1/4}(\omega^3, \, x) 
 \nonumber  \\
\hspace{-0.95in}&&  \quad  \quad \quad  \quad \quad \quad \quad   \, \,  \,\,
 \, \, = \, \, \,\,
{{ 1 } \over {(1 \,\,  -2835810000\, x \,\,  +6549518250000 \, x^2)^3 }}.
\end{eqnarray}
{\bf Remark \ref{modularcurve}.4:} The algebraic series $\, y_1$, $\, y_4$, $\, y_{1/4}(\omega^n, \, x)$, 
solutions of the modular equation $\, \Gamma_4(x, \, y)  = \,  0$, can be expressed
as linear combinations of the solutions of the three linear differential
operators $\, M_n$, $\,\, n = \, 1, \, 2, \, 3$. If one introduces the (finite) Galois group
of the polynomial associated with the modular equation $\, \Gamma_4(x, \, y)  = \,  0$,
and the differential Galois groups of the three  linear differential operators $\, M_n$,
one sees that the relation between these different Galois groups is far from being
straightforward.

\vskip .1cm 

\subsection{More correspondence series}
\label{moreqNcorresp}

Let us display\footnote[5]{For all these examples we used gfun of Bruno Salvy. We used
the following commands: algeqtodiffeq, diffeqtohomdiffeq, de2diffop, algeqtoseries, formal$\_$sols. 
} more correspondence series. More examples of correspondence series
are displayed in \ref{miscellancorresp}. 

$\bullet$ The (algebraic) series 
\begin{eqnarray}
\label{mod5}
  \hspace{-0.98in}&&  
 \tilde{X}\Bigl(\tilde{Q} (x)^5 \Bigr) \, \, =  \, \,  \, \,
 x^5 \, \,\,  +3720 \, x^6 \,\,  +9287460\, x^7 \,\, +19648405600\, x^8 \, \, +38124922672650\, x^9
\nonumber \\
 \hspace{-0.95in}&& \quad \quad   \quad  \quad \quad  \quad  \ \quad  \ \quad  \quad  
\, +70330386411705000\, x^{10}
\,  \,   \,  \, \, + \, \, \, \cdots 
\end{eqnarray}
 is solution of a modular equation $\, \Gamma_5(x, \, y) \, = \,  \Gamma_5(y, \, x) \, = \, 0$, 
that we will not write here, but can easily be obtained from its
rational parametrization~\cite{SuperMaier}:
\begin{eqnarray}
\label{gamma5}
  \hspace{-0.98in}&&    \, \, \,  \quad  \quad  \quad   \quad  \quad  
x \, \, = \, \, \,     {{t} \over {(t^2 \, +10\,t \, +5)^3}},
\quad  \quad  \quad  
y \, \, = \, \, \,    {{t^5} \over {(t^2\,  +250\,t \, +3125)^3}}. 
\end{eqnarray}
This series (\ref{mod5}) is solution of an order-six linear differential operator
$\, L_6 \, = \, \, L_1 \oplus \, L_5$, which is the direct sum of an order-one
linear differential  operator $\, L_1$
with a rational function solution  (\ref{ratio5}), 
and an irreducible order-five
linear differential  operator $\, L_5$.

\vskip .1cm 

$\bullet$ The (algebraic) series 
\begin{eqnarray}
\label{mod6}
  \hspace{-0.98in}&&   \quad   \quad   \quad \quad  
 \tilde{X}\Bigl(\tilde{Q} (x)^6 \Bigr) \, \, =  \, \,  \, \, \, \,
 x^6 \, \,  \, +4464 \, x^7 \, \,  \,+12805560 \, x^8 \,\,  \, \, +30222607872\, x^9
\nonumber \\
\hspace{-0.95in}&& \quad \quad   \quad    \quad
\quad  \quad   \quad  \quad \quad  \quad  \ \quad  \ \quad  \quad  
\, +64062187946172 \, x^{10}  \,  \,   \,  \, + \, \, \, \cdots 
\end{eqnarray}
 is solution of a modular equation $\, \Gamma_6(x, \, y) \, = \,  \Gamma_6(y, \, x) \, = \, 0$,
that will not be written here, but can easily be obtained from its
rational parametrization~\cite{SuperMaier}:
\begin{eqnarray}
\label{gamma6}
\hspace{-0.98in}&&    \, \, \quad  \quad  \quad  \quad  \quad  \quad  
 x \, \, = \, \, \,
 {{t \cdot \, (t+8)^3 \cdot \, (t+9)^2} \over {
     (t+6)^3 \cdot  \, (t^3 \, + 18 \, t^2 \, +84\,t \, +24)^3 }},
\nonumber \\
\hspace{-0.98in}&&    \, \, \quad  \quad  \quad  \quad  \quad \quad   
 y \, \, = \, \, \,
 {{t^6 \cdot \, (t+8)^2 \cdot \, (t+9)^3 } \over {
     (t+12)^3 \cdot  \, (t^3 \, + 252 \, t^2 \, +3888\,t \, +15552)^3}}. 
\end{eqnarray}
This series (\ref{mod6}) is solution of an order-twelve linear differential operator
$\, L_{12} \, = \, \, L_1 \oplus \, L_{11}$, which is the {\em direct sum}
of an order-one linear differential  operator $\, L_1$
with a rational function solution of the form 
\begin{eqnarray}
\label{rat6}
\hspace{-0.98in}&&  \quad  \quad  \quad \quad  \quad \quad  \quad \quad  \quad  \,\, \,
x \cdot \, {{p_{11}(x)} \over { (54000\,x \, -1)^3 \cdot \,
q_3(x)^3}}, 
\end{eqnarray}
where $\, p_{11}(x)$ is a polynomial of degree eleven, and where $\, q_3(x)$ reads
\begin{eqnarray}
\label{denomrat6}
\hspace{-0.98in}&& \,\,\,
1879994705688000000000\, x^3 \, \, -224179462188000000\, x^2 \, \,
+151013228706000\, x \, \, -1,
\nonumber 
\end{eqnarray}
and an order-eleven
linear differential  operator $\, L_{11}$.

\vskip .1cm

$\bullet$ We can also consider
\begin{eqnarray}
\label{mod13}
  \hspace{-0.98in}&&  \, \, \,  
\tilde{X}\Bigl(\tilde{Q} (x)^{13} \Bigr) \, \, =  \, \,  \, \,
x^{13} \,  \, +9672\, x^{14} \,  +52931268\, x^{15} \,
   +216226356320\, x^{16} \,  \, \, + \, \, \, \cdots  
\end{eqnarray}
which is solution of a modular equation\footnote[1]{The  polynomial $\, \Gamma_{13}(x, \, y)$
is of  degree 14 in $\, y$ (or $\, x$).}
$\, \Gamma_{13}(x, \, y) \, = \,  \Gamma_{13}(y, \, x) \, = \, 0$, that we will
not write here, but can easily be obtained from its rational parametrization~\cite{SuperMaier}:
\begin{eqnarray}
\label{gamma13}
  \hspace{-0.98in}&&       \quad  \quad        \quad      
 x \, \,  = \,  \,  \,
{{t} \over {(t^2 \, +5\,t \, +13) \cdot \, (t^4 \, +7\, t^3 \, +20\, t^2 \, +19\,t  \,+1)^3}},
\nonumber \\
\hspace{-0.98in}&&  \quad    \quad        \quad             
y  \, \, = \, \, \,
{{t^{13}} \over {(t^2 \, +5\,t \, +13)
    \cdot \, (t^4 \, +247\, t^3 \, +3380\, t^2 \, + 15379\,t  \,+28561)^3}}.  
\end{eqnarray} 
This series (\ref{mod13}) is solution of an order-fourteen linear differential
operator $\, L_{14} \, = \, \, L_1 \oplus \, L_{13}$, which is the direct sum of
an order-one linear differential operator $\, L_1$
with a rational function solution, and an irreducible order-thirteen
linear differential operator $\, L_{13}$.

\vskip .1cm

$\bullet$ Let us consider
\begin{eqnarray}
\label{mod9}
\hspace{-0.98in}&&  \, \, \,  \, 
 \tilde{X}\Bigl(\tilde{Q} (x)^{9} \Bigr) \, \, =  \, \,  \,\,\, 
 x^9 \, \, \, +6696 \, x^{10} \, \, \, +26681076 \, x^{11} \, \, \,+82647211104\, x^{12}
 \,   \, \, \, + \, \, \, \cdots 
\end{eqnarray} 
which is solution of a modular equation
$\, \Gamma_{9}(x, \, y) \, = \,  \Gamma_{9}(y, \, x) \, = \, 0$, that we will
not write here, but can easily be obtained from its rational parametrization~\cite{SuperMaier}:
\begin{eqnarray}
\label{mod9xy}
  \hspace{-0.98in}&&  \,  
x    = \,  
{{t \cdot \, (t^2 +9\, t \, +27))} \over {
     (t+3)^3 \cdot \, (t^3 \, +9\, t^2 \, +27\,t  \,+3)^3}},
 \, \, \, \,   
y   = \,
{{t^{9} \cdot \, (t^2 +9\, t \, +27))} \over {
    (t+9)^3 \cdot  \, (t^3 \, +243\, t^2 \, +2187\,t  \,+6561)^3}}.
\nonumber 
\end{eqnarray}
The polynomial $\, \Gamma_{9}(x, \, y)$ is of degree $\, 12$ in $\, y$ (resp. in $\, x$).
We thus have twelve algebraic solutions-series of the {\em modular equation}
$\, \Gamma_{9}(x, \, y) \, = \,  \, 0$.
This series (\ref{mod9}) is solution of an order-twelve linear differential operator
$\, L_{12} \, = \, \, L_1 \oplus \, L_{11}$, which is the direct sum of an order-one operator $\, L_1$
with a rational function solution of the form
\begin{eqnarray}
\label{gamma9rat}
\hspace{-0.98in}&&      \, \, \,      \quad    \quad  \quad   \quad  \quad \quad  \quad \quad  \quad
x \cdot \, {{  q_{11}(x)} \over {(12288000\,x \, +1)^3 \cdot \, q_3(x)^3 }}, 
\end{eqnarray}
where $\, q_{11}(x)$ is a polynomial of degree eleven, and where $\, q_3(x)$ reads
\begin{eqnarray}
\label{gamma9ratq3}
\hspace{-0.98in}&& \quad \quad \quad \quad 
3338586724673519616000000000\, x^3
\,\, -3750657365033091072000000\, x^2
\nonumber \\
\hspace{-0.98in}&& \quad \quad \quad \quad \quad  \quad 
\, +1855762905734664192000\,x \,\,\, +1, 
\end{eqnarray}
and an order-eleven linear differential  operator $\, L_{11}$.
The (nine) series which are compositional inverse of the series (\ref{mod9}),
are also solutions of the modular equation $\, \Gamma_{9}(x, \, y) \, = \,  \, 0$, 
read:
\begin{eqnarray}
 \label{mod1sur9}
\hspace{-0.98in}&&  \quad   
\tilde{X}\Bigl(\tilde{Q} (x)^{1/9} \Bigr)
\, \, \,\, =  \, \,  \,\,\,  \,\, 
\omega \cdot \, x^{1/9} \,\, \,  \,   -744 \cdot \, \omega^2 \cdot \, x^{2/9}
\,\, \, \,   +356652 \cdot \, \omega^3 \cdot \, x^{1/3}
\nonumber \\
\hspace{-0.98in}&&  \, \, \,  \quad    
\, -140361152 \cdot \, \omega^4 \cdot \, x^{4/9}\,  \,
+49336682190 \cdot \, \omega^5 \cdot \, x^{5/9}
\, \, -16114625669088 \cdot \, \omega^6 \cdot \, x^{2/3}
 \nonumber \\
  \hspace{-0.98in}&&  \quad  \quad    \quad \quad    \quad  \quad  \quad  \quad
    \, +4999042477430456 \cdot \, \omega^7 \cdot \, x^{7/9}
\,  \,  \, \, \,  + \, \, \, \, \cdots 
\end{eqnarray} 
where $\, \omega^9 \, = \, 1$. These (nine) series  (\ref{mod1sur9}) are
solutions of the order-twelve linear differential operator $\, L_{12}$. 
Note that the (two) {\em order-three} series 
\begin{eqnarray}
\label{mod9-1}
  \hspace{-0.98in}&& \quad  \quad \quad \quad
  y_{\omega}(x)\, \, =  \, \,  \, \, y_{1/3}\Bigl(y_{3}(x) \Bigr)    \, \, \, =  \, \,  \, \,  \, \,
  \omega \cdot \, x \, \, \,  \,  \, -744 \cdot \, \omega \cdot \, (\omega -1) \cdot \, x^2
    \nonumber \\
 \hspace{-0.95in}&& \, \quad \quad \quad \quad  \quad \quad  \,
 +36\cdot \, \omega \cdot  \, (\omega -1)\cdot \, (9907\, \omega -20845)\cdot \, x^3
  \\
 \hspace{-0.95in}&& \,  \quad \quad \quad\quad \quad \quad \quad  \,
 \, -32\cdot \, \omega \cdot \, (\omega -1) \cdot \,
( - 24876477  \, \omega  \, + 22887765 ) \cdot \, x^4
 \,  \, \,\, \,  \, + \, \, \, \cdots 
\nonumber
\end{eqnarray}
where $\, \omega^2 \, + \,\omega \, +1\,  = \, 0$,           
are also solutions of the modular equation $\, \Gamma_{9}(x, \, y) \,  \, = \, 0$,  
and also of the order-twelve operator $\, L_{12}$.  We thus
have $\, 1 \, +2  +\, 9 \, = \, \, 12$
{\em algebraic} solutions of the modular equation
$\, \Gamma_{9}(x, \, y) \,  \, = \, 0$, and solutions of $\, L_{12}$.

\vskip .1cm

$\bullet$ The (algebraic) series
\begin{eqnarray}
\label{mod10}
  \hspace{-0.98in}&&  \, \, \, 
 \tilde{X}\Bigl(\tilde{Q} (x)^{10} \Bigr) \, \, =  \, \,  \,
 x^{10} \,  \,+ 7440 \, x^{11} \,  +32413320 \, x^{12} \,  + 108395513600 \, x^{13} \,      
    \, \, + \, \, \, \cdots 
\end{eqnarray}
is solution of a modular equation
$\, \Gamma_{10}(x, \, y) \, = \,  \Gamma_{10}(y, \, x) \, = \, 0$,  which has the
rational parameterization~\cite{SuperMaier}:
\begin{eqnarray}
\label{mod10param}
  \hspace{-0.98in}&& \quad \quad \, \, \,
x \, \, \, = \, \,
  {{ t \cdot \, (t+4)^5 \cdot \, (t+5)^2 } \over {
 (t^6 +20 \, t^5 +160 \, t^4 +640 \, t^3 +1280 \, t^2 +1040 \, t +80)^3 }}
 \\
\hspace{-0.98in}&&  \, \, \, \quad \quad 
y \, \, \, = \, \,  \,
   {{ t^{10} \cdot \, (t +4)^2 \cdot \, (t +5)^5 } \over {
 (t^6 +260 \, t^5 +6400 \, t^4 +64000 \, t^3 +320000\, t^2 +800000 \, t +800000)^3}}.
  \nonumber 
\end{eqnarray}
The degree of the polynomial in $\, \Gamma_{10}(x, \, y) \,\, = \, 0\, $  in $\, y$
(resp. in $\, x$) is $\, 18$. 
The other (algebraic) series solutions of  $\, \Gamma_{10}(x, \, y) \,\, = \, 0\, $
are the compositional inverse of series (\ref{mod10}), namely
\begin{eqnarray}
\label{mod1sur10}
  \hspace{-0.98in}&& \quad  \quad  \quad  \quad 
  y_{1/10}(x)\,   \, \, \, =  \, \,\,  \, \, 
  \omega \cdot \, x^{1/10} \, \, \, \, \, -744 \cdot \, \omega^2 \cdot \, x^{2/10}
  \, \,\, +356652 \cdot \, \omega^3 \cdot  \, x^{3/10}
     \nonumber \\
 \hspace{-0.95in}&& \quad \quad \quad  \quad \quad  \quad   \,
 \,  - 140361152  \cdot \, \omega^4 \cdot  \, x^{4/10}
 \,  \,\, + 49336682190 \cdot \, \omega^5 \cdot  \, x^{5/10} 
  \\
 \hspace{-0.95in}&& \quad \quad \quad  \quad \quad  \quad \quad\quad \quad \quad \quad  \,
 \,  - 16114625669088  \cdot \, \omega^6 \cdot  \, x^{6/10} \, 
 \,  \,  \,  \, \, + \, \, \, \cdots 
\nonumber
\end{eqnarray}
where $\, \omega^{10} \, = \, \, 1$,
together with\footnote[1]{The series (\ref{mod1sur105sur2}) corresponds to $\, y_{1/2}(y_5(x))$. }
\begin{eqnarray}
\label{mod1sur105sur2}
  \hspace{-0.98in}&&  
  y_{5/2}(x)\,   \, \, \, =  \, \,  \, \, \,
\omega \cdot \, x^{5/2} \, \,\, \,  \, +1860 \cdot \, \omega \cdot \, x^{7/2}
\,\,\, +2913930 \cdot \, \omega \cdot \, x^{9/2} \, \,\, \, -744 \cdot \, x^{10/2} 
\nonumber \\
 \hspace{-0.98in}&& \, \, \, 
   +4404293000 \cdot   \, \omega \cdot \, x^{11/2}
\,  -2767680 \cdot \, x^6    \,  +6624982333875 \cdot   \, \omega  \cdot \, x^{13/2}
 \, \, + \, \, \, \cdots 
\end{eqnarray}
where $\, \omega^{2} \, = \, \, 1$, and
\begin{eqnarray}
\label{mod1sur102sur5}
  \hspace{-0.98in}&& \,\,
  y_{2/5}(x)\,   \, \, \, =  \, \,  \, \, \,\,
 \omega \cdot \, x^{2/5} \, \, \, \, \, -744 \cdot \, \omega^2 \cdot \, x^{4/5}
 \,\,  \, +356652 \cdot \, \omega^3 \cdot  \, x^{6/5}
 \, \,\, +{{ 1488} \over {5 }}  \cdot \, \omega \cdot \, x^{7/5} 
 \nonumber \\
 \hspace{-0.95in}&& \quad \quad \,
 -140361152  \cdot \, \omega^4 \cdot  \, x^{8/5} \, \, \,\,
   -{{ 2214144} \over {5 }}  \cdot \, \omega^2 \cdot \, x^{9/5}
  \,  \,\,  \, +49336682190 \cdot \, \omega^5 \cdot  \, x^{10/5} \,
\nonumber  \\
\hspace{-0.95in}&& \quad \quad \quad \quad\quad \quad \quad \, \,  \,
       +{{1592094528 } \over {5 }} \cdot \, \omega^3 \cdot \, x^{11/5}
 \,  \, \, \, \, \, + \, \, \, \cdots 
\end{eqnarray}
where $\, \omega^{5} \, = \, \, 1$. 
We thus have $\, 1 \, + \,  2 \, +5 \,  + \, 10 \, = \, \, 18 \, $
algebraic solutions of 
$\, \Gamma_{10}(x, \, y) \,\, = \, 0\, $  and of $\, L_{18}$.
The order-eighteen linear differential operator $\, L_{18}$ is the {\em direct sum}
of   an order-seventeen linear differential operator $\, L_{17}$,
and an order-one linear differential operator $\, L_{1}$, which has a
rational function solution,
\begin{eqnarray}
\label{10together5rat}
\hspace{-0.98in}&& \quad  \quad  \quad  \quad  \quad \quad
\quad \quad  \quad \quad  \quad  \quad  \quad 
 x \cdot \, {{p_{17}} \over {  p_6^3}},
\end{eqnarray}
where $\, p_{17}$ is a polynomial of degree 17, where $\, p_6$ reads
\begin{eqnarray}
  \label{18together5ratp6}
  \hspace{-0.98in}&& \quad \quad \quad \quad 
66661978554978958501295319312489107870472732672000\, x^6
\nonumber \\
\hspace{-0.98in}&& \quad \quad \quad \quad \quad  
+62082816308629282586712746552975312469884928000\,x^5
 \nonumber \\
\hspace{-0.98in}&& \quad \quad \quad \quad \quad \quad 
 +21122955530832902270001123584504233628467200\,x^4
\nonumber \\
\hspace{-0.98in}&& \quad \quad \quad \quad \quad \quad \quad   
-233405320133674124312518469774131200\,x^3
\nonumber \\
\hspace{-0.98in}&& \quad \quad \quad \quad \quad \quad \quad \quad 
+32278855882815402576742692253440\, x^2
\nonumber \\
\hspace{-0.98in}&& \quad \quad \quad \quad \quad \quad \quad \quad \quad 
-428244362959801779810720\, x \, \, \, \, +1. 
\end{eqnarray}

\vskip .1cm

$\bullet$ The (algebraic) series
\begin{eqnarray}
\label{mod25}
  \hspace{-0.98in}&&  \,  
 \tilde{X}\Bigl(\tilde{Q} (x)^{25} \Bigr)  \, =  \, \,  
 x^{25} \,  \,+18600 \, x^{26} \,  +184821300 \, x^{27} \,  +1304017532000 \, x^{28} \,      
     \, + \, \, \, \cdots 
\end{eqnarray}
is solution of a {\em modular equation}
$\, \Gamma_{25}(x, \, y) \, = \,  \Gamma_{25}(y, \, x) \, = \, 0$, that we will
not write here, but can easily be obtained from its rational parametrization~\cite{SuperMaier}
\begin{eqnarray}
\label{mod25rat}
  \hspace{-0.98in}&&  \,  \,  \quad \quad \quad \quad   \quad   \quad  \quad  \quad 
 x \,  \,   = \, \, \,
 t \cdot \,  {{p_{5}} \over { p_{10}^3}},
\quad \quad \,  \quad  \quad 
y  \,  \, = \,\, \,
  t^{25} \cdot \,  {{ p_{5}} \over {  q_{10}^3}}, 
\end{eqnarray}
where:
\begin{eqnarray}
\label{mod25ratp10}
  \hspace{-0.98in}&&  \,  \,\,  \,  \quad  
p_{5} \, \, = \, \, \, \, t^4 \, \, +5 \,t^3 \, \, +15\, t^2 \,\,  +25\, t \, \, +25, 
  \\
  \hspace{-0.98in}&&  \,  \, \, \, \quad  
p_{10} \, \, = \, \, \,\, 
t^{10} \,\,  +10\, t^9 \,\,  +55\, t^8 \, \, +200\, t^7 \, \, +525\, t^6 \,\,  +1010\, t^5 \,\,  +1425 \, t^4 
  \nonumber  \\
  \hspace{-0.98in}&&  \,  \, \, \,\quad \quad  \quad \quad \quad   \quad  \,
+1400\, t^3 \,\,  +875\, t^2 \,\,  +250\, t \,\,  +5,   
  \\
\hspace{-0.98in}&&  \,  \, \, \, \quad  
 q_{10} \, \, = \, \, \,\, 
 t^{10} \, \,  +250\, t^9 \,\,  +4375\, t^8 \,\,  +35000\, t^7 \,\,  +178125\, t^6 \,\,  +631250\, t^5 
 \nonumber \\
  \hspace{-0.98in}&&  \, \, \,  \,\quad \,   \quad \,   \quad
   \, +1640625\, t^4 \, \, +3125000\, t^3 \,\,  +4296875\, t^2 \,\,  +3906250\, t \,\,  +1953125.  
\end{eqnarray}
The polynomial in the modular equation
$\, \Gamma_{25}(x, \, y)  \, = \, 0\, $ is of degree $\, 30$ in $\, y$ (resp. in $\, x$), and thus
has thirty algebraic solution series, corresponding to the  series (\ref{mod25}),
together with the  $\, 25$ compositional inverse of series (\ref{mod25}), namely
\begin{eqnarray}
\label{mod25more}
  \hspace{-0.98in}&&  \,  \,  \,  \,  \,  \,  \quad
 \omega \cdot \,     x^{1/25}  \, \, \, \,  -744  \cdot  \,  \omega^2 \cdot \,  x^{2/25} \,\,  \,
 +356652  \cdot  \,  \omega^3 \cdot  \,  x^{3/25}
 \, \,  \,  \, -140361152 \cdot  \,   \omega^4 \cdot  \,  x^{4/25}
  \nonumber \\
  \hspace{-0.98in}&&  \,  \,\quad \quad  \quad  \quad  \quad 
 \, \, +49336682190 \cdot  \,   \omega^5 \cdot  \,  x^{5/25} 
     \,   \,  \, \,\, \, + \, \, \, \cdots  
\end{eqnarray}
where $\, \omega^{25} \, = \, \, 1$, together with
the four (order-five\footnote[1]{The composition of series (\ref{fourorderfive})
  with itself five times gives the identity transformation.}) series
\begin{eqnarray}
\label{fourorderfive}
  \hspace{-0.98in}&&  \,  \,  \quad  \quad   \quad  \quad    \quad  
\, y_{1/5}(y_5(x))\, \, = \, \,  \,   \, \, 
\omega \cdot \, x \, \, \,  \, \, \,  -744 \cdot \, \omega \cdot \, (\omega -1) \cdot \, x^2
 \,\, \, \, \, \, + \, \, \,  \cdots
\end{eqnarray}                  
with $\, \omega^5 \, = \, 1$ but $\, \omega \, \ne \, 1$.
This thus gives
$\, 30 \, = \, \, 1\, +4 \, +25 \, \, \, $ algebraic series.
They are solutions of an order-$30$ linear differential operator
which is the direct-sum of an order-$29$  linear differential operator,
and an order-one linear differential operator with a rational function solution
\begin{eqnarray}
\label{L1order25}
\hspace{-0.98in}&&  \, \quad \quad \quad \quad
\quad \quad \quad \quad \quad \quad \quad \quad \quad
\, x \cdot \, {{ p_{29}} \over { p_{10}^3}},  
\end{eqnarray}     
where $\, p_{29}$ is a polynomial of degree $\, 29$,
and where $\, p_{10}$ is a polynomial of degree $\, 10$.

\vskip .1cm 

\section{The one-parameter series solutions of the Schwarzian equation.}
\label{one-parameterSchwarz}

The Schwarzian equation (\ref{condition1nbis}) has more solutions than the infinite discrete set of
algebraic series (see (\ref{modularequasol1}), (\ref{modularequasol13}), (\ref{modularequasol15}),
(\ref{modularequasol14}), (\ref{mod6}), (\ref{mod13}), ...) corresponding to
{\em modular correspondences}.
One also has a series {\em depending on one parameter}, namely: 
\begin{eqnarray}
\label{family_a}
\hspace{-0.95in}&& \,  \,
 y(a, \, x) \, \, = \, \, \,\, \, \,
  a \cdot \, x \, \, \, \, \, \, -744 \cdot  \, a\cdot  \, (a-1) \cdot  \, x^2 \, \,\, 
  +36 \cdot   \, a \cdot  \, (a-1) \cdot \, (9907\, a-20845) \cdot \, x^3
 \nonumber \\
 \hspace{-0.95in}&& \,  \, \quad 
 \, \, \quad   -32 \cdot  \, a \cdot  \, (a-1) \cdot \, (4386286\, a^2-20490191\, a+27274051) \cdot \, x^4
 \nonumber \\
\hspace{-0.95in}&& \,  \, \quad  \quad  \quad 
  \,\quad   \,
   +6 \cdot  \, a\cdot  \, (a-1) \cdot \, (8222780365\, a^3-61396351027\, a^2
\nonumber \\
\hspace{-0.95in}&& \,  \, \quad \quad \quad \quad  \quad  \quad  \quad  \quad 
 +171132906629\, a -183775457147) \cdot \, x^5
\nonumber
\end{eqnarray}
\begin{eqnarray}
\label{family_a1}
 \hspace{-0.95in}&& \,  \, \, \, \quad 
 -144\cdot  \, a \cdot  \, (a-1)\cdot  \,
 (111907122702\, a^4-1162623833873\, a^3+5000493989295\, a^2
 \nonumber \\
\hspace{-0.95in}&& \,  \, \quad \quad \quad \quad  \quad  
    -10801207072185\, a+10212230113145) \cdot \, x^6 \, \,
\nonumber \\
\hspace{-0.95in}&& \,  \, \quad \,  \,
+8 \cdot \, a\cdot \, (a-1) \cdot \, (624880309678807\, a^5-8367080813672297\, a^4
\nonumber \\
  \hspace{-0.95in}&& \,  \, \quad \quad \quad \quad  \quad  
   +48909476982869878\, a^3-158792594445015178\, a^2 
\nonumber \\
  \hspace{-0.95in}&& \,  \, \quad \quad \quad \quad \quad \quad  \quad  
       +293243568886999823\, a    -254689844062110385) \cdot \, x^7
 \nonumber \\
    \hspace{-0.95in}&& \,  \,\, \, \quad 
 -192 \cdot  \, a\cdot  \, (a-1)\cdot  \, (7774319708776120\, a^6 -127824707491524999\, a^5
\nonumber \\
  \hspace{-0.95in}&& \,  \, \quad \quad \quad  \quad 
 +946950323149342341\, a^4  -4101941044701784034\, a^3
 \\
  \hspace{-0.95in}&& \,  \, \quad \quad \quad  \quad  \quad  \quad  
+11156847890086765926\, a^2-18508096006772656203\, a
\nonumber \\
\hspace{-0.95in}&& \,  \, \quad \quad \quad  \quad  \quad  \quad  \quad \quad    \quad 
+15126379507970624425) \cdot \, x^8
\, \, \,\, \, + \, \, \, \, \, \cdots      
 \nonumber 
\end{eqnarray}
Note that all the algebraic series (\ref{modularequasol14c}), (\ref{mod9-1}),
(\ref{fourorderfive}), (see also (\ref{mod1sur13omega}) below), ... associated with
modular equations, are of the form (\ref{family_a}) where the parameter
is a $\, N$-th root of unity: $\, a^N= \, 1$.

Note that this one-parameter series (\ref{family_a}) is a series of the form
\begin{eqnarray}
\label{form}
\hspace{-0.95in}&& \,  \, \quad  \quad \quad \quad \quad \quad
y(a, \, x) \, \,\, = \, \, \,\,  \,
 a \cdot \, x \, \, \, \,
 +a \cdot  \, (a-1) \cdot  \,   \sum_{n=2}^{\infty} \, P_n(a) \cdot \, x^n,      
\end{eqnarray}
where the polynomials $\,  P_n(a)\,$ are polynomials   of degree $\, n-2 \, $  in the parameter $\, a$, 
{\em with integer coefficients}\footnote[2]{This can be seen as a consequence of the fact that
$\,  y(a, \, \, x) \, = \, \, \tilde{X}\Bigl(  a \cdot \,  \tilde{Q}(x) \Bigr)$, 
  where $\, \tilde{X}(x) \, $ and $\, \tilde{Q}(x) \, $ are actually series
  {\em with integer coefficients} (see (\ref{mirror}) and (\ref{mirror2})).}. 

This one-parameter series  (\ref{family_a}), (\ref{form}) verifies the following composition rule:
\begin{eqnarray}
\label{composition}
\hspace{-0.95in}&& \,  \, \quad  \quad \quad  \quad \quad \quad
 y\Bigl(a,\,  y(a', \, x)\Bigr) \, \, = \, \,\, \,   y\Bigl(a', \, y(a, \, x)\Bigr)
 \, \, = \, \,\, \, y(a\, a', \, \, x).
\end{eqnarray}
These series commute. 
One can verify that this one-parameter series (\ref{family_a}) can, in fact, be written
\begin{eqnarray}
\label{XQ}
  \hspace{-0.95in}&& \,  \,  \, \quad  \quad \quad \quad \quad \quad \quad \quad \quad \quad
 y(a, \, \, x) \, \,\, = \, \,\, \,
 \tilde{X}\Bigl(  a \cdot \,  \tilde{Q}(x) \Bigr),
\end{eqnarray}
where 
\begin{eqnarray}
\label{mirror1}
\hspace{-0.95in}&& \quad \quad 
\tilde{X}(q) \, \, = \, \, \, \,\, q \,\,  \,\, \, -744\,\,{q}^{2} 
\,\, +356652\,\,{q}^{3}  \, \,\, -140361152\,\,{q}^{4} \,\, +49336682190\,\,{q}^{5}
 \nonumber \\ 
\hspace{-0.95in}&& \quad \quad\quad  \quad  \,   \, \, 
-16114625669088\,{q}^{6} \, \,
\, +4999042477430456\,{q}^{7} \, 
 \, \, \, \, + \, \, \cdots
\end{eqnarray}
and\footnote[1]{In Maple the $\, \tilde{X}(q)$ series (\ref{mirror}), (\ref{mirror1}) 
can be obtained substituting $\, L=EllipticModulus(q^{1/2})^2$,
 in $ 1/j \,  = \,  \,$
$ \,L^2 \cdot \,(L-1)^2/(L^2-L+1)^3/256$. See https://oeis.org/A066395 
for the series (\ref{mirror})
and https://oeis.org/A091406 for the series (\ref{mirror2}).} 
its composition  inverse:
\begin{eqnarray}
\label{mirror21}
\hspace{-0.95in}&& \quad  \quad  \quad 
\tilde{Q}(x) \, \, = \, \, \,  \,
x \,  \,  \, \, +744\,{x}^{2} \, \,\, +750420\,{x}^{3} 
\, \, \,  +872769632\,{x}^{4} \,  \,  \,
+1102652742882\,{x}^{5} 
\nonumber \\ 
\hspace{-0.95in}&& \quad \quad \quad  \quad   \,  \, +1470561136292880\,{x}^{6}
 \, +2037518752496883080\,{x}^{7} \,
\,\, \, \, \, + \, \, \cdots 
\end{eqnarray}
The  nome series (\ref{mirror21}) has a radius of convergence
$\, R \, = \, 1/1728 \, \, = \, \, 0.00057870370 \, \cdots$

In the $\, a \, \rightarrow \, 0\,\,\,$ limit
one has
\begin{eqnarray}
\label{compositiona0}
  \hspace{-0.95in}&& 
 \lim_{a \rightarrow \, 0}  {{ y(a, \, \, x) } \over {a}} \, \, = \, \, \, \, \, \,
 x \, \, \,  \,+744 \, x^2\,\, \,  +750420 \, x^3 \,\, \, +872769632 \, x^4
 \,\,\, \,  +1102652742882 \, x^5
\nonumber \\
  \hspace{-0.95in}&& \,  \, \,  \, \, 
 +1470561136292880 \, x^6\,\,  +2037518752496883080 \, x^7\,\,  +2904264865530359889600 \, x^8
\nonumber \\
  \hspace{-0.95in}&& \,  \, \quad  \quad  \quad  \quad  \quad  \quad  
 \, +4231393254051181981976079 \, x^9 \,\,\, \, \,  \, + \,\, \, \cdots 
\end{eqnarray}
which is nothing but the nome series  $\, \tilde{Q}(x)$ given by (\ref{mirror21}). 
In the $\, a \, \rightarrow \, \infty \,\, $ limit one has
\begin{eqnarray}
\label{compositionainfty}
\hspace{-0.95in}&& 
\lim_{a \rightarrow \, \infty}  \, y\Bigl(a, \, \, {{x} \over {a}}\Bigr) \, \, \, = \, \, \, \,\, 
x \, \,\,\,  -744\, x^2 \, \,  \,  \,+356652\, x^3
\,  \, \, -140361152\, x^4 \,\, \, +49336682190\, x^5
\nonumber \\
\hspace{-0.95in}&& \,  \quad  \quad  \, \, 
 \, -16114625669088\, x^6 \,\, \, +4999042477430456 \, x^7 \,\, \, \, -1492669384085015040\, x^8
 \nonumber \\
\hspace{-0.95in}&& \,  \, \quad   \quad  \quad \quad  \quad  \quad  \,
+432762759484818142437\, x^9
\,\, \,\,\,  \, + \,\, \, \cdots 
\end{eqnarray}
which is nothing but $\, \tilde{X}$, the (elliptic modulus) series (\ref{mirror1}).

Let us introduce the ratio of the polynomials in expansion (\ref{form}):
\begin{eqnarray}
\label{formratio}
 \hspace{-0.95in}&& \,  \, \quad  \quad \quad \quad \quad \quad \quad \quad \quad \quad \quad
 R_n(a) \, \, = \, \, \, {{P_n(a) }  \over {P_{n+1}(a) }}.  
\end{eqnarray}
One finds, in the  $\, n \, \rightarrow \, \infty$ and $\, a \, \rightarrow \, 0$ limit,
that the ratio (\ref{formratio}) becomes $\, 1/1728 \, = \, \,  0.00057870 \cdots$
For miscellaneous small values of the parameter $\, a$
one can see, that this ratio (\ref{formratio}) also becomes  $\, 1/1728$ in the
$\, n \, \rightarrow \, \infty\, $ limit.

In the last  $\, n \, \rightarrow \, \infty$ and $\, a \, \rightarrow \, \infty$
limit (\ref{compositionainfty}),
the ratio (\ref{formratio}) becomes\footnote[9]{Obtained with 421 coefficients.}
$\,  -0.004316810242 \, \cdots$ which corresponds to the radius of convergence 
of the series (\ref{mirror}), (\ref{mirror1}). 
This radius of convergence is according to
Vaclav Kotesovec\footnote[5]{See https://oeis.org/A066395
and https://oeis.org/A066395/b066395.txt for the reciprocal of j-function.
See also in~\cite{KrattenthalerRivoal},
$\, Q( \exp( -\, \sqrt{3} \cdot \, \pi ) \, = \, 0$
or $\, J( \exp( -\, \sqrt{3} \cdot \, \pi ) \, = \, 0$,
where $\, Q$ is the Eisenstein series $\, E_4$ and $\, J$ is the Klein modular invariant.}
\begin{eqnarray}
\label{radius}
  \hspace{-0.95in}&& \,  \,  \,  \, \quad  \quad \quad \quad  \quad \quad \quad
 \exp\Bigl( -\, \sqrt{3} \cdot \, \pi \Bigr) \, \, = \, \,  \, \,  0.004333420501 \, \cdots 
\end{eqnarray}
which is reminiscent of the selected values (see equation (55) in~\cite{Heegner}):
\begin{eqnarray}
\label{radiust}
  \hspace{-0.95in}&& \,  \, 
   t \, \, = \, \, \, \exp\Bigl( i\, \pi \, {{1 \, +i \, \sqrt{3}} \over {2}}  \Bigr)
\, \, = \, \, \,   i \cdot \, \exp\Bigl( -\, {{\sqrt{3}} \over {2}} \cdot \, \pi \Bigr) \quad
      \hbox{or:}  \quad
   j\Bigl(  {{1 \, +i \, \sqrt{3}} \over {2}} \Bigr) \, \, = \, \, 0. 
\end{eqnarray}
The nearest to $\, x= \, 0$ singularity of  $\, \tilde{X}$
is thus  $\, x_c = \, t^2 = \,  -\,  \exp( -\, \sqrt{3} \cdot \, \pi)$.
We have seen that the radius of convergence of the series
(\ref{modularequasol14c}) (i.e. $\, a \, =-1$) is
$\, R \, = \, 1/1728$, corresponding to the vanishing of the discriminant of
the modular equation  $\, \Gamma_4(x, \, y) = \Gamma_4(y, \, x) = \,  0$, and more generally,
for $|a|= \, 1$,  one can see that the radius of convergence
of the series (\ref{family_a}),  (\ref{form})
for $\, N$-th root of unity, $\, a^N= \, 1$,  is also\footnote[2]{This also corresponds to
vanishing of the discriminant of the corresponding 
modular equations.} $\, R \, = \, 1/1728$. 

More generally, the  radius of convergence of (\ref{family_a}),  (\ref{form}) corresponds to
the singularities of (\ref{XQ}), namely the $\, x = \, 1/1728\, $ singularity of
$\, \tilde{Q}(x)$, and to the values of $\, x$ such that
$\,\, a \cdot \, \tilde{Q}(x) \, = \, \, - \, \exp\Bigl( -\, \sqrt{3} \cdot \, \pi \Bigr)$, 
which corresponds to the singularity of $\, \tilde{X}(x)$, namely:
\begin{eqnarray}
\label{othersing}
\hspace{-0.95in}&& \,  \quad  \quad  \quad   \quad    \quad  \quad   \quad  \quad \,
x \, = \, \,
\tilde{X}\Bigl( - \, {{1} \over {a}} \cdot \,  \, \exp\Bigl( -\, \sqrt{3} \cdot \, \pi \Bigr)  \Bigr).
\end{eqnarray}
When the parameter $\, a \, $ is large enough ($|a|  >   \,  \simeq  7.5$), the radius of convergence
no longer corresponds to $\, R = \, 1/1728$, but to
the singularity (\ref{othersing}).

This transcendental value (\ref{radius}), for the radius of convergence of the series $\, \tilde{X}(q)$, 
is a strong incentive to understand the ``very nature'' of the one-parameter series (\ref{family_a}),
(\ref{form}), especially since it can be written in the simple form (\ref{XQ}). Generically the
one-parameter series (\ref{family_a}), being solution of a Schwarzian equation, is a
{\em differentially algebraic series}, but is it possible
that this series could be, only {\em for selected values}
of the parameter, an algebraic series, or just a $\, D$-finite series,
or possibly a  $\, D$-$D$-finite series ? 

\vskip .1cm 

\section{Trying to understand the one-parameter series solutions.}
\label{understand}

\vskip .1cm 

\subsection{When the one-parameter series becomes an algebraic series}
\label{understandalge}

For $\, a \, = \, \, -1\, $ the ({\em involutive}) series $\,  y(a, \, \, x)$
(see series (\ref{modularequasol14c})) 
\begin{eqnarray}
\label{involu}
  \hspace{-0.95in}&& \,  \, \quad  \quad \quad
  -x \, \,\,\,\, -1488\, x^2 \,\, \, \, -2214144\, x^3 \,\, \, \,
  -3337633792\, x^4 \,\, \, \,  -5094329942016\, x^5
\nonumber \\
\hspace{-0.95in}&& \,  \, \quad  \quad  \quad \quad
  \,\,\,\,\,\, -7859077093785600 \, x^6  \, \, \,  \, -12234039128005541888\, x^7
    \, \,  \,  \, \,\, + \, \,\, \cdots 
\end{eqnarray}
has a radius of convergence $\, 1/1728 \, = \, \,  0.00057870 \cdots$
Let us generalize what we have seen in subsection (\ref{modularcurve4}) with series (\ref{modularequasol14c}). 
Let us first recall the algebraic series (corresponding to $\, q \, \rightarrow \, q^3$)
$\, y_3$, given by (\ref{modularequasol13}), and $\, y_{1/3}$,  given by (\ref{modularequasol13b}),
where $\, \omega^3 \, = \, 1$,  and 
 combine $\, y_{3} $ and  $\, y_{1/3} $. We first get: 
\begin{eqnarray}
\label{mod1sur3}
  \hspace{-0.98in}&& \quad  \quad \quad  \quad   \quad   \quad \quad  \quad \quad  \quad \quad \quad
  y_{3}\Bigl(y_{1/3}(x)\Bigr)   \,  \, \, =  \, \,  \, \, x.       
\end{eqnarray}
More interestingly we also get the following algebraic series (see (\ref{mod9-1}) previously): 
\begin{eqnarray}
\label{mod1sur3omega}
  \hspace{-0.98in}&& \quad  \quad \quad \quad \, \,  \,  \, 
 y_{\omega}(x)\, \, =  \, \,   \, \, y_{1/3}\Bigl(y_{3}(x) \Bigr)
 \, \, \, =  \, \,\,   \,\, \,  \, \, 
  \omega \cdot \, x \, \, \, \, \, \, -744 \cdot \, \omega \cdot \, (\omega -1) \cdot \, x^2
    \nonumber \\
 \hspace{-0.95in}&& \quad \, \, \quad \quad\quad   \quad \quad  \,
 +36\cdot \, \omega \cdot  \, (\omega -1)\cdot \, (9907\, \omega\, -20845)\cdot \, x^3
  \\
 \hspace{-0.95in}&& \quad \, \, \quad \quad\quad \quad \quad \quad \quad  \,  \,
 \, -32\cdot \, \omega \cdot \, (\omega -1) \cdot \, (22887765 \,  - 24876477  \, \omega ) \cdot \, x^4
 \,  \,  \, \,  \, \,  \, + \, \, \, \cdots 
   \nonumber
\end{eqnarray}
where $\, \omega^3 \, = \, 1$. One can verify
that series (\ref{mod1sur3omega}) is actually series (\ref{family_a})
when  $\, a^3\, = \, 1$. One can verify that this series
is (for $\, \omega \, \ne \, 1$) a series of order $\, 3$:
\begin{eqnarray}
\label{mod1sur13order}
  \hspace{-0.98in}&& \quad  \quad \quad \quad \quad \quad \quad \quad \quad \quad \quad
  y_{\omega}( y_{\omega}(    y_{\omega}(x)))   \,  \, \, = \,  \, \, \, x.
 \end{eqnarray}

\vskip .1cm

Let us also recall the algebraic series (corresponding to $\, q \, \rightarrow \, q^5$)
$\, y_5$, given by (\ref{modularequasol15}), and its compositional inverse $\, y_{1/5}$, 
given by (\ref{modularequasol15b}),
where $\, \omega^5 \, = \, 1$,  and let us compose $\, y_{5} $ and  $\, y_{1/5} $.
We first get:
\begin{eqnarray}
\label{mod1sur5}
  \hspace{-0.98in}&& \quad  \quad \quad  \quad   \quad   \quad \quad  \quad \quad  \quad \quad \quad
  y_{5}\Bigl(y_{1/5}(x)\Bigr)    \, \, =  \, \,  \, \, x.       
\end{eqnarray}
More interestingly, we also get the following series (see series (\ref{fourorderfive}) previously): 
\begin{eqnarray}
\label{mod1sur5omega}
  \hspace{-0.98in}&& \quad  \quad \quad \quad
  y_{\omega}(x)\, \, =  \, \,  \, \, y_{1/5}\Bigl(y_{5}(x) \Bigr)
  \, \, \, =  \, \, \,   \, \,  \, 
  \omega \cdot \, x \, \,  \, \,  \, -744 \cdot \, \omega \cdot \, (\omega -1) \cdot \, x^2
    \nonumber \\
 \hspace{-0.95in}&& \quad \quad \quad  \quad \quad \quad \quad  \quad \quad  \,
 +36\cdot \, \omega\cdot  \, (\omega -1)\cdot \, (9907\, \omega -20845)\cdot \, x^3
 \nonumber \\
 \hspace{-0.95in}&& \quad \quad \quad \quad \quad \quad \quad \, \quad
\, -32\cdot \, \omega \cdot \, (\omega -1) \cdot \,
(4386286\, \omega ^2-20490191\, \omega +27274051) \cdot \, x^4
   \nonumber \\
  \hspace{-0.95in}&& \quad \quad  \quad \,\quad \quad
  +6 \cdot \, \omega \cdot \, (\omega -1)\cdot \, (8222780365\, \omega ^3 -61396351027\, \omega ^2
\nonumber \\
  \hspace{-0.95in}&& \quad \quad  \quad \quad \quad \quad \quad \quad  \quad \,\quad \quad  \quad 
    +171132906629\, \omega -183775457147)\cdot \, x^5
 \nonumber \\
  \hspace{-0.95in}&& \quad \quad \,\quad
  -144\cdot \, \omega \cdot \,
    (\omega -1)\cdot \, (-1274530956575 \, \omega ^3 +4888586866593 \, \omega ^2
 \nonumber \\
  \hspace{-0.95in}&& \quad \quad \quad \quad \quad \,\quad \, \,  \quad 
 -10913114194887\, \omega +10100322990443) \cdot \, x^6
     \, \,  \,  \,\, \, \, + \, \, \cdots 
\end{eqnarray}
where $\, \omega^5 \, = \, 1$.

One can verify that (\ref{mod1sur5omega})   is actually (\ref{family_a}) when  $\, a^5 \, = \, 1$.
One can verify that this series is (for $\, \omega \, \ne \, 1$) a series of order $\, 5$:
\begin{eqnarray}
\label{mod1sur15order}
  \hspace{-0.98in}&& \quad  \quad \quad \quad \quad \quad \quad \quad \quad \quad
y_{\omega}( y_{\omega}(    y_{\omega}(   y_{\omega}(    y_{\omega}(x)))))
    \,  \, \, = \,  \, \, \, x.
\end{eqnarray}
This is a straight consequence of (\ref{XQ}) with $\, a^5 = \, 1$.
Similarly, let us now consider
\begin{eqnarray}
\label{mod13bis}
  \hspace{-0.98in}&& \quad  \, \,  \, \, 
 y_{13}  \, \, =  \, \,  \,  \tilde{X}\Bigl(\tilde{Q} (x)^{13} \Bigr) \, \, =  \, \,  \,  \, \,  \,  
x^{13} \,  \,   \,  +9672\, x^{14} \,  \,   \,  +52931268\, x^{15}  \, \,    \, +216226356320\, x^{16}
\nonumber \\
\hspace{-0.95in}&& \quad \,  \quad  \quad  \quad   \, \, \, \,
 +735033166074714\, x^{17} \,   \,  +2200510278533887632\, x^{18}
  \, \,   \,   \,  \, + \cdots 
 \end{eqnarray}            
Its compositional inverse (Puiseux) series reads
\begin{eqnarray}
\label{mod1sur13}
  \hspace{-0.98in}&& \quad  \, \,   \, \, 
y_{1/13}  \, \, =  \, \,  \, \tilde{X}\Bigl(\tilde{Q} (x)^{1/13} \Bigr)
    \, \, =  \, \,\,  \,\,\,
\omega \cdot \,  x^{1/13} \,\, \,\,\,  -744 \omega^2 \cdot \,  x^{2/13}
\, \, \,  +356652 \cdot \,  \omega^3 \cdot \,  x^{3/13} 
\nonumber \\
\hspace{-0.95in}&& \quad \quad  \quad \quad  \quad \quad
 \, \, -140361152  \cdot \,  \omega^4 \cdot \,  x^{4/13}
    \, \, \, +49336682190  \cdot \,  \omega^5 \cdot \,  x^{5/13}
 \nonumber \\
  \hspace{-0.95in}&& \quad \quad  \quad \quad \quad  \quad  \quad  \quad \quad   \, \,  \, \,
 -16114625669088  \cdot \,  \omega^6 \cdot \,  x^{6/13}
      \, \, \, \, \,  \, \,+ \, \, \, \cdots 
 \end{eqnarray}   
where $\, \omega ^{13} \, = \, \, 1$. 
Let us compose $\, y_{13} $ and  $\, y_{1/13} $. We first get 
\begin{eqnarray}
\label{mod1sur13}
  \hspace{-0.98in}&& \quad  \quad \quad  \quad   \quad   \quad \quad  \quad \quad  \quad \quad \quad
  y_{13}\Bigl(y_{1/13}(x)\Bigr)    \, \, =  \, \,  \, \, x,       
\end{eqnarray}
which corresponds to: 
$\,  \,  \, \, q \,  \, \longrightarrow \,\,   \, \, \omega \,  q^{1/13}$
$ \, \,  \longrightarrow \, \,\,   \,\Bigl(\omega \,  q^{1/13}\Bigr)^{13} \, \, = \, \,\,  \, q$.

More interestingly, we also get the following algebraic series:
\begin{eqnarray}
\label{mod1sur13omega}
\hspace{-0.98in}&& \quad  \quad
y_{\omega}(x)\, \, =  \, \,  \, \, y_{1/13}\Bigl(y_{13}(x) \Bigr)
 \, \, \, =  \, \,  \, \,  \, \, 
  \omega \cdot \, x \, \, \, \,  \, \,  -744 \cdot \, \omega \cdot \, (\omega -1) \cdot \, x^2
\nonumber \\
\hspace{-0.95in}&& \quad \quad \quad \quad  \,
 +36\cdot \, \omega\cdot  \, (\omega -1)\cdot \, (9907\, \omega -20845)\cdot \, x^3
  \nonumber \\
  \hspace{-0.95in}&& \quad \quad \quad \quad \quad \quad  \,
  \, -32\cdot \, \omega \cdot \, (\omega -1)
  \cdot \, (4386286\, \omega ^2-20490191\, \omega +27274051)\cdot \, x^4
   \nonumber \\
  \hspace{-0.95in}&& \quad \quad  \quad \,\quad 
 +6\cdot \, \omega \cdot \, (\omega -1)\cdot \, (8222780365\, \omega ^3-61396351027\, \omega ^2
\nonumber \\
  \hspace{-0.95in}&& \quad \quad \quad  \quad \quad  \quad \,\quad \quad  \quad 
    +171132906629\, \omega -183775457147)\cdot \, x^5
 \nonumber \\
  \hspace{-0.95in}&& \quad  \,  \, \, \,\,
  -144\cdot \, \omega \cdot \,
  (\omega -1)\cdot \, (111907122702\, \omega^4
   -1162623833873\, \omega^3 +5000493989295\, \omega^2
 \nonumber \\
  \hspace{-0.95in}&& \quad \quad \quad  \quad \quad  \quad \quad \,
 -10801207072185\, \omega +10212230113145) \cdot \, x^6
 \,\, \,  \,  \,\, \, + \, \, \cdots 
\end{eqnarray}
where $\, \omega ^{13} \, = \, \, 1$.  This series corresponds to
$\, q \, \rightarrow \, q^{13} \, \rightarrow \, \omega \cdot \, (q^{13})^{1/13} \, = \, \omega \cdot \, q$.
One can verify that this series is (for $\, \omega \, \ne \, 1$) a series of order $\, 13$:
\begin{eqnarray}
\label{mod1sur13order}
  \hspace{-0.98in}&& \quad  \quad \quad
 y_{\omega}( y_{\omega}( y_{\omega}( y_{\omega}( y_{\omega}( y_{\omega}( y_{\omega}( y_{\omega}( y_{\omega}(y_{\omega}( y_{\omega}( y_{\omega}( y_{\omega}( x))))))))))))))
   \,  \, \, = \,  \, \, \, x.
 \end{eqnarray}
 If we compare the series (\ref{mod1sur13omega}) with the one-parameter series (\ref{family_a}), we see that
\begin{eqnarray}
\label{mod1sur13order}
  \hspace{-0.98in}&& \quad  \quad \, \quad \quad \quad \quad \, \, 
y_{\omega}(x)\, \, =  \, \,  \, \,  y(\omega, \, x)
\quad \quad \quad \quad \hbox{where:}
\quad \quad   \quad \, \, \,  \quad  \omega^{13} \, = \, \, 1. 
\end{eqnarray}
Conversely, denoting  $ y(a, \, x) \, \, = \, \, \, y_a(x)\, $
we see that the $\, N$-times  composition of $\, y_a(x)$
becomes the identity transformation when $\,   \omega^{N} \, = \, \, 1$: 
\begin{eqnarray}
\label{mod1sur13orderN}
  \hspace{-0.98in}&& \quad  \quad \quad \quad
 y_{a}( y_{a}( \, \cdots \, ( y_{a}( x)) \, \cdots \, )) \, \, = \, \, \, x    \,  \, 
  \quad \quad  \quad  \iff \quad \quad  \, \quad \omega^{N} \, = \, \, 1.
 \end{eqnarray}
The series (\ref{mod1sur13omega})  is the solution series of the {\em modular equation}
associated with the nome-transformation $\, q \, \rightarrow \, q^{13^2}$,
 corresponding to the composition of the modular equation
 associated with $\, q \, \longrightarrow \, q^{13}$ with itself.

\subsubsection{$\, q \, \longrightarrow \, q^{N^2}, \,  \,  \, q^{M^2 \, N}$ \\}
\label{moreqN}

The previous results can be generalized for any prime number $\, N$. 
For instance, the series (\ref{family_a})
\begin{eqnarray}
\label{family_atrunc}
\hspace{-0.95in}&& \,  \,\quad  \quad  \quad  \quad  \quad 
 y(a, \, x) \, \, = \, \, \,\, \, \,
 a \cdot \, x \, \, \, \, \, \, -744 \cdot  \, a \cdot  \, (a-1) \cdot  \, x^2
 \,\,\,\,\, + \,\,\, \cdots 
\end{eqnarray}
 with $\, a^N \, = \, \, 1$,
 is the solution series of the  {\em  modular equation} $\, \Gamma_{N^2}(x, \, y) \, \, = \, \, \, 0$
 associated with $\, q \, \longrightarrow \, q^{N^2}$, which can be obtained from the
 elimination of $\,z$  between 
 $\, \Gamma_N(x, \, z)$ and  $\, \Gamma_N(z, \, y)$,
 where $\, \Gamma_N(x, \, y) \, = \, \, 0$ is the {\em modular equation} associated with 
 $\, q \, \longrightarrow \, q^{N}$. The series (\ref{family_atrunc}), with $\, a^N \, = \, \, 1$,
 corresponds to the following composition of  nome transformations:
 $\,\, q \, \rightarrow \,\, q^N \, \rightarrow \,\, \omega \cdot \, (q^N)^{1/N}$, with
 $\,\, \omega^N = \, 1$.

The modular equation
$\, \Gamma_{N^2}(x, \, y) \, \, = \, \, \, 0$, corresponding to  $\, q \, \rightarrow \, q^{N^2}$,
will have $\, 1 \, +(N-1) \, + \, N^2 \, = \, \, N \cdot \, (N \, +1)\, \, $
algebraic solution-series, corresponding respectively to the  series 
\begin{eqnarray}
  \label{mod1sur13orderNsquare}
 \hspace{-0.98in}&& \,  \quad \quad \quad \quad \quad \quad  \quad \quad \quad 
  y \, \, = \, \, \, \,\,
 x^{N^2}  \, \,\, + 744\cdot \, N^2 \cdot \, x^{N^2+1}
  \,  \, \, \, \,    + \, \, \, \cdots 
\end{eqnarray}
together with the $\, N-1$ (order-$\,N$) series (\ref{family_atrunc}), namely
\begin{eqnarray}
  \label{NN}
  \hspace{-0.98in}&& \quad  \quad \quad \quad \quad \,\,\,\,
  y_{1/N}(y_N(x)) \,\,   = \, \,  \,  \, \,   \,  
  \omega \cdot \, x \, \, \,  \,\,  \,  -744 \cdot \, \omega \cdot \, (\omega -1) \cdot \, x^2
  \,\, \,\,   \,   + \, \, \,  \cdots
\end{eqnarray}
with $\, \omega^N \, = \, 1$ but $\, \omega \, \ne \, 1$, and the $\, N^2$
compositional inverse (Puiseux) series of the series  (\ref{mod1sur13orderNsquare}).
Series (\ref{modularequasol14c}), (\ref{mod1sur3}) and (\ref{mod1sur5omega}) are such examples. 

One (modular correspondence) series solution of
$\, \Gamma_{N}(x, \, y) \, \, = \, \, \, 0 \, $ is of the form:
\begin{eqnarray}
  \label{mod1sur13orderN}
 \hspace{-0.98in}&& \quad  \quad \quad \quad \quad \quad \quad\quad  \quad \quad 
  y \, \, = \, \, \, \,\,
  x^{N}  \, \,  \, +   744\cdot \, N \cdot \, x^{N+1}
  \,   \, \,\, \, \,  + \, \, \, \cdots 
\end{eqnarray}
One can also compose such an algebraic (modular correspondence) series (\ref{mod1sur13orderN})
with the $\, M$-th root algebraic series (\ref{family_atrunc}) (here $\, a^M = \, 1$),
to get more  (modular correspondence) {\em algebraic}  series:
\begin{eqnarray}
\label{morealgebraic}
\hspace{-0.98in}&& \quad  \quad 
 y \, \, = \, \, \, \,\,
 a \cdot \, x^{N}  \, \,  \, +   744\cdot \, N \cdot \, a \cdot \, x^{N+1} \,\,\, + \, \, \cdots
 \quad   \quad \hbox{with:}  \quad   \quad  \quad  a^M = \, 1. 
\end{eqnarray}
Series (\ref{morealgebraic}) is a (modular correspondence) series solution of the modular equation 
$\, \Gamma_{N \cdot \, M^2}(x, \, y) \, \, = \, \, \, 0$.
The series (\ref{morealgebraic}) corresponds, in the nome, to transformation
$\, q \, \rightarrow \,  q^N  \, \rightarrow \,  \, \omega \cdot \, ((q^N)^M)^{1/M} \, = \, \,  \omega \cdot \, q^N$,
where $\, \omega^M = \, 1$.

\vskip .1cm 
\vskip .1cm 

\subsubsection{The one-parameter series  (\ref{family_a}) is not generically a $\, D$-finite series\\}
\label{isnot}

The one-parameter series (\ref{family_a}) becomes an {\em algebraic series}
when the parameter is a $\, N$-th root of unity. 
 All the previous {\em algebraic series} associated with  {\em modular equations},
 can also be seen as $\, D$-finite series as displayed in the previous section (\ref{linearoper}).
 Along this line it is crucial to note that these
 series are solutions of a linear differential operator
 (like $\, M_3$ in the previous section (\ref{linearoper}))
{\em of order increasing with} $\, N$.
Therefore, we see that {\em one cannot expect the one-parameter series} (\ref{family_a})
 {\em  to be generically $\, D$-finite}, being  solution a finite order linear differential  operator
with coefficients polynomial in $\, x$ and in
the parameter $\, a$, since the order of this linear differential operator  {\em grows
with} $\, N$ when the parameter is a $\, N$-th root of unity.

\vskip .1cm 

\subsection{When the one-parameter series becomes a globally bounded series}
\label{understandglob}

Note that, for {\em integer} values of the parameter $\, a$,  the  series $\,  y(a, \, \, x)$
are series with  {\em integer} coefficients.
More generally,  one can see easily that
{\em such series are globally bounded}~\cite{Christol,ChristolUna}    {\em for any rational number}
$\, a \, = \, P/Q$: the series (\ref{family_a}) can be recast
into a series with {\em integer} coefficients
if one rescales $\, x$ as follows: $ \,  \, x \, \longrightarrow \, \, Q \cdot \, x$.

If one of these series is $\, D$-finite, the series should be,
according to Christol's conjecture~\cite{Chrisconj},
a diagonal of a rational (or algebraic) function~\cite{Christol}. In particular this series should
{\em reduce to algebraic function modulo any prime number}~\cite{Christol,ChristolUna}. Let
us focus, for instance, on the particular value $\, a= \, 3$. 
For $\, a \, = \, \, 3\, $ the  series $\,  y(a, \, \, x)\, $ is
a series with {\em integer} coefficients
\begin{eqnarray}
\label{a3}
\hspace{-0.95in}&& \,  \, \quad  \quad \quad
S \, = \, \,  \,\, 
3 \, x \, \,  \, -4464\, x^2 \,\,  \, +1917216\, x^3 \, \,  \, \,
-1013769984\, x^4 \,  \, \,  \, -33437759328\, x^5
\nonumber \\
\hspace{-0.95in}&& \,  \, \quad  \quad \quad \quad \quad
 \, \, -420498625999104\, x^6 \, \,  \, -452363497164804864\, x^7
 \, \,  \,  \, \, + \, \, \, \cdots           
\end{eqnarray}
which has a radius of convergence $\, 1/1728 \, = \, \,  0.00057870 \cdots$
If one considers the series (\ref{a3}) modulo different primes $\, p$, it is
very difficult to see (for $\, p$ large enough)
if this series (\ref{a3}) is an {\em algebraic series}
modulo  $\, p$, or, even, is $\, D$-finite modulo  $\, p$.
We have, however, found the following result. Introducing 
\begin{eqnarray}
\label{a3relation}
  \hspace{-0.95in}&& \,  \,   \, \,  \quad
 \sigma \, \,\, = \,  \, \, \,\,
 {{S \, -3\, x} \over { 3 \cdot \, 2^5 \cdot \, x }} \,\,\, \,  +{{99 } \over {2}} \cdot \, x
       \,\, \,  \, +1 \, 
  \, \, = \, \,  \, \, \,
  1  \,\,\, + \,\, 3  \, x\, \,+ \, 19971\, x^2 \,\, \,-10560104\, x^3 
 \nonumber \\
 \hspace{-0.95in}&& \,  \, \,   \, \quad \quad
  -348309993\, x^4\,\, -4380194020824 \, x^5 \, \, -4712119762133384\, x^6
    \, \, \,\, + \,  \,\cdots                  
\end{eqnarray}
this series reduces, modulo $\,p =\, 2$, to the algebraic series
\begin{eqnarray}
\label{a3relationsigma}
  \hspace{-0.95in}&& \,  \,  \,  \,  \, \, 
  \sigma(x) \, \, = \, \, \,
  1 \, \, \,  +x  \, \,  +x^2  \,  \, +x^4  \, +x^8
  \, +x^{16} \,  +x^{32} \,  +x^{64} \,  +x^{128} \,  +x^{256}
   \,  \,\,  + \, \, \, \cdots                
\end{eqnarray}
solution, modulo $\,p=\, 2$,  of the algebraic polynomial:
\begin{eqnarray}
\label{a3relationsigma}
  \hspace{-0.95in}&& \,  \, \quad  \quad \quad \quad \quad
 \sigma(x^2)  \, \,  - \sigma(x)  \,  \, +x \, \, \, \,  = \, \, \, \,
 \sigma(x)^2  \, \,  - \sigma(x)  \,  \, +x
 \,\, \,  = \, \, \, \, \, 0.
\end{eqnarray}
The nature of the series (\ref{a3}), or more generally
of (\ref{family_a}) for integer, or rational values of the parameter $\, a$,
remains an open question. It seems that
such globally bounded series are not $\, D$-finite. At least, one has an
{\em infinite number of  differentially algebraic series}. Are these
globally bounded series  $\, D$-$D$-finite series~\cite{DD,DD1}  ?

\subsection{Miscellaneous calculations.}
\label{miscell}

The  nome series (\ref{mirror21})
and the mirror map series (\ref{mirror1}), 
are, respectively, solutions of the following Schwarzian equations 
\begin{eqnarray}
\label{Harnad214bis}
\hspace{-0.95in}&& \quad  \quad  \quad   \quad \quad \quad 
 \, \{\tilde{Q}(x), \, x \} \, \, \, \, 
 + {{1} \over {2  \cdot \, \tilde{Q}(x) ^2 }} 
 \cdot  \Bigl({{ d \tilde{Q}(x) } \over {d x}} \Bigr)^2
\,  \, +  \, W(x) \, \,  \, =  \, \, \, \, \, 0,
\end{eqnarray}
and
\begin{eqnarray}
\label{Harnad214ter}
\hspace{-0.95in}&& \,  \,    \quad  \quad \quad \quad \quad 
 \, \{\tilde{X}(x) , \, x \} \, \, \, \, \, 
 - {{1} \over {2  \cdot \, x^2 }} 
 \, \,  \,
 -  \, W\Bigl( \tilde{X}(x)  \Bigr) \cdot  \,  \Bigl({{ d \tilde{X}(x) } \over {d x}} \Bigr)^2
 \,  \, =  \, \, \, \, \, 0,
\end{eqnarray}
where:
\begin{eqnarray}
\label{whereW}
\hspace{-0.95in}&& \quad  \,  \quad \quad \quad  \quad  \quad \quad  \quad 
 W(x) \, \, = \, \, \,
 - {{1} \over {2}} \cdot \,
 {{ 1 \, -1968\, x  \, +2654208\, x^2} \over { x^2 \cdot \, (1 \, - 1728\, x)^2 }}.
\end{eqnarray}
Let us introduce the hypergeometric function:
\begin{eqnarray}
\label{Fx}
\hspace{-0.95in}&& \,  \, \quad  \,  \quad \quad  \quad 
 F(x) \, \, = \, \, \,
 x \cdot \, (1\, -1728 \cdot \, x)^{1/2} \cdot \,
 _2F_1\Bigl([{{1} \over {12}}, \, {{5} \over {12}}], \, [1], \, 1728 \cdot \, x   \Bigr)^2.
\end{eqnarray}
Note that the Schwarzian equation (\ref{Harnad214bis}), on $\,  {\tilde{Q}(x)}$,
can be seen to be a consequence of (see (\ref{wherecondn1}) below): 
\begin{eqnarray}
\label{whereWQ}
  \hspace{-0.95in}&& \, \,
F(x) \, \, = \, \, \, {{ {\tilde{Q}(x)} } \over { {\tilde{Q}(x)'}}}
  \, \, \quad  \quad \hbox{together with:} \quad  \quad \, \,
 W(x)   \, \, = \, \, \,
  {{F''(x)} \over {F(x)}} \, \, -{{1} \over {2}} \cdot \, \Bigl({{F'(x)} \over {F(x)}}\Bigr)^2.  
\end{eqnarray}

Therefore the nome $\,  \, \tilde{Q}(x)\, $ is {\em also} solution
of the order-one linear differential operator:
\begin{eqnarray}
\label{whereW}
\hspace{-0.95in}&& \quad  \,  \quad \quad \quad   {\cal L}_1
\, \, = \, \, \,  F(x) \cdot \, D_x \,\, -1
\quad \quad \quad  \quad \quad \hbox{where:} 
  \\
\hspace{-0.95in}&& \quad  \,  \quad  \quad  \quad \quad \quad 
F(x) \, \,  \, = \, \, \,  \,
x \cdot \, (1 \, -1728 \cdot \, x)^{1/2} \cdot \,  
  _2F_1\Bigl([{{1} \over {12}}, \, {{5} \over {12}}], \, [1], \,  1728 \cdot \,  x  \Bigr)^2.
\nonumber 
\end{eqnarray}
It is thus $\, DD$-finite\footnote[1]{See~\cite{DD,DD1}.}:
\begin{eqnarray}
\label{DD-finiteexample}
  \hspace{-0.95in}&& \quad  \,  \, \quad \quad \quad
 {{  \tilde{Q}(x)' }  \over {  \tilde{Q}(x)  }} \, \, = \,  \, \, {{ 1}  \over { F(x)}}
 \quad \quad \quad  \,  \,  \hbox{or:} \quad \quad \, \,  \,  \quad
   \tilde{Q}(x) \, \, = \, \, \,  \exp\Bigl( \int^x \, {{dx} \over {F(x)}}  \Bigr).   
\end{eqnarray}
The one-parameter series $\, y(x) \, = \, \, y(a, \, x)$, given by (\ref{family_a}),
is solution of the rank-two equation (see (\ref{cas2})) 
\begin{eqnarray}
\label{rota1}
\hspace{-0.95in}&& \quad \quad  \quad \quad  \quad  \quad   \quad 
 A_R(x) \, \, \, \,  -A_R(y(x)) \cdot  \, y'(x)  \, \,  \, \, + {{y''(x)} \over {y'(x)}}
  \,\,  = \,\,  \,\,  0, 
\end{eqnarray}
with
\begin{eqnarray}
\label{rota1}
\hspace{-0.95in}&& \quad \quad  \quad \quad  \quad  \quad \quad  \quad   \quad  \quad  \quad   \quad  
  A_R(x)    \, \, = \, \, \, {{F'(x)} \over {F(x)}},    
\end{eqnarray}
and also solution of  the Schwarzian condition
\begin{eqnarray}
\label{condition1n1}
\hspace{-0.95in}&& \quad \quad  \quad \quad  \quad \quad  
 W(x)  \, \, \,  \, -W(y(x)) \cdot  \, y'(x)^2 \, \, \,  \,+ \,  \{ y(x), \, x\} 
\, \,\, \, = \,\, \, \,  \, 0, 
\end{eqnarray}
where:
\begin{eqnarray}
\label{wherecondn1}
\hspace{-0.95in}&& \quad   \quad \quad  \quad  \quad  
W(x)  \, \, = \, \,  \, \, \,
 {{F''(x)} \over {F(x)}} \, \, -{{1} \over {2}} \cdot \, \Bigl({{F'(x)} \over {F(x)}}\Bigr)^2
 \, \, = \, \,  \, \, \,   A_R'(x) \, \, \,  + \, \,  {{A_R(x)^2} \over {2 }}
\nonumber \\
\hspace{-0.95in}&& \quad   \quad \quad \quad \quad \quad \quad \quad \quad
 \, \, = \, \,  \, \, \,
 - {{1} \over {2}} \cdot \,
 {{ 1 \, -1968\, x  \, +2654208\, x^2} \over { x^2 \cdot \, (1 \, - 1728\, x)^2 }}.                               
\end{eqnarray}
Note that $\, W(x)$ is a {\em rational function}, but this is
far from being the case for $\, A_R(x)$.  
We will see, in the following, that the one-parameter series $\, y(x) \, = \,  \, y(a, \, x)$,
given by (\ref{family_a}), {\em is also solution of}:
\begin{eqnarray}
\label{wherecondn1partial}
  \hspace{-0.95in}&& \quad   \quad \quad \quad \quad \quad
 a \cdot \, {{\partial y(a, \,x)} \over { \partial a}}
 \, \, = \, \, \, F(y(a,  \, x))
 \, \, = \, \, \,
 F(x) \cdot \,  {{\partial y(a, \, x)} \over { \partial x}}.
\end{eqnarray}

\vskip .1cm 
\vskip .1cm 

\subsection{More one-parameter series solutions.}
\label{moreoneparam}

If one combines $\, y_2$, the ``correspondence'' series (\ref{modularequasol1}) solution of the
modular equation (\ref{modularequa}), with the one-parameter series (\ref{family_a}),
one gets a one-parameter series
\begin{eqnarray}
\label{oneparamy2}
  \hspace{-0.95in}&& \quad  
y_2^{(a)} \,   \, \, =  \, \,  \,   y(a, \, y_2) \,   \, \, =  \, \,  \, \,
\tilde{X}\Bigl( a \cdot \, \tilde{Q} (x)^2 \Bigr) \, \, =  \, \,  \, \, \,
  a \cdot \, x^2   \, \,\, \, +1488 \cdot \, a \cdot \, x^3 \,  
 \nonumber \\
  \hspace{-0.95in}&& \quad   \quad  \quad  
\,  -24 \cdot \, a \cdot \, (31\, a-85599)\cdot \, x^4 \, \,  \,  \,
 -256 \cdot \, a\cdot \, (8649\, a-11180329)\cdot \, x^5
\nonumber \\
 \hspace{-0.95in}&& \quad   \quad  \quad  \quad   \,
  +12\cdot \, a\cdot \, (29721\, a^2-392019552\, a+338926406215)\cdot \, x^6
 \\
  \hspace{-0.95in}&& \quad   \quad  \quad  \quad  \quad  
 \, +192 \cdot \, a\cdot \, (8292159\, a^2-45872836768\, a+30686235044193)\cdot \, x^7
   \, \,  \, \, + \, \, \, \cdots \nonumber 
\end{eqnarray}
This series (\ref{oneparamy2}) is also solution of the Schwarzian equation (\ref{EQSchwar}).
Furthermore we have:
\begin{eqnarray}
\label{modularform2explicitfollow2further}
\hspace{-0.95in}&& \quad \,  \, 
2 \cdot \,  y_2^{(a)}  \cdot \, (1 \, -1728 \cdot \, y_2^{(a)} )^{1/2} \cdot \,
 _2F_1\Bigl([{{1} \over {12}}, \, {{5} \over {12}}], \, [1], \, 1728 \cdot \, y_2^{(a)}   \Bigr)^2
\nonumber \\
 \hspace{-0.95in}&& \quad \quad  \, \,  \quad  
\, = \, \, \, \,\, 
 x \cdot \, (1 \, -1728 \cdot \, x)^{1/2} \cdot \,  
  _2F_1\Bigl([{{1} \over {12}}, \, {{5} \over {12}}], \, [1], \,
                    1728 \cdot \,  x  \Bigr)^2 \cdot \,  {{d y_2^{(a)}  } \over {d x}}. 
\end{eqnarray}

When $\, a \, = \, 1$, the radius of convergence of (\ref{oneparamy2})
is $\, 1/1728 \, = \,  0.000578703703 \, \cdots$, and this is also
the case for any $\, a$, $\, N$-th root of unity $\, a^N=\, 1$.
Similarly to what has been sketched in section (\ref{one-parameterSchwarz})
(see equation (\ref{othersing})), let us remark that the {\em one-parameter} series
(\ref{oneparamy2}) can be written $\,  \tilde{X}\Bigl( a \cdot \,  \tilde{Q}(x)^2 \Bigr)$.
For generic value of the parameter $\, a$, the radius of convergence of (\ref{oneparamy2})
will correspond, for $\, a$ small enough,  to the singularity of $\, \tilde{Q}(x)$,
namely  $\, 1/1728$, and for  $\, a$ large enough,  to the values of $\, x$ such that
$\,\, a \cdot \, \tilde{Q}(x)^2 \, = \, \, - \, \exp\Bigl( -\, \sqrt{3} \cdot \, \pi \Bigr)$, 
which correspond to the singularity of $\, \tilde{X}(x)$, namely:
\begin{eqnarray}
\label{othersing2}
\hspace{-0.95in}&& \,  \quad  \quad  \quad   \quad   \quad   \quad    \quad  \quad   \quad  \quad \,
x \, = \, \,
\tilde{X}\Bigl(\Bigl( - \, {{1} \over {a}} \cdot \,  \, \exp\Bigl( -\, \sqrt{3} \cdot \, \pi \Bigr)  \Bigr)^{1/2}  \Bigr).
\end{eqnarray}
More generally, all the series
\begin{eqnarray}
\label{na}
  \hspace{-0.95in}&& \,  \, \quad \quad  \quad \quad  \quad \quad  \quad \quad \quad  \quad
  \tilde{X}\Bigl( a \cdot \,  \tilde{Q}(x)^n \Bigr)
  \, \, = \, \, \, \,  \, a \cdot \, x^n \, \, \,  + \, \, \, \cdots 
\end{eqnarray}
have a radius of convergence corresponding, for $\, a$ small enough, to the occurrence of the
singularity of the nome-like series $\, \tilde{Q}(x)$, namely  $\, x \, = \, 1/1728$.

Similarly to (\ref{oneparamy2}),
if one combines $\, y_3$, the ``correspondence''
series (\ref{modularequasol13}) solution of the
modular equation (\ref{modularequa3}),  with the one-parameter series (\ref{family_a}),
one gets a one-parameter series
\begin{eqnarray}
\label{oneparamy3}
  \hspace{-0.95in}&& \, \, \,
 y_3^{(a)}  \, \, =  \, \,   \,  y(a, \, y_3)   \, \, =  \, \,  \, 
\tilde{X}\Bigl( a \cdot \, \tilde{Q}(x)^3 \Bigr) \, \, =  \, \,  \, \,  \, 
 a \cdot \, x^3  \, \, \,  +2232 \cdot \, a \cdot \, x^4  \,\, \, +3911868 \cdot \, a \cdot \, x^5
  \nonumber \\
  \hspace{-0.95in}&& \quad   \quad  \quad  \,\,
  -24\cdot  \, a\cdot  \, (31\, a -265833940) \cdot \, x^6 \,\, \, \,
  -54 \cdot \, a \cdot \, (61504\, a -187769367601) \cdot \, x^7
 \nonumber \\
  \hspace{-0.95in}&& \quad   \quad  \quad  \quad  \,\, \quad \quad
 -1296 \cdot \, a\cdot \, (7351340\, a-12322394107529)\cdot \, x^8
   \, \,  \, \, \,  + \, \, \, \cdots 
\end{eqnarray}
This series (\ref{oneparamy3}) is also solution of the Schwarzian equation (\ref{EQSchwar}).
Furthermore we have:
\begin{eqnarray}
\label{modularform2explicitfollow3further}
\hspace{-0.95in}&& \quad \, \,   
3 \cdot \,  y_3^{(a)}  \cdot \, (1 \, -1728 \cdot \, y_3^{(a)} )^{1/2} \cdot \,
  _2F_1\Bigl([{{1} \over {12}}, \, {{5} \over {12}}], \, [1], \, 1728 \cdot \, y_3^{(a)}   \Bigr)^2
\nonumber \\
 \hspace{-0.95in}&& \quad \, \,   \quad  \,  \quad  
\, = \, \, \, \,\, 
 x \cdot \, (1 \, -1728 \cdot \, x)^{1/2} \cdot \,  
  _2F_1\Bigl([{{1} \over {12}}, \, {{5} \over {12}}], \, [1], \,  1728 \cdot \,  x  \Bigr)^2
 \cdot \,  {{d y_3^{(a)}  } \over {d x}}. 
\end{eqnarray}
Similarly:
\begin{eqnarray}
\label{oneparamy5}
  \hspace{-0.95in}&& \quad  \quad
y_5^{(a)}(x) \,   \, \, =  \, \,  \,  y\Bigl(a, \, y_5(x)\Bigr)
 \,   \, \, =  \, \,  \,
\tilde{X}\Bigl( a \cdot \, \tilde{Q}(x)^5 \Bigr) \, \, =  \, \, \,  \,\,
a \cdot \, x^5  \, \, \, \,+3720 \cdot \, a \cdot \, x^6
\nonumber \\
 \hspace{-0.95in}&& \quad  \quad  \quad \quad
\, +9287460 \cdot \, a \cdot \, x^7 \, \, \, +19648405600 \cdot \, a \cdot \, x^8
   \,\, \, \, +38124922672650 \cdot \, a \cdot \, x^9
\nonumber \\
 \hspace{-0.95in}&& \quad   \quad   \quad  \quad  \quad \,\,
 -24 \cdot \, a \, \cdot \, (31\, a -2930432767154406) \cdot \, x^{10}
    \\
 \hspace{-0.95in}&& \quad    \quad  \quad    \quad   \quad   \quad     \, \,
 -40 \cdot \, a \cdot \, (138384\, a -3142471028063763509) \cdot \, x^{11}
  \nonumber \\
 \hspace{-0.95in}&& \quad   \quad  \quad\,  \quad   \quad  \quad  \quad  \,
 -960 \cdot \, a \cdot \, (25120323\, a -229208433006295134073) \cdot \, x^{12}
 \,\,\, \, \, \, + \, \, \, \cdots
   \nonumber 
\end{eqnarray}
This series (\ref{oneparamy5}) is also solution of the Schwarzian equation (\ref{EQSchwar}).
Furthermore we have:
\begin{eqnarray}
\label{modularform2explicitfollow5further}
\hspace{-0.95in}&& \quad \, \, 
5 \cdot \,  y_5^{(a)}  \cdot \, (1 \, -1728 \cdot \, y_5^{(a)} )^{1/2} \cdot \,
_2F_1\Bigl([{{1} \over {12}}, \, {{5} \over {12}}], \, [1], \, 1728 \cdot \, y_5^{(a)}   \Bigr)^2
\nonumber \\
 \hspace{-0.95in}&& \quad \, \,   \quad  \,  \quad  
\, = \, \, \, \,\, 
 x \cdot \, (1 \, -1728 \cdot \, x)^{1/2} \cdot \,  
_2F_1\Bigl([{{1} \over {12}}, \, {{5} \over {12}}], \, [1], \,  1728 \cdot \,  x  \Bigr)^2
  \cdot \,  {{d y_5^{(a)}  } \over {d x}}. 
\end{eqnarray}

\vskip .1cm

One also easily   gets:
\begin{eqnarray}
\label{modularform2explicitfollow5furthereasily}
\hspace{-0.95in}&& \quad  \quad  \, \,   \quad \quad \,\,\,
 5 \cdot \, F(y_5^{(a)} (x)) \, \, = \, \, \,\, F(x) \cdot \, {{ d y_5^{(a)} (x)} \over { dx}}
 \, \, = \, \, \,\,
 5 \cdot \,  a \cdot \, {{\partial y_5^{(a)} (x)} \over { \partial a}}.
\end{eqnarray}

\vskip .1cm

More generally, let us introduce the modular correspondence series
$\, y_n(x) \, = \, \, x^n \, + \, 744 \cdot \, n \cdot \, x^{n+1} \, \, + \, \, \, \cdots$
(for $\, n\, \ge \, 2$), one can verify {\em that these  series commute}.
These  modular correspondences $\, y_n(x)$ can easily
be generalized to {\em one-parameter} series
$\, y\Bigl(a, \,  y_n(x)  \Bigr)$  which are also solutions of the Schwarzian equations:
\begin{eqnarray}
\label{compo1}
 \hspace{-0.98in}&& \quad  \quad  \quad  \quad  \quad  \, \, \,  
   y\Bigl(a, \,  y_n(x)  \Bigr)  \, \, = \, \, \, \,  \, \,
   a \cdot \, x^n \, \,  \, \, +  744 \cdot \, n \cdot \, a  \cdot \, x^{n+1}
   \, \,   \, \,  \, + \, \, \, \cdots 
 \end{eqnarray}              
Let us recall the
one-parameter series $\, y(a, \, x)$ given by  (\ref{family_a}),
we have the following relation: 
\begin{eqnarray}
\label{compo1u}
\hspace{-0.98in}&& \quad  \, \, \,  
y\Bigl(a^n, \,  y_n(x)  \Bigr)  \, \, = \, \, \,    y_n\Bigl( y(a, \, x)  \Bigr)
\, \, = \, \, \, \,
a^n \cdot \, x^n \, \, \,\, + \, 744 \cdot \, n \cdot \, a^n  \cdot \, x^{n+1}
 \,  \, \, \, + \, \, \, \cdots 
\end{eqnarray}

\vskip .1cm 

\subsection{Composition in general}
\label{MoreGenerallysub}
The one-parameter series (\ref{compo1}) can be written 
\begin{eqnarray}
\label{introducingmoregener}
 \hspace{-0.95in}&& \quad \quad \quad   \quad  \quad   \quad  \, \,   \quad \quad \quad \quad\quad
y_n^{(a)}(x) \, \, = \, \,  \tilde{X}\Bigl( a \cdot \, \tilde{Q}(x)^n \Bigr).        
\end{eqnarray}
We have the following composition:
\begin{eqnarray}
\label{introducingmoregenercompo}
\hspace{-0.95in}&& \quad    \quad  \, \,\,   
y_n^{(a)}\Bigl(  y_m^{(b)}(x)   \Bigr) \, \, = \, \, \,
\tilde{X}\Bigl( a \cdot \, \tilde{Q}\Bigl(\tilde{X}\Bigl( b \cdot \, \tilde{Q}(x)^m \Bigr) \Bigr)^n \Bigr)
\, \, = \, \, \, \tilde{X}\Bigl( a \cdot \, \Bigl( b \cdot \, \tilde{Q}(x)^m  \Bigr)^n \Bigr)
\nonumber \\
\hspace{-0.95in}&& \quad \quad  \,\, \,   \quad \quad \quad \quad
\, \, = \, \, \, \tilde{X}\Bigl( a \cdot \, b^n \cdot \, \tilde{Q}(x)^{m \, n}  \Bigr)
\, \, = \, \, \,    \,    y_{m\, n}^{(a \, b^n)}(x).          
\end{eqnarray}
Note that the condition to have series solutions of the Schwarzian equation
of the form  $\, y_n^{(a)}(x) \, \, = \, \,  a \cdot \, x^n \, \, + \, \, \, \cdots $, 
with $\, n \, \ge \, 2 $, 
amounts to having~\cite{Schwarz1,Schwarz2} 
$\, W(x)$ of the form  $\, W(x) \, = \, \, -1/2/x^2 \, \, \, + \, \, \cdots$
which is satisfied when  $\, F(x) \, = \, \, \alpha \cdot \, x \, \, + \, \, \cdots$, 
or $\, \tilde{Q}(x) \,\, = \, \, \,  \rho \cdot \, x^{1/\alpha} \, \,\,  + \, \, \cdots $

\vskip .2cm 

\section{The one-parameter series (\ref{one-parameterSchwarz}) seen as a $\, \epsilon$-expansion.}
\label{epsexp}

In the $\, a \, \rightarrow \, 1\, $ limit, let us denote $\, \epsilon \, = \, a \, -1$.
The  one-parameter series $\, y(x) \, = \,  \, y(a, \, x)$,  given by (\ref{family_a}),
can, thus, be seen as an $\, \epsilon$-expansion: 
\begin{eqnarray}
\label{seriesmodcurve1aeps1}
\hspace{-0.95in}&& \quad  \quad  \quad  \quad  \quad  \quad \quad \quad \quad \, 
y(a, \, x) \, \, \, = \, \, \,  \,  \, 
x \,  \, \,\,   +\sum_{n=1}^{\infty} \, \epsilon^n \cdot \, B_n(x), 
\end{eqnarray}
where $\, B_1(x) \, = \, F(x)$, with $\, F(x)$ given by (\ref{Fx}), and where  $\, B_2(x)$
reads (see equation (115) in~\cite{Schwarz1}): 
\begin{eqnarray}
\label{holo1eps11}
\hspace{-0.95in}&&  \quad \quad \quad \quad \quad \quad  \quad \quad \,\, 
 B_2(x) \,  = \, \,\, 
 {{1} \over {2}}  \cdot \, F(x) \cdot \, \Bigl( {{d B_1(x)} \over {dx }}  \, -1\Bigr).
\end{eqnarray} 
Assuming that (\ref{seriesmodcurve1aeps1}) is solution of the Schwarzian
condition (\ref{condition1n1}) (with $\, W(x)$ given by (\ref{whereW})),   
we actually obtained the next $\, B_n(x)$'s:
\begin{eqnarray}
\label{holo1eps1}
\hspace{-0.98in}&& \, \,   
 B_3(x) \,  = \, \,\, 
 {{1} \over {3}}  \cdot \, F(x) \cdot \, \Bigl(  {{d B_2(x)} \over {dx }}
          \, - {{d B_1(x)} \over {dx }}  \, +1 \Bigr),
\nonumber \\
  \hspace{-0.98in}&& \, \,    
 B_4(x) \,  = \, \,\, 
{{1} \over {4}}  \cdot \, F(x) \cdot \, \Bigl( {{d B_3(x)} \over {dx }}
\, -{{d B_2(x)} \over {dx }}  +{{d B_1(x)} \over {dx }}  \, -1 \Bigr),
 \\
  \hspace{-0.98in}&& \, \,    
 B_5(x) \,  = \, \,\, 
 {{1} \over {5}}  \cdot \, F(x) \cdot \,\Bigl(  {{d B_4(x)} \over {dx }}  \, -{{d B_3(x)} \over {dx }}
 + {{d B_2(x)} \over {dx }} -{{d B_1(x)} \over {dx }}  \, +1 \Bigr),
\nonumber \\
  \hspace{-0.98in}&& \, \,    
 B_6(x) \,  = \, \,\, 
 {{1} \over {6}}  \cdot \, F(x) \cdot \,\Bigl(  {{d B_5(x)} \over {dx }}  \, -{{d B_4(x)} \over {dx }}
 + {{d B_3(x)} \over {dx }} -{{d B_2(x)} \over {dx }}   +{{d B_1(x)} \over {dx }} \, -1 \Bigr),
  \quad \cdots     \nonumber               
\end{eqnarray}
More generally, one easily discovers the recursion
\begin{eqnarray}
\label{holo1eps1recur}
  \hspace{-0.98in}&& \, \, \quad  \quad \quad \quad \quad \quad \quad 
  (n +1) \cdot \, B_{n+1} \,  \,\,  + n \cdot \, B_n
  \, \, = \, \, \, \, F(x) \cdot \, {{d B_n(x)} \over {dx }}, 
\end{eqnarray}
which yields on the series (\ref{seriesmodcurve1aeps1})
\begin{eqnarray}
\label{holo1eps1recursum}
  \hspace{-0.98in}&& \, \, \quad 
  \sum_{n} \,  (n +1) \cdot \, B_{n+1}  \cdot \,  \epsilon^n
  \,  \, \,  +  \sum_{n} \,   n \cdot \, B_n  \cdot \,  \epsilon^n
  \, \, = \, \, \,   \,
  F(x) \cdot \, \Bigl(  \sum_{n} {{d B_n(x)} \over {dx }} \cdot \,  \epsilon^n \Bigr), 
\end{eqnarray}
or
\begin{eqnarray}
\label{holo1eps1recursum2}
  \hspace{-0.98in}&& \, \,  \quad 
  {{ \partial \sum_{n} \, B_{n+1}  \cdot \,  \epsilon^{n+1} } \over {\partial \epsilon }}
 \,  \,  + \epsilon \cdot \,  {{ \partial \sum_{n} \,  B_n  \cdot \,  \epsilon^n} \over {\partial \epsilon }}  
 \, \, = \, \, \,   \,
 F(x) \cdot \, \Bigl( {{ \partial  \sum_{n} \, B_n(x) \cdot \,  \epsilon^n } \over {\partial x }} \Bigr), 
\end{eqnarray}
yielding finally
\begin{eqnarray}
\label{wherecondn2}
  \hspace{-0.95in}&& \quad   \quad \quad \quad \quad \quad \quad \quad \, \, \, 
 (1\, +\epsilon) \cdot \, {{\partial y(a, x)} \over { \partial \epsilon}}
        \, \, = \, \, \,  F(x) \cdot \,  {{\partial y(a, \, x)} \over { \partial x}}, 
\end{eqnarray}
namely:
\begin{eqnarray}
\label{wherecondn3}
  \hspace{-0.95in}&& \quad   \quad \quad \quad \quad \quad \quad \quad \quad \,  \, 
 a \cdot \, {{\partial y(a, x)} \over { \partial a}}
        \, \, = \, \, \,  F(x) \cdot \,  {{\partial y(a, \, x)} \over { \partial x}}. 
\end{eqnarray}
Note that $\, y(a, \, x)\, $ is also solution of:
\begin{eqnarray}
\label{wherecondn3also}
  \hspace{-0.95in}&& \quad   \quad \quad \quad \quad \quad \quad \quad \quad
 F\Bigl(   y(a, \, x) \Bigr) 
        \, \, = \, \, \,  F(x) \cdot \,  {{\partial y(a, \, x)} \over { \partial x}}. 
\end{eqnarray}
Recalling some relation on the nome $\ q$ (see equation (33) in~\cite{Schwarz1}):
\begin{eqnarray}
\label{nomerecall}
  \hspace{-0.95in}&& \quad   \quad \quad \quad \quad \quad \, 
 {{q'} \over {q}}   \, \, = \, \, \, {{1} \over {F(x)}}
        \quad   \quad \quad \hbox{or:} \quad \quad \quad \quad
  q \cdot \, {{d } \over {dq}}  \, \, = \, \, \,  F(x) \cdot \, {{d } \over {dx}}, 
\end{eqnarray}
we see that relation  (\ref{wherecondn3}) also reads more simply: 
\begin{eqnarray}
\label{wherecondn4}
  \hspace{-0.95in}&& \quad   \quad \quad \quad \quad \quad \quad \quad  \quad \quad
 a \cdot \, {{\partial y(a, x)} \over { \partial a}}
        \, \, = \, \, \, q \cdot \, {{\partial y(a, x)} \over { \partial q}}. 
\end{eqnarray}
which is reminiscent of the fact that changing $\,\, x \, \rightarrow \,  y(a, x)\,$
just amounts, on the nome, to  changing $\,\, \, q \, \rightarrow \,  a \cdot \, q$.
Equation (\ref{wherecondn3}) means that $\, y(a, \, x)$ is a function of
\begin{eqnarray}
\label{functionof}
\hspace{-0.95in}&& \quad   \quad \quad \quad \quad \quad \quad 
\int\Bigl( {{da} \over {a}} \, \, + {{dx} \over {F(x)}} \Bigr)
\, \, = \, \, \,\,
\ln(a) \,\, + \int\Bigl({{dx} \over {F(x)}} \Bigr), 
\end{eqnarray}
or, recalling (\ref{DD-finiteexample}), a function of:  
\begin{eqnarray}
\label{functionofequiv}
\hspace{-0.95in}&& \quad   \quad \quad \quad \quad \quad \quad \quad \quad 
\exp\Bigl(\int\Bigl( {{da} \over {a}} \,\, +{{dx} \over {F(x)}} \Bigr)\Bigr)
\, \, = \, \, \, \, a \cdot \, \tilde{Q}(x).
\end{eqnarray}
This is actually the case since  $\, y(a, \, x)\, $  is nothing but
$\, \tilde{X}\Bigl(a \cdot \, \tilde{Q}(x)\Bigr)$ (see (\ref{XQ})).  

\vskip .1cm

{\bf Remark \ref{epsexp}.1:} Do note that the previous calculations {\em are still valid}
when $\, F(x)$ {\em is not given by} (\ref{Fx}). One can verify, for {\em any function} 
$\, F(x)$, that the $\,\epsilon$-expansion (\ref{seriesmodcurve1aeps1})
with coefficients $\, B_n$  given by (\ref{holo1eps11}),  (\ref{holo1eps1}), (\ref{holo1eps1recur}),   
{\em is actually solution of the Schwarzian relation} (\ref{condition1n1}),
with $\, W(x)$ given by:
\begin{eqnarray}
\label{defW}
\hspace{-0.95in}&& \quad   \quad \quad \quad \quad \quad \quad \quad
W(x)  \, \, = \, \,  \, \, \,
 {{F''(x)} \over {F(x)}} \, \, -{{1} \over {2}} \cdot \, \Bigl({{F'(x)} \over {F(x)}}\Bigr)^2.
\end{eqnarray}

\vskip .1cm 

\section{Generalization of $\, W(x)$ in the Schwarzian equation: adding an extra parameter $\, \alpha$.}
\label{EvenmoreGenerally}

For a given function $\, F(x)$ let us consider the relation
\begin{eqnarray}
\label{relation}
  \hspace{-0.95in}&& \quad   \quad \quad \quad \quad \quad \quad \quad \quad \quad \quad 
   F(y(x)) \, \, = \, \, \, \, F(x) \cdot \, {{d y(x)} \over {d x}}, 
\end{eqnarray}
which corresponds  to:
\begin{eqnarray}
\label{relationcorresp}
  \hspace{-0.95in}&& \quad   \quad \quad \quad \quad \quad \quad \quad \quad \quad 
{{d y} \over {F(y)}} \, \, = \, \, \,  \, \,  {{d x} \over {F(x)}}
  \, \, = \, \, \,   \,  \, {{d q} \over { q }}. 
\end{eqnarray}
From (\ref{relation}), namely $\, F(y) \, = \, F(x) \cdot \, y'$, one gets
\begin{eqnarray}
\label{relation11}
\hspace{-0.95in}&& \quad   \quad \quad \quad \quad \quad \quad  \quad 
F'(y)\cdot \, y' \, \, = \, \, \, \,
F'(x) \cdot \, y' \, \,  \, + F(x) \cdot \, y'', 
\end{eqnarray}
or
\begin{eqnarray}
\label{relation7}
\hspace{-0.95in}&& \quad   \quad \quad \quad \quad \quad \quad \quad \quad \quad 
{{F'(y) } \over {F(y)}} \cdot \, y'
  \, \, = \, \, \, {{F'(x)} \over {F(x)}} \, \,  \, + {{ y''} \over {y'}}. 
\end{eqnarray}
or, more generally, using (\ref{relation}) in order to introduce an extra parameter $\, \alpha$:
\begin{eqnarray}
\label{relation9}
  \hspace{-0.95in}&& \quad  \quad \quad \quad \quad \quad \,
 \Bigl( {{F'(y) } \over {F(y)}}   \, \, + {{\alpha } \over {F(y)}}  \Bigr) \cdot \, y'
\, \, = \, \, \, \,
\Bigl(  {{F'(x)} \over {F(x)}}  \, \, + {{\alpha } \over {F(x)}}  \Bigr)
\, \,  \, + \,{{ y''} \over {y'}}. 
\end{eqnarray}
Let us introduce
\begin{eqnarray}
  \label{relation5}
\hspace{-0.95in}&& \quad  \quad \quad \quad \quad \quad \quad \quad \quad \quad 
  A_R(x) \, \, = \,  \, \, {{F'(x)} \over {F(x)}}  \, \,  \, + {{\alpha } \over {F(x)}}, 
\end{eqnarray}
we see that (\ref{relation9}) can be written
\begin{eqnarray}
\label{relationAR}
  \hspace{-0.95in}&& \quad   \quad \quad \quad \quad \quad  \quad  \, \quad \quad 
 A_R(x) \, \, \, \,   -  A_R(y) \cdot \, y'\,\, \,   \,   +{{ y''} \over {y'}}
                   \, \, \, = \, \,  \, \, 0. 
\end{eqnarray}
which is (\ref{cas2})  of section (\ref{Questions}).
From (\ref{relation11}), that we  rewrite
\begin{eqnarray}
\label{relation111}
 \hspace{-0.95in}&& \quad   \quad \quad \quad \quad \quad \quad \quad \quad \quad 
F'(y)  \, \, = \, \, \,   \, F'(x) \,  \,   \, \,  + F(x) \cdot \, {{y''} \over {y'}}, 
\end{eqnarray}
one gets
\begin{eqnarray}
\label{relation111}
  \hspace{-0.95in}&& \quad  \quad \quad \quad \quad \quad 
 F''(y) \cdot \, y' \, \, = \, \, \, \, F''(x) \,  \, \, 
 + F'(x) \cdot \, {{y''} \over {y'}}
 \, \,  \,  \, + F(x) \cdot \, \Bigl({{y''} \over {y'}}\Bigr)', 
\end{eqnarray}
or, using (\ref{relation}), written $\, F(x) \, = \, \, F(y)/y'$:
\begin{eqnarray}
\label{relation1111}
  \hspace{-0.95in}&& \quad  \quad \quad \quad \quad \quad \quad \quad 
{{F''(y)} \over {F(y)}}  \cdot \, y'^2  \, \, \, = \, \,  \, \,
{{F''(x)} \over {F(x)}} \,  \,
+ {{F'(x)} \over {F(x)}}  \cdot \, {{y''} \over {y'}}\, \,  \,  + \Bigl({{y''} \over {y'}}\Bigr)'.
\end{eqnarray}
Taking the square of  (\ref{relation7}) one gets (up to a factor 2):
\begin{eqnarray}
\label{relation92}
\hspace{-0.95in}&&   \, \,  \,  \quad \quad 
 {{1} \over {2}} \cdot \, \Bigl( {{F'(y) } \over {F(y)}}  \Bigr)^2 \cdot \, y'^2
   \, \, = \, \, \, \, {{1} \over {2}} \cdot \,  \Bigl(  {{F'(x)} \over {F(x)}}  \Bigr)^2
  \, +  {{1} \over {2}} \cdot \,  \Bigl({{ y''} \over {y'}}\Bigr)^2
  +  {{F'(x)} \over {F(x)}}   \cdot \,{{ y''} \over {y'}}.
\end{eqnarray}
From (\ref{relation1111}) and (\ref{relation92}) we deduce:
\begin{eqnarray}
\label{relation1111b}
 \hspace{-0.95in}&& 
 \Bigl(  {{F''(y)} \over {F(y)}}  \,
 - {{1} \over {2}} \cdot \, \Bigl( {{F'(y) } \over {F(y)}}  \Bigr)^2 \Bigr)  \cdot \, y'^2
 \, \, = \, \, \, \, {{F''(x)} \over {F(x)}} \,\,\, 
 -{{1} \over {2}} \cdot \,  \Bigl(  {{F'(x)} \over {F(x)}}  \Bigr)^2   \,\, \,
  + \Bigl({{y''} \over {y'}}\Bigr)' -  {{1} \over {2}} \cdot \,  \Bigl({{ y''} \over {y'}}\Bigr)^2,
 \nonumber
\end{eqnarray}
or, recalling the Schwarzian derivative,
\begin{eqnarray}
\label{relation1111b}
 \hspace{-0.95in}&& \quad  \,\,  \,\, 
 {{F''(x)} \over {F(x)}} \,\,  -{{1} \over {2}} \cdot \,  \Bigl(  {{F'(x)} \over {F(x)}}  \Bigr)^2\,\,  \,  
 -\Bigl(  {{F''(y)} \over {F(y)}}
 \, - {{1} \over {2}} \cdot \, \Bigl( {{F'(y) } \over {F(y)}}  \Bigr)^2 \Bigr)  \cdot \, y'^2
  \,\, \, \,  + \{ y(x), \, x\} \, \, = \, \, \, 0,
 \nonumber
\end{eqnarray}
or, more generally, using  (\ref{relation}), which allows to introduce
an extra parameter $\, \alpha$
\begin{eqnarray}
\label{relation1111ab}
 \hspace{-0.95in}&& \quad  \quad  \,  \quad   \,
 {{F''(x)} \over {F(x)}} \,\,  -{{1} \over {2}} \cdot \,  \Bigl(  {{F'(x)} \over {F(x)}}  \Bigr)^2
 \,\,   +{{1} \over {2}} \cdot \,   {{\alpha^2 } \over {F(x)^2}}
  \\
 \hspace{-0.95in}&& \quad  \quad  \quad  \,  \quad \quad \quad  
 -\Bigl(  {{F''(y)} \over {F(y)}}
 \, - {{1} \over {2}} \cdot \, \Bigl( {{F'(y) } \over {F(y)}}  \Bigr)^2
 \, \,  +{{1} \over {2}} \cdot \,   {{\alpha^2 } \over {F(y)^2}}  \Bigr)  \cdot \, y'^2
\,  \,\, \,  + \{ y(x), \, x\}       \, \, = \, \, \, 0.
\nonumber 
\end{eqnarray}
Note that (\ref{relation1111ab}) is actually of the form
\begin{eqnarray}
\label{condition1n1sec6}
\hspace{-0.95in}&& \quad  \quad \quad  \quad \quad  \quad  
 W(x)  \, \, \,  \, -W(y(x)) \cdot  \, y'(x)^2 \, \, \,  \,+ \,  \{ y(x), \, x\} 
\, \,\, \, = \,\, \, \,  \, 0, 
\end{eqnarray}
where ($A_R$ given by (\ref{relation5})):
\begin{eqnarray}
\label{wherecondn1alpha}
\hspace{-0.95in}&& \,   \, \,
W(x)  \, \, = \, \,  \, \, \,
{{F''(x)} \over {F(x)}} \,\,  \, -{{1} \over {2}} \cdot \, \Bigl({{F'(x)} \over {F(x)}}\Bigr)^2
\,\, \,   +{{1} \over {2}} \cdot \,   {{\alpha^2 } \over {F(x)^2}}
\, \, \, = \, \,  \, \, \,
A_R'(x) \, \, \,  + \, \,  {{A_R(x)^2} \over {2 }}.
\end{eqnarray}
{\bf Remark \ref{EvenmoreGenerally}.1:} Note that these calculations {\em also work} with 
\begin{eqnarray}
\label{relationmu}
  \hspace{-0.95in}&& \quad   \quad \quad \quad \quad \quad \quad \quad \quad \quad \quad 
  \mu \cdot \,   F(y(x)) \, \, \, = \, \, \,\, \, F(x) \cdot \, {{d y(x)} \over {d x}}, 
\end{eqnarray}
which corresponds to (\ref{modularform2explicitnothingbut}), (\ref{modularform3explicitnothingbut}),
(\ref{modularform5explicitnothingbut}), (\ref{modularform4explicitnothingbut}). 

\vskip .1cm 

\section{An ``academical'' Schwarzian equation: $\, W(x)$ is no longer a rational function}
\label{academical}

Recalling
\begin{eqnarray}
\label{Fxbis}
\hspace{-0.95in}&& \,  \, \quad  \quad  \quad  \quad 
 F(x) \, \, = \, \, \, \,
 x \cdot \, (1\, -1728 \cdot \, x)^{1/2} \cdot \,
 _2F_1\Bigl([{{1} \over {12}}, \, {{5} \over {12}}], \, [1], \, 1728 \cdot \, x   \Bigr)^2,
\end{eqnarray}
the one-parameter series $\, y(x) \, = \, \, y(a, \, x)$, given by (\ref{family_a}), is,
for any value of $\, \alpha$, solution of the rank-two equation
\begin{eqnarray}
\label{rota1bis}
\hspace{-0.95in}&& \quad \quad  \quad  \quad  \quad   \quad   \quad   \quad 
 A_R(x) \, \,  \,  -A_R(y(x)) \cdot  \, y'(x)  \, \,  \,  + {{y''(x)} \over {y'(x)}}
\, \,\,  = \,\, \,  \,  0, 
\end{eqnarray}
with
\begin{eqnarray}
\label{rota1alpha}
\hspace{-0.95in}&& \quad \quad  \quad    \quad   \quad \quad  \,  \quad  \quad  \quad   \quad  
A_R(x)    \, \, \, = \, \, \, \, \, {{F'(x)} \over {F(x)}}
\, \, \, \, +{{\alpha} \over {F(x)}},     
\end{eqnarray}
but is also solution of  the Schwarzian condition
\begin{eqnarray}
\label{condition1n1bis}
\hspace{-0.95in}&& \quad \quad  \quad \,  \quad \quad  \quad  \quad  
 W(x)  \, \, \,  \, -W(y(x)) \cdot  \, y'(x)^2 \, \, \,  \,+ \,  \{ y(x), \, x\} 
\, \,\, \, = \,\, \, \,  \, 0, 
\end{eqnarray}
where:
\begin{eqnarray}
\label{wherecondn1bis1}
\hspace{-0.95in}&& \quad   \quad \quad   
W(x)  \, \, = \, \,  \, \, \,
{{F''(x)} \over {F(x)}} \, \, -{{1} \over {2}} \cdot \, \Bigl({{F'(x)} \over {F(x)}}\Bigr)^2
\, \, +{{1} \over {2}} \cdot \, {{\alpha^2} \over {F(x)}}
 \, \, = \, \,  \, \, \,   A_R'(x) \, \, \,  + \, \,  {{A_R(x)^2} \over {2 }}
\nonumber \\
\hspace{-0.95in}&& \quad   \quad \quad \quad \quad \quad \quad
 \, \, = \, \,  \, \, \,
 - {{1} \over {2}} \cdot \,
 {{ 1 \, -1968\, x  \, +2654208\, x^2} \over { x^2 \cdot \, (1 \, - 1728\, x)^2 }}
 \,  \,\, \, +{{1} \over {2}} \cdot \, {{\alpha^2} \over {F(x)}}.                               
\end{eqnarray}
For {\em generic} values of $\, \alpha$, the solution-series
of the form $\, \, a \cdot \, x \, \, + \cdots$, 
of the rank-two equation (\ref{rota1bis}) with (\ref{rota1alpha}),
as well as the Schwarzian equation (\ref{condition1n1bis}),
with $\, W(x)$ given by (\ref{wherecondn1bis1}), 
is just the one-parameter series $\, y(x) \, = \, \, y(a, \, x)$, given by (\ref{family_a}).
However, for a selected set of values of $\, \alpha$, namely (non-zero) {\em  integer values},
the solution-series of the form $\, \, a \cdot \, x \, \, + \cdots$, becomes a {\em two-parameters series}.
For instance, for $\, \alpha \, = \, \pm \, 1$, the extra parameter occurs with the coefficient of $\, x^2$,
for $\, \alpha \, = \, \pm \, 2\, $ the extra parameter occurs with the coefficient of $\, x^3$, ...
and, more generally, for $\, \alpha \, = \, \pm \, N\, $ the extra parameter occurs
with the coefficient of $\, x^{N+1}$. Let us display the $\, \alpha =\, 1\, $ case in detail. 

\vskip .1cm 

\subsection{The $\alpha =1$ case: two-parameters series}
\label{alpha1}

Let us consider the case $\, \alpha \, = \, 1$ in the Schwarzian equation (\ref{condition1n1bis}) with
$\, W(x)$ given by (\ref{wherecondn1bis1}), 
or in the rank-two relation (\ref{rota1bis}) with (\ref{rota1alpha}). 

The {\em two-parameter} series
\begin{eqnarray}
\label{family_ab}
\hspace{-0.98in}&& \,  \, \quad \quad  \, 
 y(a, \, b, \, x) \, \, = \, \, \,\, \,\,
  a \cdot \, x \, \, \, \, \, +\Bigl(1728 \cdot \, b \, -744 \cdot  \, a\cdot  \, (a-1)\Bigr) \cdot  \, x^2
 \nonumber \\
  \hspace{-0.98in}&& \quad \quad \quad \quad \quad  \quad  \quad 
\, \,  +\Bigl(2985984 \cdot \, a  \cdot \, b^2 \, \, -2571264 \cdot \, a\cdot \, (a-1) \cdot \, b
 \\
\hspace{-0.98in}&& \quad \quad \quad \quad \quad \quad \quad \quad  \quad  \quad 
    \, \,  +36 \cdot   \, a \cdot  \, (a-1) \cdot \, (9907\, a-20845) \Bigr)\cdot \, x^3
 \nonumber \\
  \hspace{-0.98in}&& \,  \, \quad \quad \, 
  \, \, +\Bigl( 5159780352\cdot \, a\cdot \, b^3 \,\, \, -6664716288\cdot \, a \cdot \, (a-1) \cdot \, b^2
\nonumber \\
  \hspace{-0.98in}&& \quad \quad \quad  \quad \quad
   \, +186624 \cdot \, (9907\,a^2-30752\,a +19022) \cdot \, a \cdot \, b
\nonumber    \\
\hspace{-0.98in}&& \quad \quad \quad \quad  \quad \quad 
 -32 \cdot  \, a \cdot  \, (a-1) \cdot \, (4386286\, a^2-20490191\, a+27274051)  \Bigr)\cdot \, x^4
 \, \,\, \, \, + \, \, \, \cdots
\nonumber 
\end{eqnarray}
is {\em actually solution of the  Schwarzian equation} (\ref{condition1n1bis}) with
$\, W(x)$ given by (\ref{wherecondn1bis1}),
or of the rank-two relation (\ref{rota1bis}) with (\ref{rota1alpha}), for  $\, \alpha \, = \, 1$.
Note that the {\em two-parameter} series (\ref{family_ab}) is also
solution\footnote[1]{However it is {\em not} solution of $\, F(x) \cdot \, y' = \, F(y)\, $
  or  $\, F(x) \cdot \, y' = \, a \, {{\partial y} \over {\partial a}}$.} of
\begin{eqnarray}
\label{wherecondn3b}
  \hspace{-0.95in}&&  \quad \quad \, \, \, \, \,  \quad \quad \quad
  a \cdot \, {{\partial y(a, \, b, \, x)} \over { \partial a}}  \,\,\,
  +b \cdot \, {{\partial y(a, \, b, \, x)} \over { \partial b}}
  \, \, = \, \, \,  F\Bigl(y(a, \, b, \, x)\Bigr),
\end{eqnarray}
with $\, F(x)$ given by (\ref{Fxbis}).
We have the following composition rules for the {\em two-parameter} series (\ref{family_ab}):
\begin{eqnarray}
\label{family_abother}
\hspace{-0.98in}&& \,  \,  \quad \quad \quad  \quad \quad \quad \quad
y\Bigl(a', \, b', \,  y(a, \, b, \, x)\Bigr)
\, \, = \, \, \,\, \, y\Bigl(a\,a', \, a^2 \, b' \, +a' \, b, \, x\Bigr). 
\end{eqnarray}

Let us introduce an {\em alternative parametrization} of the two-parameter
series (\ref{family_ab}), changing $\, b$ into $\, a \, b$,  in (\ref{family_ab}):
\begin{eqnarray}
\label{family_abY}
\hspace{-0.98in}&& \,  \, \quad \quad \quad  \, 
 Y(a, \, b, \, x) \, \, = \, \, \,\, \,\,
 a \cdot \, x \, \, \, \, \, +\Bigl(1728 \cdot \, a  \, b \,
 -744 \cdot  \, a\cdot  \, (a-1)\Bigr) \cdot  \, x^2
 \nonumber \\
  \hspace{-0.98in}&& \quad \quad \quad \quad \quad  \quad  \quad 
\, \,  +\Bigl(2985984 \cdot \, a^3  \cdot \, b^2 \, \, -2571264 \cdot \, a^2 \cdot \, (a-1) \cdot \, b
 \\
\hspace{-0.98in}&& \quad \quad \quad \quad \quad \quad \quad \quad  \quad  \quad 
\, \,  +36 \cdot   \, a \cdot  \, (a-1) \cdot \, (9907\, a-20845) \Bigr)\cdot \, x^3
\,\,\, \,\, + \, \, \, \cdots
\nonumber
\end{eqnarray}
We have the following composition rules for the {\em two-parameter} series (\ref{family_abY})
\begin{eqnarray}
\label{family_abotherY}
\hspace{-0.98in}&& \,  \,  \quad \quad \quad  \quad \quad \quad \quad
Y\Bigl(a', \, b', \,  Y(a, \, b, \, x)\Bigr)
 \, \, = \, \, \,\, \, Y\Bigl(a\,a', \, a\, b' \, +b, \, x\Bigr). 
\end{eqnarray}
The series (\ref{family_abY}) is, now, solution of:
\begin{eqnarray}
\label{wherecondn3b}
  \hspace{-0.95in}&&  \quad \quad \quad  \quad \quad \quad \quad \quad \quad
  a \cdot \, {{\partial Y(a, \, b, \, x)} \over { \partial a}}  \,
    \, \, = \, \, \,  F\Bigl(Y(a, \, b, \, x)\Bigr),
\end{eqnarray}
with $\, F(x)$ given by (\ref{Fxbis}).
Let us introduce the  $\, a \, \rightarrow \, 0\, $  limit:
\begin{eqnarray}
\label{compositiona0ab}
  \hspace{-0.95in}&& 
  Q_b(x) \, \, = \, \, \,
  \lim_{a \rightarrow \, 0}  {{  Y(a, \, b, \, x) } \over {a}}
\nonumber \\
  \hspace{-0.95in}&& \,  \, \quad  \,  \, \, 
  \, \, = \, \, \, \, \, \,
 x \, \, \,  \,+(744 +1728\, b)  \cdot \, x^2\,\, \,
 +(750420 +2571264\,b +2985984\,b^2) \cdot \, x^3
\nonumber \\
  \hspace{-0.95in}&& \,  \, \quad  \,  \, \, 
  +(872769632 +3549961728\,b \, +6664716288\,b^2 +5159780352\,b^3) \cdot \, x^4
  \nonumber \\
  \hspace{-0.95in}&& \,  \, \, \quad  \, \,  \,\,\, \,
  +(1102652742882 +4945819779072\,b +11680775258112\,b^2
  \nonumber \\
\hspace{-0.95in}&& \,  \quad  \quad  \quad  \, \,  \, \, 
  +15355506327552\,b^3 +8916100448256\,b^4) \cdot\, x^5
\nonumber \\
  \hspace{-0.95in}&& \,  \, \,  \, \, 
  +(1470561136292880 +7027977959274240\, b +19050621395927040\, b^2
  \nonumber \\
\hspace{-0.95in}&& \,   \quad  +32624754548539392\, b^3
+33167893667512320\, b^4 +15407021574586368\, b^5) \cdot \, x^6
 \nonumber \\
\hspace{-0.95in}&& \,  \quad  \quad  \quad
\,\,  
 \,\,\, \, \,  \, + \,\, \, \cdots 
\end{eqnarray}
In the $\, b\, \rightarrow \, 0$ limit, this series (\ref{compositiona0ab})
reduces to the nome series (\ref{mirror21}) or (\ref{compositiona0}).

In the $\, a \, \rightarrow \, \infty \,\, $ limit one gets:
\begin{eqnarray}
\label{compositionainftyab}
\hspace{-0.95in}&& \,\,\,\,
X_b(x)  \, \, = \, \, \,
\lim_{a \rightarrow \, \infty}  \, Y\Bigl(a, \, b, \, {{x} \over{a}} \Bigr) \, \, \, = \, \, \, \,\, 
x \, \,\,\,  -744\, x^2 \, \,  \,+356652 \, x^3\, \,\, -140361152\, x^4
\nonumber \\
\hspace{-0.95in}&& \,  \quad  \quad  \, \, 
\,\, \, +49336682190\, x^5 \,\, -16114625669088\, x^6 \,\, \, +4999042477430456 \, x^7
 \nonumber \\
 \hspace{-0.95in}&& \,  \,  \quad \quad  \quad  \quad
 -1492669384085015040\, x^8 \,
+432762759484818142437\, x^9
\,\, \,\,\,  \, + \,\, \, \cdots 
\end{eqnarray}
This series (\ref{compositionainftyab}) is nothing but (\ref{mirror1}) or (\ref{compositionainfty}),
and, thus, {\em does not depend} on the second parameter $\, b$.

One actually finds that the two parameter series (\ref{family_abY}) is {\em nothing but}:
\begin{eqnarray}
\label{onefindsout}
\hspace{-0.95in}&& \quad  \quad  \quad   \quad  \quad   \quad  \quad \quad  \quad 
 Y(a, \, b, \, x) \, \, = \, \, \,   X_b\Bigl( a \cdot \,   Q_b(x) \Bigr).   
\end{eqnarray}
From (\ref{onefindsout}) we can also deduce that (\ref{wherecondn3b}), is,
in fact, nothing but  equation:
\begin{eqnarray}
\label{wherecondn3brewrite}
\hspace{-0.95in}&&  \quad \quad \quad  \quad \quad \quad \quad \quad \quad
 a \cdot \, {{\partial X_b(a \cdot  x)} \over { \partial a}}  \,
 \, \, = \, \, \,  F\Bigl(X_b(a \cdot  x)\Bigr).
\end{eqnarray}
Furthermore, since
$\, \, \, a \cdot \, {{\partial X_b(a \cdot  x)} \over { \partial a}}  \,  \, \, = \, \, \,
x \cdot \, {{\partial X_b(a \cdot  x)} \over { \partial x}}$,
relation (\ref{wherecondn3brewrite}) also gives:
\begin{eqnarray}
\label{wherecondn3brewritefurthermore}
  \hspace{-0.95in}&&  \quad \quad \quad  \quad \quad \quad \quad \quad \quad
  x \cdot \, {{\partial X_b(a \cdot  x)} \over { \partial x}}
  \, \, = \, \, \,  F\Bigl(X_b(a \cdot  x)\Bigr).
\end{eqnarray}

In contrast with the $\, b=\, 0 \,$ case, the two functions, $\, Q_b$ and $\, X_b$,
given by the two limits (\ref{compositiona0ab}),  (\ref{compositionainftyab}),
are {\em not compositional inverse}. 
In the $\, a \, \rightarrow \, \, 1\, $ limit, the decomposition (\ref{onefindsout}) becomes:
\begin{eqnarray}
\label{onefindsouta1}
\hspace{-0.95in}&&   \quad 
Y(1, \, b, \, x) \, \, = \, \, \, X_b\Bigl(  Q_b(x) \Bigr)
\nonumber \\
 \hspace{-0.95in}&& \,  \,  \quad \quad
 = \, \, \,
 x \, \,\, +1728 \cdot \, b \cdot \, x^2 \,
+2985984 \cdot \, b^2 \cdot \, x^3 \,
+186624 \cdot \, (27648\,b^2 \, -1823) \cdot \, b \cdot \, x^4
\nonumber \\
 \hspace{-0.95in}&& \,  \,  \quad \quad \quad \quad 
\, + 110592 \cdot \, (80621568\,b^3 \, -15947604\,b \, -5249233) \cdot \, b \cdot \, x^5
\, \,  \, \, \, + \, \, \, \cdots 
\end{eqnarray}
The series (\ref{onefindsouta1}) is a one-parameter family of commuting series:
\begin{eqnarray}
\label{onefindsouta1YY}
\hspace{-0.95in}&&   \quad  \quad  \quad  \, \,
Y\Bigl(1, \, b  , \, Y(1, \, b', \, x) \Bigr) \, \, = \, \, \,
Y\Bigl(1, \, b'  , \, Y(1, \, b, \, x) \Bigr) \, \, = \, \, \,
Y(1, \, b \, +b', \, x).
\end{eqnarray}
In particular the compositional inverse of $\, \,  Y(1, \, b, \, x)\, $
is  $\,\,   Y(1, \, -b, \, x)$:
\begin{eqnarray}
\label{onefindsouta1compoinverse}
\hspace{-0.95in}&&   \quad  \quad  \quad \quad \quad 
Y\Bigl(1, \, b  , \, Y(1, \, -b, \, x) \Bigr) \, \, = \, \, \,
Y\Bigl(1, \,  -b , \, Y(1, \, b, \, x) \Bigr) \, \, = \, \, \, x. 
\end{eqnarray}
Note that:
\begin{eqnarray}
\label{onefindsouta1QX}
\hspace{-0.95in}&&   \quad   \quad  \quad 
Q_b\Bigl(  X_b(x) \Bigr) \, \, = \, \, \,
{{x } \over { 1 \, \, -1728 \cdot  \, b \cdot \, x}} 
\, \, \, = \, \, \, \,\,\,
 x \, \, \, \,+1728 \cdot \, b \cdot \, x^2\, \,\, \, + \,  \cdots 
\end{eqnarray}
From (\ref{onefindsouta1QX}) we deduce an alternative expression for $\, Q_b(x)$
in terms of the nome (\ref{mirror2})
(i.e. the compositional inverse of (\ref{compositionainftyab}), or, equivalently
$\, Q_b(x)$ for $\, b= \, 0$):   
\begin{eqnarray}
\label{onefindsouta1QXalternat}
\hspace{-0.95in}&&   \quad   \quad  \quad \quad  \quad  \quad  \quad   \quad  \quad   \quad 
Q_b(x) \, \, = \, \, \,
{{Q_0(x)  } \over { 1 \, \, -1728 \cdot  \, b \cdot \, Q_0(x)}}. 
\end{eqnarray}
Conversely, the nome (\ref{mirror2}), {\em which does not depend on the parameter} $\, b$,
can be simply expressed in terms of the series (\ref{compositiona0ab}):
\begin{eqnarray}
\label{onefindsouta1QXalternatconversely}
\hspace{-0.95in}&&   \quad   \quad  \quad \quad  \quad   \quad  \quad \quad \quad   \quad 
Q_0(x) \, \, = \, \, \,
{{Q_b(x)  } \over { 1 \, \, +1728 \cdot  \, b \cdot \, Q_b(x)}}. 
\end{eqnarray}
Note that the composition rule relation (\ref{family_abotherY}) can, now,
be seen as a straightforward consequence of
relation (\ref{onefindsouta1QXalternat}).  
From relation (\ref{onefindsouta1QXalternat}) one can see that the radius of convergence of
the series (\ref{compositiona0ab}) corresponds, for small enough values of the
additional parameter $\, b$, to the singularity of $\, Q_0(x)$, (i.e. $\, R = \, 1/1728$),
and for large
enough values of the parameter $\, b$, to the singularity $\, Q_0(x) \, = \, 1/1728/b$,
namely:
\begin{eqnarray}
\label{secondsingul}
\hspace{-0.95in}&&   \quad   \quad   \quad   \quad  \quad  \quad  \quad \quad  \quad
x \, \, = \, \, \, X_b\Bigl( {{1} \over { 1728 \, b}}  \Bigr)
\, \, = \, \, \,  \tilde{X}\Bigl( {{1} \over { 1728 \, b}}  \Bigr). 
\end{eqnarray}

\vskip .1cm 

{\bf Remark \ref{academical}.1:} Do note that, in contrast
with the $\, \alpha\, = \, \, 0\, $ case,
{\em there is no} solution-series of the form
$\, a \cdot \, x^2 \, + \, \, \cdots $ or, more generally, of the form
$\, a \cdot \, x^N \, + \, \, \cdots\,\,  $
with $\, N\, \ne \, 1$, of the   Schwarzian equation (\ref{condition1n1bis}),
when $\, W(x)$ is given by  (\ref{wherecondn1bis1}). This corresponds
to the fact that, when $\, \alpha \, \ne \, 0$,  $\, W(x)$ is no
longer of the form $\, W(x) = \, -1/2/x^2 \, + \, \cdots$ (see~\cite{Schwarz1,Schwarz2}). 

\vskip .1cm 

\section{Polynomial examples for  $\, F(x)$.}
\label{polyexamplesF}

Modular correspondences, modular curves, correspond to a (transcendental)
function $\, F(x)$ associated to elliptic functions like (\ref{Fx}), (\ref{Fxbis}).

\ref{beyond} provides a (non globally bounded) Heun function example showing that the previous results
and calculations also work, mutatis mutandis
with Shimura curves~\cite{Shimura} (and their associated automorphic forms~\cite{Ford}).

Let us now recall
the general results of section (\ref{epsexp}),  which describes
the {\em one-parameter} solution-series (\ref{seriesmodcurve1aeps1}) of
the Schwarzian equation (\ref{condition1n1}), and also the partial differential
equations (\ref{wherecondn3}), (\ref{wherecondn3also}), and the fact that
these equations {\em are actually valid for any function} $\, F(x)$.

Let us consider, here, one-parameter functions $\, y(a, \, x)$, corresponding to
miscellaneous {\em polynomial} examples of functions $\, F(x)$,
that are, thus, far from being associated with the previous ``classical''
modular forms~\cite{Koblitz,1-2-3,SuperMaier} and
hypergeometric/elliptic functions~\cite{Jacobi,Nova}, or even
Shimura curves/automorphic forms examples, possibly with Heun functions~\cite{GlobNilp}
(see \ref{beyond}). Even
more simple polynomial examples are
given in \ref{miscellancorresp}.  

From the  general results of the previous
section (\ref{epsexp})
we will, thus, get a set of miscellaneous examples.
All the corresponding  one-parameter series, below, will verify the  composition rule:
\begin{eqnarray}
\label{composition2}
\hspace{-0.95in}&& \,  \, \quad  \quad \quad  \quad  \quad  \quad \quad \quad \quad \quad
 y\Bigl(a, y(a', \, x)\Bigr) \, \, = \, \,  \, y(a\, a', \, \, x).
\end{eqnarray}
All these one-parameter series will also verify: 
\begin{eqnarray}
  \label{family_a6also}
\, 
  F \Bigl(y(a, \, x) \Bigr) \, \, = \, \, \,  a \cdot \,  {{\partial y(a, \, x) } \over { \partial a}}
\, \, = \, \, \, F(x) \cdot  {{\partial y(a, \, x) } \over { \partial x}}.
\end{eqnarray}

One will also consider a polynomial that will be the truncation of the hypergeometric
function  (\ref{Fx}). One will, then, get a one-parameter solution-series,
very similar\footnote[2]{The three first terms are the same.}
to (\ref{family_a}),  which also verifies the  composition rule (\ref{composition}),
but does not correspond to globally bounded series~\cite{Christol}. 

\subsection{A first simple polynomial example for $\, F(x)$}
\label{another}

Let us first consider the following polynomial expression for $\, F(x)$:
\begin{eqnarray}
  \label{family_another}
   \hspace{-0.95in}&& \quad  \quad \quad \quad \quad \quad   \quad  \quad  \quad
  F(x) \, \, = \, \, \, \, x \cdot \, (1 \, - 2 \, x) \cdot \, (1 \, -3\, x). 
\end{eqnarray}
One deduces, from (\ref{whereWQ}), (\ref{defW}), 
the following rational expression for $\, W(x)$:
\begin{eqnarray}
  \label{family_anotherW}
  \hspace{-0.95in}&& \quad   \quad  \quad \quad   \quad \quad  \quad
  W(x) \, \, = \,\,  \, \,
  -{{1} \over {2}} \cdot \, {{1 \, -36\, x^2 \, +120\, x^3 \, -108\, x^4 } \over {
      x^2 \cdot \, (1 \, - 2 \, x)^2 \cdot \, (1 \, -3\, x)^2 }}. 
\end{eqnarray}
The Schwarzian condition
\begin{eqnarray}
\label{condition1n1another}
\hspace{-0.95in}&& \quad \quad  \quad \,  \quad \quad  \quad  \quad  
 W(x)  \, \, \,  \, -W(y(x)) \cdot  \, y'(x)^2 \, \, \,  \,+ \,  \{ y(x), \, x\} 
\, \,\, \, = \,\, \, \,  \, 0, 
\end{eqnarray}
has the following one-parameter solution
\begin{eqnarray}
  \label{condition1n1anothersol1}
  \hspace{-0.95in}&& \quad  
  y(a, \, x) \, \, = \, \, \,  \,  \,  a \cdot \, x \, \,  \,  \,\,
  -5 \cdot \, a \cdot \, (a-1) \cdot \, x^2 \, \,\,
  +2 \cdot \, a \cdot \, (14\, a-11)\, (a-1) \cdot \, x^3
 \nonumber \\
 \hspace{-0.95in}&& \quad \quad 
  \, -15 \cdot \, a \cdot \, (a-1)^2\,  (11\, a-6) \cdot \, \, x^4  \,\, \,
 +a \cdot \, (a-1)^2 \cdot \,  (1001\, a^2-1298\, a+351) \cdot \, x^5
 \nonumber \\
 \hspace{-0.95in}&& \quad \quad \, \quad
  -7 \cdot \, a \cdot \,(a-1)^3 \cdot  (884\, a^2-923\, a+189) \cdot \, x^6
 \\
  \hspace{-0.95in}&& \quad \quad \quad \quad
 \, +60 \cdot \, a \cdot \, (a-1)^3 \cdot \, (646\, a^3-1156\, a^2+600\, a-81)  \cdot \, x^7
  \nonumber \\
  \hspace{-0.95in}&& \quad \quad \, \quad \quad \quad
  -3 \cdot \, a \cdot \, (a-1)^4 \cdot \, (81719\, a^3-125324\, a^2+54162\, a-5832) \cdot \,\, x^8
\, \, \,\, \, + \, \, \, \cdots
 \nonumber 
\end{eqnarray}
as well as modular equation-like series (with no parameter) like
\begin{eqnarray}
  \label{condition1n1anothersol2}
  \hspace{-0.95in}&& \quad \quad \quad
  y_2 \, \, = \, \, \,\,\,
  x^2 \,\,  \, +10\, x^3 \,\, \, +64\, x^4 \,\,  \, +300\, x^5 \, \, \,
  +924\, x^6 \, \,\, +56\, x^7 \, \, \, -24140\, x^8
\nonumber \\
  \hspace{-0.95in}&& \quad \quad \quad \quad \quad \quad
 \, -209856\, x^9 \,\, -1158600\, x^{10}
  \,\, \,\, \, + \, \, \, \cdots 
\end{eqnarray}
or:
\begin{eqnarray}
\label{condition1n1anothersol3}
\hspace{-0.95in}&& \quad \quad \, \, 
y_3 \, \, = \, \, \,\,\,\,  x^3 \,\, \, \, +15\, x^4 \,\, +141\, x^5 \,\,
+1050\, x^6 \,\,  +6705\, x^7 \,\, +37854 \, x^8 \,\,  +189603\, x^9
\nonumber \\
  \hspace{-0.95in}&& \quad \quad \quad  \quad \quad \quad
  \, +820584\, x^{10} \, \, +2777004\, x^{11} \,\,  +4024890\, x^{12}
 \,\, \,\,\, + \, \, \, \cdots 
\end{eqnarray}                   
More generally, for the solution series of the form
$\, x \, \rightarrow \, \, x^p \, + \, \, \cdots \,\,\,$ we have:
\begin{eqnarray}
\label{condition1n1anotherdx}
  \hspace{-0.95in}&& \quad \quad  \quad \quad  \quad \quad  \quad \quad \quad \quad \quad \quad
 p \cdot \,  {{dx} \over {F(x)}} \, \, \, = \, \, \,  {{dy} \over {F(y)}}, 
\end{eqnarray}
From (\ref{family_another}) we get:      
\begin{eqnarray}
\label{condition1n1anotherdxexpint}
\hspace{-0.95in}&& \quad \,   \quad \quad  \quad \quad \quad \quad \quad
\exp\Bigl(\int {{dx} \over {F(x)}}\Bigr)
 \, \, \, = \, \, \,  \mu \cdot \, {{ x \cdot \, (1 \, -2 \, x)^2} \over { (1 \, -3 \, x)^3}}.
\end{eqnarray} 
From (\ref{condition1n1anotherdx}), and (\ref{condition1n1anotherdxexpint}), we get the {\em algebraic curve}: 
\begin{eqnarray}
\label{condition1n1anotherdxcond}
  \hspace{-0.95in}&& \quad \quad  \quad \quad  \quad \quad \quad \quad \quad
   \rho \cdot \,    \Bigl({{ x \cdot \, (1 \, -2 \, x)^2} \over { (1 \, -3 \, x)^3}} \Bigr)^p
  \, \, \, = \, \, \,    {{ y \cdot \, (1 \, -2 \, y)^2} \over { (1 \, -3 \, y)^3}}. 
\end{eqnarray} 
One finds that $\, y_2$, given by (\ref{condition1n1anothersol2}),
is actually solution of the algebraic condition (\ref{condition1n1anotherdxcond}) 
for $\, p\, = \, 2$ and $\, \rho \, = \, \, 1$,
namely:
\begin{eqnarray}
\label{condition1n1anotherdxcond2}
  \hspace{-0.95in}&& \quad \quad  \quad    \quad \quad \quad \quad\quad  \quad
    \Bigl({{ x \cdot \, (1 \, -2 \, x)^2} \over { (1 \, -3 \, x)^3}} \Bigr)^2
  \, \, \, = \, \, \,    {{ y \cdot \, (1 \, -2 \, y)^2} \over { (1 \, -3 \, y)^3}}. 
\end{eqnarray}
One also finds that 
 $\, y_3$, given by (\ref{condition1n1anothersol3}), is actually solution
of the algebraic condition (\ref{condition1n1anotherdxcond}), 
for $\, p\, = \, 3$ and $\, \rho \, = \, \, 1$,
namely:
\begin{eqnarray}
\label{condition1n1anotherdxcond3}
  \hspace{-0.95in}&& \quad \quad  \quad \quad  \quad \quad \quad \quad \quad  \quad
    \Bigl({{ x \cdot \, (1 \, -2 \, x)^2} \over { (1 \, -3 \, x)^3}} \Bigr)^3
  \, \, \, = \, \, \,    {{ y \cdot \, (1 \, -2 \, y)^2} \over { (1 \, -3 \, y)^3}}. 
\end{eqnarray}

On the other hand one finds that the one-parameter series
$\, y(a, \, x)$, given by (\ref{condition1n1anothersol1}),
is actually solution of the algebraic condition (\ref{condition1n1anotherdxcond}) 
for $\, p\, = \, 1\, $ and $\, \rho \, = \, \, a$,
namely:
\begin{eqnarray}
\label{condition1n1anotherdxconda}
  \hspace{-0.95in}&& \quad \quad  \quad \quad  \quad  \quad \quad \quad \quad \quad
   a  \cdot  \Bigl({{ x \cdot \, (1 \, -2 \, x)^2} \over { (1 \, -3 \, x)^3}} \Bigr)
  \, \, \, = \, \, \,    {{ y \cdot \, (1 \, -2 \, y)^2} \over { (1 \, -3 \, y)^3}}.        
\end{eqnarray}
{\bf Remark \ref{polyexamplesF}.1:}  Note, from (\ref{condition1n1anotherdxconda}),
 that the one-parameter series $\, y(a, \, x)$,
given by  (\ref{condition1n1anothersol1}),
is actually an {\em algebraic series} for {\em any value of} the parameter
$\, a$ (and not only $ \, N$-th root of unity).
The algebraic equations (\ref{condition1n1anotherdxcond2}) and  (\ref{condition1n1anotherdxcond3}), 
and their corresponding algebraic series solutions
(\ref{condition1n1anothersol2}) and  (\ref{condition1n1anothersol3}), 
could be seen to be the ``equivalent'' of the modular equations (\ref{modularequa}) and  (\ref{modularequa3}),
and their corresponding algebraic series solutions (\ref{modularequasol1}) and  (\ref{modularequasol13}).
However, one should note that the {\em modular equations} (\ref{modularequa}) and  (\ref{modularequa3})
are $\, x \, \leftrightarrow \, y$ symmetric, and, consequently,  the modular equation (\ref{modularequa})
 represents $\, q \, \rightarrow \, q^2 \, $ {\em and}  $\, q \, \rightarrow \, q^{1/2}\, $
{\em in the same time} (see series (\ref{modularequasol1}) {\em but also} (\ref{modularequasol2})).
Similarly the modular equation (\ref{modularequa3}) represents
$\, q \, \rightarrow \, q^3 \, $ {\em and}  $\, q \, \rightarrow \, q^{1/3} \, $ {\em in the same time}
(see  series (\ref{modularequasol13}) {\em and also} (\ref{modularequasol13b})).
In contrast (\ref{condition1n1anotherdxcond2}) and (\ref{condition1n1anotherdxcond3})  {\em break the} 
$\, x \, \leftrightarrow \, y$ {\em  symmetry}. Therefore,
the ``equivalent'' of the modular equation (\ref{modularequa})
is rather
\begin{eqnarray}
\label{rather1}
  \hspace{-0.95in}&& 
\Bigl( \Bigl({{ x \cdot \, (1 \, -2 \, x)^2} \over { (1 \, -3 \, x)^3}} \Bigr)^2
   \, \, - {{ y \cdot \, (1 \, -2 \, y)^2} \over { (1 \, -3 \, y)^3}}\Bigr) \cdot \,
\Bigl( \Bigl({{ y \cdot \, (1 \, -2 \, y)^2} \over { (1 \, -3 \, y)^3}} \Bigr)^2
  \, \, - {{ x \cdot \, (1 \, -2 \, x)^2} \over { (1 \, -3 \, x)^3}}\Bigr) 
 \, \, \, = \, \, \,  0,
 \nonumber 
\end{eqnarray}
when the equivalent of the modular equation (\ref{modularequa3})
is rather:
\begin{eqnarray}
\label{rather2}
  \hspace{-0.95in}&& 
 \Bigl( \Bigl({{ x \cdot \, (1 \, -2 \, x)^2} \over { (1 \, -3 \, x)^3}} \Bigr)^3
  \, \, - {{ y \cdot \, (1 \, -2 \, y)^2} \over { (1 \, -3 \, y)^3}}\Bigr) \cdot \,
\Bigl( \Bigl({{ y \cdot \, (1 \, -2 \, y)^2} \over { (1 \, -3 \, y)^3}} \Bigr)^3
  \, \, - {{ x \cdot \, (1 \, -2 \, x)^2} \over { (1 \, -3 \, x)^3}}\Bigr) 
 \, \, \, = \, \, \,  0.
   \nonumber 
\end{eqnarray}
The series (\ref{condition1n1anothersol1}) verifies the composition rule:  
\begin{eqnarray}
\label{composition2bb}
  \hspace{-0.95in}&& \,  \, \quad  \quad \quad  \quad \quad \quad \quad  \quad   \quad   \quad 
 y\Bigl(a, y(a', \, x)\Bigr) \, \, = \, \, \, y(a\, a', \, \, x).
\end{eqnarray}
The series (\ref{condition1n1anothersol1}) also verifies the relations:
\begin{eqnarray}
\label{wherecondnalsoverif}
  \hspace{-0.95in}&& \quad   \quad \quad \quad \quad \quad \quad
  F\Bigl(y(a, \, x) \Bigr)  \, \, = \, \, \,
  F(x) \cdot \,  {{\partial  y(a, \, x)} \over { \partial x}}
 \, \, = \, \, \,  \, a \cdot \, {{ \partial  y(a, \, x)} \over { \partial a}}.
\end{eqnarray}
Let us introduce the two limits:
\begin{eqnarray}
\label{compositiona0bis1}
  \hspace{-0.95in}&& \quad 
\tilde{Q}(x) \, \, = \, \, \,
 \lim_{a \rightarrow \, 0}  {{ y(a, \, \, x) } \over {a}} \, \, = \, \, \, \, \,
    x \, \, \, \, +5 \, x^2\,\, \,  +22 \, x^3 \, \, \, +90 \, x^4 \,\,   +351\, x^5 \,  \, +1323 \, x^6
 \nonumber  \\
  \hspace{-0.95in}&& \,\quad   \, \,  \, \, \, \quad  \quad  \quad 
 \,\,  + 4860 \, x^7 \, \, \, + 17496 \, x^8 \,\,  +61965 \, x^9 \,\,  +216513 \, x^{10}
 \,\,  \, \,\,\, \,  + \,\, \, \cdots
\end{eqnarray}
\begin{eqnarray}
\label{compositionainftybis1}
  \hspace{-0.95in}&&  \quad 
 \tilde{X}(x) \, \, = \, \, \,
     \lim_{a \rightarrow \, \infty}  \, y\Bigl(a, \, \, {{x} \over {a}}\Bigr)  \, \, = \, \, \,\,
 x \,\,\,  -5 \, x^2 \, \,\, +28 \, x^3 \,\, \, -165\, x^4 \,\, \, +1001 \, x^5 \,\, \,  -6188 \, x^6
 \nonumber  \\
  \hspace{-0.95in}&& \, \quad  \, \,  \, \, \, \, \,  \quad  \quad 
\, \, +38760 \, x^7
\,  \,  \,-245157 \, x^8 \, \,  \, +1562275 \, x^9 \,\,  \, -10015005 \, x^{10}
\, \,\,\,  \,\,  + \,\, \, \cdots
\end{eqnarray}
One can verify that these two series $\,\tilde{Q}(x)\, $ and  $\,\tilde{X}(x) \,$
are {\em compositional inverse}.
The radius of convergence of the ``nome-like'' series $\, \tilde{Q}(x)$, given by
(\ref{compositiona0bis1}), is $\, R \, = \, 1/3$. The radius of convergence
of the series $\, \tilde{X}(x)$,
given by (\ref{compositionainftybis1}), is $\, R \, = \, 4/27$. 

These two series, with {\em integer} coefficients,
are solutions of the two Schwarzian equations
\begin{eqnarray}
\label{Harnad214bisrecall}
\hspace{-0.95in}&& \quad  \quad  \quad   \quad \quad \quad 
 \, \{\tilde{Q}(x) , \, x \} \, \, \, \, 
 +{{1} \over {2  \cdot \, \tilde{Q}(x) ^2 }} 
 \cdot  \Bigl({{ d \tilde{Q}(x) } \over {d x}} \Bigr)^2
\,  \,\, +  \, W(x) \, \,  \, =  \, \, \, \, \, 0,
\end{eqnarray}
and
\begin{eqnarray}
\label{Harnad214terrecall}
\hspace{-0.95in}&& \,  \,    \quad  \quad \quad \quad \quad 
 \, \{\tilde{X}(x) , \, x \} \, \, \, \, \, 
 - {{1} \over {2  \cdot \, x^2 }} 
 \, \, \, \,
 -  W\Bigl( \tilde{X}(x)  \Bigr) \cdot  \,  \Bigl({{ d \tilde{X}(x) } \over {d x}} \Bigr)^2
 \,  \, =  \, \, \, \, \, 0,
\end{eqnarray}
where $\, W(x)$ is given by (\ref{family_anotherW}).  In fact using
the {\em explicit algebraic form}
of $\, y(a, \, x)$,  given by (\ref{condition1n1anotherdxconda}), one can find
a closed exact expression for the ``nome-like'' series $\, \tilde{Q}(x)$, namely:
\begin{eqnarray}
\label{nome}
\hspace{-0.95in}&& \,  \,    \quad  \quad \quad \quad \quad \quad \quad \quad \quad \quad \quad
 \tilde{Q}(x)    \,  \,  =  \,  \,   \, \,
  {{ x \cdot \, (1 \, -2 \, x)^2} \over { (1 \, -3 \, x)^3}},    
\end{eqnarray}
in agreement with series (\ref{compositiona0bis1}). 
Relation (\ref{condition1n1anotherdxconda}) is nothing but:
\begin{eqnarray}
\label{nomea}
  \hspace{-0.95in}&& \,  \,    \quad  \quad \quad \quad \quad  \quad \quad \quad \quad \quad \quad
 a \cdot \, \tilde{Q}(x)
 \,  \,  =  \,  \, \,  \, \tilde{Q}\Bigl( y(a, \, x)    \Bigr).   
\end{eqnarray}                
The one-parameter series (\ref{condition1n1anothersol1}) can thus be written:
\begin{eqnarray}
\label{compositiona0bis1first}
\hspace{-0.95in}&& \, \, \,
\quad  \quad  \quad  \quad   \quad  \quad  \quad  \quad  \quad \quad \quad
  y(a, \, x)     \, \, = \, \, \,  \, \tilde{X}\Bigl( a \cdot \, \tilde{Q}(x)   \Bigr).        
\end{eqnarray}
From (\ref{compositiona0bis1first}), and from the fact that  $\, y(a, \, x)= \, x$ for $\, a= \, 1$,
one can deduce that 
 $\, \tilde{X}(x)$ {\em must be the compositional inverse} of the ``nome-like'' series (\ref{nome}). 
Note that  $\, \tilde{X}(x)$ is an algebraic function. It is solution of the
polynomial equation:
\begin{eqnarray}
  \label{polynom}
\hspace{-0.95in}&&  \, \,
(27 \, x \, +4) \cdot \, \tilde{X}(x)^3  \,\, \,  -(27 \, x \, +4) \cdot \, \tilde{X}(x)^2
\,\, \, \, +(9 \, x \, +1)\cdot \,  \tilde{X}(x) \,\,  \,\, -x   \,\,\, \, = \, \, \,\, \, 0,
\end{eqnarray}
in agreement with the $\, R= \, 4/27$ radius of convergence of the series $\, \tilde{X}(x)$.

\vskip .1cm

\subsection{Another simple polynomial example for $\, F(x)$}
\label{badtruncation}

Let us now consider the polynomial 
\begin{eqnarray}
\label{trunc0moins1}
 \hspace{-0.95in}&& \,  \, \, \,  \,  \, \,
F(x) \, \, = \, \, \, x \cdot \, (1 \, - 373 \cdot \,x) \cdot \, (1 \,  - 371 \cdot  \, x)
  \, \, = \, \, \,   \,  \,  x \, \, \,\, -744 \, x^2 \, \,\,  +138383 \, x^3, 
\end{eqnarray}
which has the {\em same first two terms} as the series expansion
of the hypergeometric function (\ref{Fx}). 
The function $\, W(x)$ in the Schwarzian equation is given by (\ref{defW}):
\begin{eqnarray}
\label{trunc0moins1W}
\hspace{-0.95in}&& \,  \, \, \,  \,  \, \, \quad \quad 
W(x) \, \, = \, \, \,
  -{{1} \over {2}} \cdot \,
  {{ 1 \, -830298\, x^2 \, +411827808\, x^3 \, -57449564067\, x^4} \over {
  x^2 \cdot \, (1 \, -373\, x)^2 \cdot \, (1 \, -371\, x)^2 }}. 
\end{eqnarray}

A solution of the Schwarzian equation, with $\, W(x)$ given by (\ref{trunc0moins1W}), reads:
\begin{eqnarray}
\label{trunc0moins1relatFya}
  \hspace{-0.98in}&& 
y(a, \, x) \, \, = \, \, \,
 a \cdot \, x \, -744 \cdot\, a \cdot\, (a-1) \cdot \, x^2  \,
 +{{1} \over {2}}  \cdot\, a \cdot\, (1245455\, a \, -968689) \cdot\, (a-1) \cdot \, x^3
\nonumber \\
 \hspace{-0.95in}&& \,  \, \quad \quad  \quad \quad  \, \,
-620 \cdot \, a \cdot \, (885656\, a \, -470507) \cdot \, (a-1)^2 \cdot \, x^4
 \, \,  \,  \, \, \, + \, \, \, \cdots 
\end{eqnarray} 
The functional relation  $\, F(y) \, \, = \, \, F(x) \cdot \, y'(x)$, gives
$\, dy/F(y) \, = \, dx/F(x)\, $, and thus 
\begin{eqnarray}
\label{trunc0moins1relatF}
  \hspace{-0.95in}&& \,  \, \quad  \quad 
\mu \cdot \,  \exp\Bigl(   \int \, {{dy} \over {F(y)}} \Bigr)
 \, \, = \, \, \tilde{Q}(x)
    \, \, = \, \, \,
  \exp\Bigl( \int \,  {{dx } \over {F(x)}}  \Bigr)
  \, \, = \, \,  \, \,
  x \cdot \, {{ (1  \, -371 \, x)^{371/2} } \over { (1  \, -373 \, x)^{373/2}}}
   \nonumber \\
  \hspace{-0.95in}&& \,  \, \quad  \quad  \quad \quad 
 \, \, = \, \, \,  \,\, \,
 x \, \,\,\, \, +744 \, x^2 \, \, \,\, +{{968689} \over  {2}} \, x^3  \, \,\, \, +291714340\, x^4
 \,  \, \,\,  \, + \, \, \, \cdots 
\end{eqnarray}              
We actually have the relation:
\begin{eqnarray}
\label{qaQ}
  \hspace{-0.95in}&& \,  \, \quad  \quad  \quad \quad  \quad \quad  \quad \quad \quad \quad  \quad
  a \cdot \, \tilde{Q}(x)     \, \, = \, \, \,  \,  \,  \tilde{Q}\Bigl(y(a, \, x)  \Bigr).
\end{eqnarray}  
In that case, since $\, \tilde{Q}(x)\, $ is an {\em  algebraic function},
we see that the one-parameter series
$\, y(a, \, x)$, given by (\ref{trunc0moins1relatFya}),
{\em is actually an algebraic series for any value of} the parameter $\, a$.
The series $\, y \, = \, y(a, \, x)$ is actually solution of:
\begin{eqnarray}
\label{qaQpol}
  \hspace{-0.95in}&& \,  \,   \,   \quad \quad \quad \quad  \quad
 a^2 \cdot \,  x^2 \cdot \, {{ (1  \, -371 \, x)^{371} } \over { (1  \, -373 \, x)^{373}}}
 \, \, \, \,\,  -  y^2 \cdot \, {{ (1  \, -371 \, y)^{371} } \over { (1  \, -373 \, y)^{373}}}
\, \,\, \, \,\, = \, \, \, \,\, 0.
\end{eqnarray} 
Taking into account the large degree in $\, x$
or $\, y$ of the polynomial condition  (\ref{qaQpol}), one should note that it can
actually be quite difficult to get this polynomial equation
from\footnote[5]{Using, for instance, the command seriestoalgeq of gfun
of Bruno Salvy.} a large series (\ref{trunc0moins1relatFya}). 
The {\em compositional inverse} of $\, \tilde{Q}(x) \, $ is:
\begin{eqnarray}
\label{compoinverse}
\hspace{-0.95in}&& \,  \, \,  \,  \,  \,
\tilde{X}(x)    \, \, = \, \, \,  \,  \,
x \,\, \,\, -744\, x^2 \,\, \,\, +{{1245455} \over {2}} \cdot \, x^3
\,\, \,\,  -549106720\, x^4 \,\,\,\, +{{3989599188003} \over {8}} \cdot \, x^5
\nonumber \\
\hspace{-0.95in}&& \,  \,  \quad  \quad \quad  \quad 
 \, \,  -461623555588416\, x^6 \,\,\,\,
 +{{6928370820171415659} \over {16}} \cdot \, x^7
\nonumber \\
  \hspace{-0.95in}&& \,  \,  \quad \, \quad \quad \quad \quad  \quad  \quad \,
 -410201463628637176320\, x^8
   \, \, \, \,\, \, + \, \, \, \cdots 
\end{eqnarray}

This is an {\em algebraic series} $\,\, y \, = \,  \tilde{X}(x)$,  solution  of:
\begin{eqnarray}
\label{compoinversePol}
  \hspace{-0.95in}&& \,  \,   \quad  \quad    \quad \quad \quad \quad
  x^2 \cdot \, (1 \,  -373 \cdot \, y)^{373}
  \,\,\, \, - \, \, y^2 \cdot \, (1 -371 \cdot \, y)^{371}
\, \, \,\, = \, \, \, 0.
\end{eqnarray} 
Note that, even with a very large series (\ref{compoinverse}), it is also quite hard, because
of the high degree in $\, y$ of (\ref{compoinversePol}), 
to find the algebraic expression (\ref{compoinversePol}) even if it is really simple. 

\vskip .1cm

\subsubsection{Two-parameter family.}
\label{twopara}

Following the calculations displayed in subsection (\ref{alpha1}), let us generalize $\, W(x)$
given by (\ref{trunc0moins1W}), to the form (\ref{wherecondn1bis1}):
\begin{eqnarray}
\label{wherecondn1bis1recall}
\hspace{-0.95in}&& \quad   \quad \quad    \quad \quad      \quad \quad   
W(x)  \, \, = \, \,  \, \, \,
{{F''(x)} \over {F(x)}} \, \, -{{1} \over {2}} \cdot \, \Bigl({{F'(x)} \over {F(x)}}\Bigr)^2
\, \, +{{1} \over {2}} \cdot \, {{\alpha^2} \over {F(x)}}
\end{eqnarray}
For $\, \alpha= \, 1$ with $\, F(x)$ given by (\ref{trunc0moins1}), $\, W(x)$ reads:
\begin{eqnarray}
\label{Wrecall}
\hspace{-0.95in}&& \quad   \, \quad  \quad \quad \quad
W(x)  \, \, = \, \,  \, \, \,
{{ 415149 } \over {2}} \cdot \,
  {{138383\,x^2 \, -992\, x \, +2 } \over {
  x^2 \cdot \, (1 \, -373\, x)^2 \cdot \, (1 \, -371\, x)^2 }}. 
\end{eqnarray}
The {\em two-parameter} series (generalizing (\ref{trunc0moins1relatFya})) 
\begin{eqnarray}
\label{family_abPol}
\hspace{-0.98in}&& \,  \, \quad \quad \quad  \, 
 y(a, \, b, \, x) \, \, = \, \, \,\, \,\,
 a \cdot \, x \, \, \, \, \,
 +\Bigl(1728 \cdot \, b \, -744 \cdot  \, a\cdot  \, (a-1)\Bigr) \cdot  \, x^2
 \nonumber \\
 \hspace{-0.98in}&&
 \quad \quad \quad   \quad  \quad 
\, \,  +\Bigl( 2985984 \cdot \, {{b^2} \over {a}} \, \, -2571264 \cdot \, (a-1) \cdot \,b 
\\
\hspace{-0.98in}&& \quad \quad \quad \quad \quad \quad \quad \quad  \quad  \quad 
    \, \,  +{{1} \over {2}}  \cdot\, a \cdot\, (1245455\, a \, -968689) \cdot\, (a-1)  \Bigr)\cdot \, x^3
\nonumber \\
\hspace{-0.95in}&& \,  \, \quad \quad  \quad \quad  \, \,
+\Bigl( 5159780352 \cdot \, {{b^3} \over {a^2}} \,  \, -6664716288 \cdot \,(a-1) \cdot \, {{b^2} \over {a}}
\nonumber \\
\hspace{-0.95in}&& \,  \, \quad\quad  \quad \quad  \quad \quad  \, \,
  \, +864\cdot \, (3736365\,a^2-6642432\,a+3044450) \cdot \, b
  \nonumber \\
  \hspace{-0.95in}&& \,  \, \quad \quad \quad  \quad \quad \quad  \quad \quad  \, \,
  \, -620 \cdot \, a \cdot \, (885656\, a \, -470507) \cdot \, (a-1)^2 \Bigr) \cdot \, x^4
  \, \,\, \, \, \, \, + \, \, \, \cdots
\nonumber 
\end{eqnarray}
is {\em actually solution of the  Schwarzian equation} (\ref{condition1n1bis}) with
$\, W(x)$ given by (\ref{Wrecall}).
Note that, again, the {\em two-parameter} series (\ref{family_abPol}) is also
solution of
\begin{eqnarray}
\label{wherecondn3bbis}
  \hspace{-0.95in}&&  \quad \quad \, \, \, \, \,  \quad \quad \quad
  a \cdot \, {{\partial y(a, \, b, \, x)} \over { \partial a}}  \,\,\,
  +b \cdot \, {{\partial y(a, \, b, \, x)} \over { \partial b}}
  \, \, = \, \, \,  F\Bigl(y(a, \, b, \, x)\Bigr),
\end{eqnarray}
with $\, F(x)$ given by (\ref{trunc0moins1}).
We have, again,  the  {\em two-parameters composition rules}  (\ref{family_abother}):
\begin{eqnarray}
\label{family_abotherbis}
\hspace{-0.98in}&& \,  \,  \quad \quad \quad  \quad \quad \quad \quad
y\Bigl(a', \, b', \,  y(a, \, b, \, x)\Bigr)
\, \, = \, \, \,\, \, y\Bigl(a\,a', \, a^2 \, b' \, +a' \, b, \, x\Bigr). 
\end{eqnarray}
similarly to subsection (\ref{alpha1}), let us introduce
an {\em alternative parametrization} of the two-parameter
series (\ref{family_abPol}), changing $\, b$ into $\, a \, b$,  in (\ref{family_abPol}):
\begin{eqnarray}
\label{family_abYPol}
\hspace{-0.98in}&& \,  \, \quad \quad \quad \quad  \, 
 Y(a, \, b, \, x) \, \, = \, \, \,\, \,\,
 a \cdot \, x \, \, \, \, \, +\Bigl(1728 \cdot \, a  \, b \,
 -744 \cdot  \, a\cdot  \, (a-1)\Bigr) \cdot  \, x^2
 \nonumber \\
 \hspace{-0.98in}&& \,  \, \quad \quad \quad \quad \quad \quad  \,
 +{{1} \over {2}} \cdot \,
 \Bigl(  5971968 \cdot \, a\cdot \, b^2 \, \, -5142528\cdot \, (a-1)\cdot \, a\cdot \, b
 \nonumber \\
 \hspace{-0.98in}&& \,  \, \quad \quad \quad \quad \quad \quad \quad \quad \quad  \,
 \, \, + a \cdot\, (1245455\, a \, -968689) \cdot\, (a-1)\Bigr) \cdot \, x^3
 \, \,\, \, \, + \, \, \, \cdots 
\end{eqnarray}
Again, we have the following composition rules for the {\em two-parameter} series (\ref{family_abYPol})
\begin{eqnarray}
\label{family_abotherYbis}
\hspace{-0.98in}&& \,  \,  \quad \quad \quad  \quad \quad \quad \quad
Y\Bigl(a', \, b', \,  Y(a, \, b, \, x)\Bigr)
 \, \, = \, \, \,\, \, Y\Bigl(a\,a', \, a\, b' \, +b, \, x\Bigr), 
\end{eqnarray}
The series (\ref{family_abYPol}) is, now, solution of:
\begin{eqnarray}
\label{wherecondn3bNEW}
  \hspace{-0.95in}&&  \quad \quad \quad  \quad \quad \quad \quad \quad \quad
  a \cdot \, {{\partial Y(a, \, b, \, x)} \over { \partial a}}  \,
    \, \, = \, \, \,  F\Bigl(Y(a, \, b, \, x)\Bigr),
\end{eqnarray}
with $\, F(x)$ given by (\ref{trunc0moins1}).

\vskip .1cm

\subsection{Truncation of the hypergeometric function $\, F(x)$.}
\label{truncation}

The hypergeometric function $\, F(x)$ given by (\ref{Fx}),
expands as $\, x \, -744\, x^2 -393768\, x^3 \, + \cdots$
Let us consider a simple truncation of this hypergeometric function:
\begin{eqnarray}
\label{trunc0}
 \hspace{-0.95in}&& \,  \, \quad  \quad \quad  \quad \quad  \quad \quad 
F(x) \, \, = \, \, \, \,  x \,\, \, \, -744 \, x^2 \, \,\,  -393768\, x^3. 
\end{eqnarray}
From (\ref{defW}) this  gives:
\begin{eqnarray}
\label{trunc4}
  \hspace{-0.95in}&& \,  \,\,  \,\,  \, \quad  \,\,
  W(x) \, \, = \, \, \,
- {{ 1} \over { 2}}  \cdot \,
{{ 1 \, +2362608 \, x^2 \, -1171853568\, x^3 \, -465159713472\, x^4} \over {
   x^2 \cdot \, (1 \, -744\,x \, -393768\, x^2)^2 }}. 
\end{eqnarray}
The Schwarzian equation (\ref{condition1n1}) with the previous $\, W(x)$,
namely (\ref{trunc4}), has the following {\em one-parameter} solution-series:
\begin{eqnarray}
\label{trunc5}
\hspace{-0.95in}&& \,  \, 
y(a, x) \, \, = \, \, \, \,\,
a \cdot \, x \,\, \,\, \, -744 \cdot \, a \cdot \, (a-1) \cdot \, x^2
\, \, \, \, +36 \cdot \, a \cdot \, (a-1) \cdot \, (9907\, a -20845) \cdot \, x^3
 \nonumber \\
 \hspace{-0.95in}&& \,  \,  \quad  \quad   \quad 
 -80352 \cdot \, a \cdot \, (a-1)^2 \cdot \, (264\, a -9379) \cdot \, \, x^4
   \\
  \hspace{-0.95in}&& \,  \,  \quad  \quad \,  \quad 
    -648 \cdot \, a \cdot \, (a-1)^2 \cdot \, (250310357\, a^2 +598043050\, a -1207272939) \cdot \, x^5
  \nonumber \\
  \hspace{-0.95in}&& \,  \,  \, \,  \,
  +{{482112} \over {5}}  \cdot \,
  a \cdot \, (a-1)^3 \cdot \, (1944308192\, a^2 -424834349\, a -8498464743) \cdot \, x^6
\, \, \,\, + \, \, \, \cdots \nonumber 
\end{eqnarray}
This one-parameter series (\ref{trunc5}) is quite
similar\footnote[1]{The first three coefficients are actually the same.} to the 
one-parameter series (\ref{family_a}).   
The series (\ref{trunc5}) actually verifies the composition rule:  
\begin{eqnarray}
\label{composition2}
  \hspace{-0.95in}&& \,  \, \quad  \quad \quad  \quad \quad \quad \quad  \quad   \quad   \quad 
 y\Bigl(a, y(a', \, x)\Bigr) \, \, = \, \, \, y(a\, a', \, \, x).
\end{eqnarray}
Let us introduce the two limits
\begin{eqnarray}
\label{compositiona0bis}
  \hspace{-0.95in}&& 
\tilde{Q}(x) \, \, = \, \, \,
  \lim_{a \rightarrow \, 0}  {{ y(a, \, \, x) } \over {a}} \, \, = \, \, \, \, \,\, 
 x \, \, \, \, +744 \, x^2 \,\, \, \,  +750420 \, x^3 \, \, \, +753621408\, x^4 
 \\
  \hspace{-0.95in}&& \,  \,  \quad 
  +782312864472\, x^5  \,\,  +{{4097211834177216} \over {5}} \, x^6
  \,\,\,   +{{4331866321367059104} \over {5}} \, x^7
  \,\,  \, \,  + \,\, \, \cdots
  \nonumber 
\end{eqnarray}
and:
\begin{eqnarray}
\label{compositionainftybis}
  \hspace{-0.95in}&& \, \, \, 
\tilde{X}(x) \, \, = \, \, \,
\lim_{a \rightarrow \, \infty}  \, y\Bigl(a, \, \, {{x} \over {a}}\Bigr)  \, \, = \, \, \, \, \,
 x \,  \,  \, \, -744\, x^2 \, \, \,  \,  +356652\, x^3 \, \, \,  \,  -21212928\, x^4\, 
 \\ 
\hspace{-0.95in}&& \,  \quad  \, 
  \, -162201111336\, x^5 \, \, \,  +{{937374311061504} \over {5}} \, x^6
  \, \,  \,\,    -{{563689525139743392} \over {5}} \, x^7
 \, \,\,\, \,  + \,\, \, \cdots \nonumber 
\end{eqnarray}
One  verifies that the one-parameter series (\ref{trunc5}) {\em is actually of the form}:
\begin{eqnarray}
\label{oftheform}
 \hspace{-0.95in}&&  \quad  \quad \quad  \quad \quad \quad   \quad \quad \quad \quad \quad 
y(a, x) \, \, = \, \, \,    \tilde{X}\Bigl( a \cdot \, \tilde{Q}(x) \Bigr). 
\end{eqnarray}
Again, from (\ref{oftheform}) and from the fact that $\, y(a, \, x) = \, x$
for $\, a= \, 1$, we see that 
the series (\ref{compositionainftybis}) is actually the {\em compositional inverse}
of the ``nome-like'' series (\ref{compositiona0bis}):
\begin{eqnarray}
\label{particular}
 \hspace{-0.95in}&&  \quad  \quad \quad  \quad \quad \quad   \quad \quad \quad \quad \quad 
y(1, x) \, \, = \, \, \,  x \,\, \, = \, \, \,    \tilde{X}\Bigl(\tilde{Q}(x) \Bigr). 
\end{eqnarray}
The {\em one-parameter} series (\ref{trunc5}) is also solution of 
\begin{eqnarray}
\label{wherecondn1parti}
  \hspace{-0.95in}&& \quad   \quad \quad \quad \quad   \quad \quad
 a \cdot \, {{\partial y(a, \,x)} \over { \partial a}}
 \, \, = \, \, \,\, F(y(a,  \, x))   \, \, = \, \, \,\,
 F(x) \cdot \,  {{\partial y(a, \, x)} \over { \partial x}},
\end{eqnarray}
and one can verify that:
\begin{eqnarray}
\label{yamoremore}
  \hspace{-0.95in}&& \quad \quad \quad \quad \quad \quad \quad \quad \quad \quad
 \tilde{Q}\Bigl( y(a, \, x) \Bigr)   \, \, = \, \, \,   \,   a \cdot \,  \tilde{Q}(x).    
\end{eqnarray}
Conversely, from (\ref{yamoremore}), we get, recalling (\ref{relationcorresp}) 
\begin{eqnarray}
\label{yamoremore-d}
\hspace{-0.95in}&& \quad \quad \quad  \quad \quad \quad   \quad  \quad \quad \quad
{{dy} \over {  F(y(a,  \, x))}} \,  \, \, = \, \, \,  \,\,
  {{dx } \over { F(x)}} \, \, \, \, + {{da} \over {a}}, 
\end{eqnarray}
which gives for $\, a$ fixed
\begin{eqnarray}
\label{yamoremoregives1}
\hspace{-0.95in}&& \quad \quad \quad\quad \quad  \quad \quad \quad  \quad
F(y(a,  \, x))   \, \, = \, \, \,  F(x) \cdot \, {{\partial y(a, \, x)} \over { \partial x}}, 
\end{eqnarray}
and  for $\, x$ fixed:
\begin{eqnarray}
\label{yamoremoregives2}
  \hspace{-0.95in}&& \quad   \quad \quad \quad \quad \quad \quad \quad  \quad \quad
 a \cdot \, {{\partial y(a, \,x)} \over { \partial a}}      \, \, = \, \, \, F(y(a,  \, x)). 
\end{eqnarray}
Let us introduce the series
\begin{eqnarray}
\label{y2more}
\hspace{-0.95in}&& \quad   
y_2 \, \, = \, \, \,    \tilde{X}\Bigl(\tilde{Q}(x)^2 \Bigr)
  \, \, = \, \, \, \,\,
 x^2 \, \,\, \,\,   +1488\, x^3 \,   \,\, +2053632\, x^4 \, \, \,  \, +2621653632\, x^5
 \nonumber \\
 \hspace{-0.95in}&& \quad   \quad  \,     \,            
 \, +3244440682476\, x^6 \,\, \, +{{19627900112688192} \over {5}} \, x^7
\,\,  \,  +{{23401843163094440736} \over {5}} \, x^8
\nonumber \\
 \hspace{-0.95in}&& \quad    \quad    \quad \quad \quad \quad    \quad    
 \,  +{{193179165341208747259392} \over {35 }} \, x^9
 \,  \,  \, \, \, + \, \, \cdots 
\end{eqnarray}
This series (\ref{y2more}) is solution of the Schwarzian equation (\ref{condition1n1}),
with $\, W(x)$ given by (\ref{trunc4}), and is also solution of
\begin{eqnarray}
\label{wherecondn1}
  \hspace{-0.95in}&& \quad\, \,  \quad \quad \quad \quad \quad  \quad  \quad  \quad 
2 \cdot \,    F(y_2)   \, \, = \, \, \,  F(x) \cdot \,  {{\partial y_2} \over { \partial x}}, 
\end{eqnarray}
i.e.
\begin{eqnarray}
\label{wherecondn1bis}
  \hspace{-0.95in}&& \quad   \quad  
 2 \cdot \,  \Bigl(y_2 \, \,  -744 \, y_2^2 \,  \,  -393768\, y_2^3 \Bigr)
 \, \, \, = \, \, \, \,
 \Bigl(x \,  \, -744 \, x^2 \,  \,  -393768\, x^3 \Bigr)
  \cdot \,  {{\partial y_2} \over { \partial x}}, 
\end{eqnarray}
and one also has:
\begin{eqnarray}
\label{wherecondn1Q}
 \hspace{-0.95in}&& \quad   \quad \quad   \quad \quad \quad \quad \quad  \quad  \quad  \quad
 \tilde{Q}\Bigl(y_2(x) \Bigr)     \, \, = \, \, \,  \tilde{Q}(x)^2.        
\end{eqnarray}                 
Let us introduce the {\em one-parameter} series
\begin{eqnarray}
\label{y2morebisa}
  \hspace{-0.95in}&& \quad \quad  \, \,
y_2^{(a)} \, \, = \, \, \,  y(a, \, y_2)  \, \, = \, \, \,\,
\tilde{X}\Bigl(a \cdot \, \tilde{Q}(x)^2 \Bigr)
\, \, = \, \, \,\, \,
 a\cdot \, x^2 \, \,\,\,  +1488\,\, a \cdot \,x^3 \, 
  \nonumber \\
\hspace{-0.95in}&& \quad \quad \quad \quad 
 -24 \cdot \, a \cdot \,(31\,a -85599) \cdot \,x^4
\,\, -35712 \cdot \,a \cdot \,(62\,a-73473) \cdot \,x^5
\nonumber \\
\hspace{-0.95in}&& \,\quad \quad \quad \quad  \, \, \,
+36 \cdot \,a \cdot \,(9907\,a^2-130673184\,a+90254015568) \cdot \,x^6
 \\
\hspace{-0.95in}&& \,\quad \quad \quad \quad \quad  \, \,
\, +{{160704} \over {5}} \cdot \,a \cdot \,(49535\,a^2-262999040\,a+122399922528) \cdot \,x^7               
 \, \, \, \,  \, \, + \, \, \, \cdots
\nonumber             
\end{eqnarray}
This one-parameter series (\ref{y2morebisa}) is solution of the
Schwarzian equation (\ref{condition1n1}), with $\, W(x)$
given by (\ref{trunc4}).
It is also solution  of
\begin{eqnarray}
\label{wherecondn1Qa}
  \hspace{-0.95in}&& \quad   \quad \quad \quad \quad \quad \quad \quad  \quad  \quad  \quad
\tilde{Q}\Bigl(y_2^{(a)}(x) \Bigr)     \, \, = \, \, \,  a \cdot \, \tilde{Q}(x)^2,        
\end{eqnarray}    
and  also solution  of
\begin{eqnarray}
\label{wherecondn1}
  \hspace{-0.95in}&& \quad  \, \,  \quad \quad \quad \quad \quad \quad
2 \cdot \,    F(y_2^{(a)})   \, \, = \, \, \,
F(x) \cdot \,  {{\partial  y_2^{(a)}} \over { \partial x}}
\, \, = \, \, \,\,
2 \cdot \, a \cdot \, {{ \partial y_2^{(a)}} \over { \partial a}}.
\end{eqnarray}
where
\begin{eqnarray}
\label{trunc0again1}
 \hspace{-0.95in}&& \,  \, \quad  
F(x) \, \, = \, \, \, \, \,  x \, \, \, \, -744 \, x^2\, \,\,  -393768\, x^3
  \, \, \,= \,\,  \, \,
 x \cdot \, \Bigl(1 \, -p \cdot \, x\Bigr)
 \cdot \, \Bigl(1 \, -q \cdot \, x\Bigr), 
\end{eqnarray}
with:
\begin{eqnarray}
\label{trunc0again2}
  \hspace{-0.95in}&& \,  \, \, \,  \,   \quad \quad  \quad \quad 
  p \, \, = \, \, \, 372 \,\,\, +6\cdot \, 14782^{1/2},
  \quad \quad \, \,  \,  \, 
 q \, \, = \, \, \, 372 \,\,\, -6\cdot \, 14782^{1/2},
\end{eqnarray}
Let us denote
\begin{eqnarray}
\label{denote}
  \hspace{-0.95in}&& \,  \, \quad \quad \quad \quad 
 \alpha \, \, = \, \, \,\,
{{1} \over {2}} \cdot \, {{p+q } \over {q-p  }}\,\, \, = \, \, \,\,
 -{{31} \over {14782}} \cdot 14782^{1/2}
 \, \,\, = \, \, \, \, -0.25497\, \, \cdots 
\end{eqnarray}
Following the previous calculations in section (\ref{badtruncation}), one
easily finds that the ``nome-like'' series (\ref{compositiona0bis}) reads:
\begin{eqnarray}
\label{nomereads} 
\hspace{-0.95in}&& \,  \, 
 \tilde{Q}(x) \,  \,  =  \,  \,   \,
    {{ x \cdot \, (1 \, -p \cdot \, x)^{p/(q-p)}} \over { (1 \, -q \cdot \, x)^{q/(q-p)}}}
 \,  \,  =  \,  \,   \,  \,  \,
    {{ x } \over { \Bigl((1 \, -p \cdot \, x) \cdot \, (1 \, -q \cdot \, x) \Bigr)^{1/2} }}  \cdot \,
     \Bigl( {{1 \, -p \cdot \, x } \over {1 \, -q \cdot \, x }} \Bigr)^{\alpha}
\nonumber  \\
\hspace{-0.95in}&& \,  \, \, \, \, \,  \quad
   \,  \,  =  \,  \,   \,  \,  \,
   x \, \, \, \, \,  +744 \, x^2\,\, \,  \, +750420 \,\, x^3 \,\, \, \, 
   +753621408\, x^4 \, \,\,  \, +782312864472 \, \, x^5  
\nonumber  \\
 \hspace{-0.95in}&& \,  \, \,  \, \, \,  \,  \,  \,\quad  \quad  \quad 
 \,  \, +{{4097211834177216} \over {5}} \,\, x^6
 \,\, \,\,\,  +{{4331866321367059104} \over {5}}\, \, x^7
 \,\,  \, \,\,\, \,  + \,\, \, \cdots  
\end{eqnarray}

This ``nome-like'' series (\ref{nomereads}) is {\em actually $\, D$-finite}.
It is solution of the order-one linear differential operator ($\theta \, = \, x \cdot \, D_x$
is the homogeneous derivative):
\begin{eqnarray}
\label{order-one}
  \hspace{-0.95in}&& \,  \, \quad \quad \quad\quad
 {\cal L}_1 \, \, = \, \,   \,   F(x) \cdot D_x \, \, \, -1
\,\, \, = \, \,  \, \,
 ( x \, \, \, -744 \, x^2\, \,  -393768\, x^3) \cdot D_x \, \,\, -1
\nonumber   \\
  \hspace{-0.95in}&& \,  \, \quad \quad \quad \quad \quad\quad \quad
\, \, = \, \,  \, \,
( 1 \, \, \, -744 \, x\, \,  -393768\, x^2) \cdot \, \theta \,\,  \, \, -1.
\end{eqnarray}
Do note, however, that this ``nome-like'' series (\ref{nomereads}) is {\em not} globally bounded.
The radius of convergence of  the ``nome-like'' series (\ref{nomereads})  is $\, 1/p$, with $\, p \, $
given by (\ref{trunc0again2}):
\begin{eqnarray}
\label{order-onevalue}
  \hspace{-0.95in}&& \,  \, \quad \, \, \,  \quad \quad \quad
 R \, \, = \,  \,\,  \, {{1} \over {p}}  \, \,\,  = \,  \, \,
   {{14782^{1/2} } \over {65628 }} \,\, \, - {{31} \over {32814 }} 
 \, \,\, = \,  \, \, \,  0.0009078632370\,  \cdots      
\end{eqnarray}
This ``nome'' series (\ref{nomereads}) is   $\, D$-finite,  with
a finite radius of convergence, {\em but it is not globally bounded}.
Note that  $\, \tilde{X}(x)$ is only a {\em differentially algebraic function}.

Note that the order-one linear differential operator $\,  {\cal L}_1 $,
given by (\ref{order-one}), is {\em not globally nilpotent}~\cite{GlobNilp}.
The corresponding $\, p$-curvatures are null (or nilpotent that is the same
for order-one linear differential operators) for the following primes: 
\begin{eqnarray}
\label{pcurvature}
\hspace{-0.95in}&& \,  \, \quad \quad \quad \quad \quad
3, \, \,  11, \,\,  13, \, \, 17, \,\,  23, \, \, 31, \,\,  47,
\,\,  61, \,\,  73, \,\,  79, \,\,  89, \, \, 101,  \, \, \, \cdots                 
\end{eqnarray}
but non-zero for the following primes: 
\begin{eqnarray}
\label{pcurvature2}
\hspace{-0.95in}&& \,  \,  \quad \quad \quad \quad \quad
5, \, \,  7, \, \,  19, \, \,  29, \,\,  37, \,  \, 41,
\,\,   43, \, \, 53, \,\,  59, \,\,   67,  \,\,  71,   \,\,  \, \cdots                 
\end{eqnarray}
Note that, since $\,  14782 \, = \, 2 \cdot \, 19 \cdot \, 389$,  we could
have expected that one does not see the transcendence of the ``nome'' mod. $\, 19$,
the ``nome'' reducing to an algebraic function (see (\ref{trunc0again2}), (\ref{denote})),
and thus one could expect a zero $\, p$-curvature. This is not the case.

Note that the exponent of the ``nome-like'' series (\ref{nomereads}), at the singularity
$\,\, x \, = \, 1/p$,  is
\begin{eqnarray}
\label{order-oneexponent}
 \hspace{-0.95in}&& \,  \, \quad \quad \quad \quad \quad \quad \quad
 {{p} \over {q \, -p}}  \, \, \, = \, \, \,   \,
 -{{1} \over {2}} \,\,  \,   -{{31} \over { 14782^{1/2}}}
 \,\, \, = \, \, \, \,   -0.7549735291 \,\,  \cdots 
\end{eqnarray}
 {\em which is not a rational number}. This rules out the fact that the
    order-one linear differential operator
    (\ref{order-one}) could be  globally nilpotent~\cite{GlobNilp}.
 
\vskip .1cm

Let us consider the simplest example of  series $\, y(a, \, x)$, namely
the (involutive) series (\ref{trunc5}) for $\, a \, =-1$:
\begin{eqnarray}
\label{trunc5invol}
 \hspace{-0.98in}&& \,  \, \, 
  y(-1, \, x) \, \, = \, \, \, \,
  -x  \,\, \, -1488\, x^2  \, \,\, -2214144\, x^3
  \,\, \, -3099337344\, x^4 \,\,\,  -4030574598144\, x^5
\nonumber \\
\hspace{-0.95in}&& \,  \, \quad  \quad \quad  \quad 
\,  -{{23640158283604992} \over {5 }}\, x^6 \, \,\,\, \,
-{{23310435220175683584} \over {5 }} \, x^7
 \\
\hspace{-0.95in}&& \,  \, \quad  \quad 
-{{20590422517553304526848} \over {7 }}\, x^8 \,\,\,\,
+{{12494610391145690921435136} \over {7}} \, x^9  
\,\, \,\,\, + \, \, \, \cdots
\nonumber
\end{eqnarray}
Calculating the first fifty coefficients of this series, one can see
that this (involutive) series is {\em not globally bounded}.

\vskip .2cm

{\bf Remark \ref{polyexamplesF}.2:} Following subsection (\ref{twopara}),
the generalization to two-parameter series
can be performed on this last polynomial example, mutatis mutandis.

\vskip .2cm

\section{Comments and speculations on differentially algebraic series.}
\label{CommentsandSpecul}

We have displayed  miscellaneous series 
solutions of Schwarzian equations (and thus having
a compositional property~\cite{Schwarz1,Schwarz2}), which can be seen 
to be, or to generalize, {\em  modular correspondences}~\cite{Goro}.
We remark that we have the following situation:
we have series depending on {\em one} parameter (sometimes {\em two parameters}
for slightly ``academical'' examples like in subsection (\ref{alpha1})),  which reduce to series
with {\em integer} coefficients for an  {\em infinite set} of values of the parameter(s), namely
the {\em integer values}\footnote[2]{More generally, for {\em rational values}
of the parameters we have globally bounded differentially algebraic series.}.
These one-parameter  series are {\em generically only differentially algebraic},
even for {\em  integer} values
of the parameter (where they are probably not even $D$-finite, see for instance (\ref{a3})).  In
contrast, and remarkably, when the  parameter
is a $\, N$-th root of unity, the generically differentially algebraic one-parameter  series
become {\em algebraic functions}.  We thus have an  {\em infinite number} of {\em algebraic functions}. 

It is interesting to note that a totally and utterly similar situation have been seen to occur in
other very interesting situations in physics, or enumerative combinatorics. 
Along this line, differentially algebraic series with {\em integer}
coefficients\footnote[5]{Not simply reducible to ratio of
  globally bounded  $D$-finite series, or composition of globally bounded  $D$-finite series.}
exist, and correspond to remarkable solutions of {\em differentially algebraic}
equations in physics,
or enumerative combinatorics, like
{\em $\, \lambda$-extensions of Ising correlation functions}~\cite{extension,lambdaPainl},
or solution of a differentially algebraic Tutte equation~\cite{Selected}.
We have an  {\em infinite set} of differentially
algebraic series with {\em integer} coefficients that are {\em not}
$\, D$-finite~\cite{Selected,extension,lambdaPainl}.
We also have the occurence
of an {\em infinite number} of {\em algebraic series} for an {\em infinite set}
of Tutte-Beraha values of the $ \, \lambda$ parameter. Note that these selected values
can also be seen as $\, N$-th root of unity situation.

At first sight, these Tutte-Beraha examples~\cite{Selected},
or $\, \lambda$-extension
of correlation functions of the Ising model~\cite{extension,lambdaPainl},
{\em are not related to Schwarzian equations}
with their composition function properties\footnote[1]{These $\, \lambda$-extension
  of Ising correlation functions are solutions of {\em Painlev\'e equations}~\cite{extension,lambdaPainl}.}. Is it
possible that such differentially algebraic series
could also reduce (in a more or less involved way ...)
to exact decompositions like $ \, X\Bigl( \omega \cdot \, Q^n(x)  \Bigr)$, that we found
systematically through this paper, since many of the results of this paper
are, in fact, consequences of such exact decompositions  ? 

\vskip .1cm 

$\bullet$  One motivation of this paper was to understand the very nature of the one-parameter
series $\, y(a, \, x)$ : we have seen that this series cannot be solution of an order-$\, N$
linear differential operator (for some integer $\, N$ independent of the parameter $\, a$)
with coefficients polynomials in $\, x$ and in the parameter $\, a$.

\vskip .1cm 
\vskip .1cm 

$\bullet$ The relation between the Schwarzian equations (such that
$\, W(x) \, = \, -1/2/x^2 \, + \, \, \cdots$, see~\cite{Schwarz1,Schwarz2}), 
and {\em modular correspondences} was also an important motivation.  The solutions of the
Schwarzian equations are larger
that just the (infinite) set of ``modular correspondences'', {\em precisely} because of the occurrence
of one-parameter series $\, y(a, \, x)$.
 Along this line we have
first seen that the solution of the Schwarzian equations  can actually correspond to series with
{\em more than one parameter}.
Modular correspondences  are associated with
modular curves and modular forms~\cite{Koblitz,1-2-3,Berndt}. Consequently,
another question was to know if one can generalize these concepts {\em beyond}
the elliptic curves and modular forms framework. 

\vskip .1cm 

We have also shown, with very simple (polynomial) examples for the function
$\, F(x)$, that these structures can actually be generalized {\em far beyond}
the elliptic curve (modular curve, Shimura curves, modular form, automorphic form) framework.
Along this line, a first polynomial example (\ref{badtruncation})
provides an example of {\em one-parameter series} $\, y(a, \, x)$,
{\em algebraic for any value of the parameter}. We also found that the
equivalent of the nome is a simple algebraic function (square root of a rational function).
With that example
one also understands why it can be extremely hard to see that some series
are algebraic, even if the algebraic function to guess is of a quite simple form. 
Furthermore, a ``truncated'' example (\ref{truncation})
shows that the ``modular equation-like'' series (see for instance (\ref{y2more}), (\ref{trunc5invol}))  
can actually be {\em  non globally bounded}. The  ``nome-like''
series is a {\em  non globally bounded}, 
but {\em still} $\, D$-finite, series (see (\ref{order-one})),  the corresponding
 linear differential operator being {\em non globally nilpotent}.

\vskip .1cm 

\section{Conclusion}
\label{Conclus}

This paper provides a simple, and pedagogical, illustration of 
exact non-linear symmetries in physics (exact representations of the 
renormalization group transformations like the {\em Landen transformation} for 
the square Ising model~\cite{Heegner,bo-ha-ma-ze-07b},  ...) and is a strong incentive
to discover more {\em differentially algebraic} equations involving fundamental symmetries, and to 
develop more  differentially algebraic series analysis 
in physics~\cite{Selected,IsTheFull}, beyond examples 
like the full susceptibility of the 
square-lattice Ising model~\cite{IsTheFull,extension,lambdaPainl,Automat,bo-gu-ha-je-ma-ni-ze-08}.

In this paper we first focused, essentially, on identities relating the {\em same} 
hypergeometric function with {\em two different} algebraic pullback 
transformations related by   modular equations. This corresponds
to the ``classical''  {\em modular forms}~\cite{SuperMaier} (resp. automorphic forms)
that emerged so many times in physics~\cite{IsingCalabi,IsingCalabi2,Christol}:
these algebraic transformations can be seen as  simple illustrations 
of exact representations of the renormalization group
of some Yang-Baxter integrable models~\cite{Heegner,bo-ha-ma-ze-07b,Hindawi}.
These transformations are seen to be solutions of some Schwarzian relation.

The Schwarzian relation is seen 
to ``encapsulate'', in one differentially algebraic (Schwarzian) 
equation, all the {\em modular forms} 
and {\em modular equations} of the theory of elliptic curves. 
The Schwarzian condition can thus be seen as some 
quite fascinating ``pandora box'', which encapsulates an {\em  infinite number}
of highly remarkable modular equations, and a whole ``universe'' of  
{\em Belyi-maps}\footnote[2]{Belyi-maps~\cite{Belyi3,Belyi,Belyi2,Belyi4,Belyi5}
 are central to Grothendieck's program 
 of ``dessins d'enfants''.}.
It is however important to underline that these Schwarzian conditions
are  actually richer than just elliptic curves, and 
go beyond\footnote[9]{See
  the two Heun functions given by (164) in~\cite{Shimura_Una}.} 
``simple'' restrictions~\cite{maier-05}
to pullbacked $\, _2F_1$ hypergeometric functions.
In a more general perspective, such Schwarzian conditions
occur in Malgrange's pseudo-group approach~\cite{Casale,Casale2,Casale3,Malgrange}
of $\, {\cal D}$-enveloppes. At this level of mathematical abstraction,
the question of a {\em modular correspondence interpretation}
of these ``Schwarzian'' series was clearly an open question. This paper sheds some light on
this open question. It shed some light on the very nature of the
one-parameter series solution of the Schwarzian equation, which is {\em not} generically
a modular correspondence series, but {\em actually reduces} to an
{\em infinite set of modular correspondence series} for an infinite set
of ($N$-th root of unity) values of the parameter. This paper also provides (polynomial)
examples that are very similar to modular correspondence series, but are far beyond the
elliptic curves framework.

\vskip .1cm
\vskip .4cm 
\vskip .4cm 
\vskip .4cm 

{\bf Acknowledgments:} 
We would like to thank
 A. Bostan and  G. Casale,  for very fruitful 
 discussions.  We thank  A. Bostan for some $\, p$-curvature calculations.
 This work has been performed 
without any ERC, ANR, PES or MAE financial support. 

\vskip .5cm

\pagebreak 

\vskip .5cm

\appendix

\vskip .2cm

\section{Miscellaneous modular correspondences.}
\label{miscellancorresp}

$\bullet$ Let us consider
\begin{eqnarray}
\label{mod7}
  \hspace{-0.98in}&&  \quad  \quad  \,  \,\, \,
\tilde{X}\Bigl(\tilde{Q} (x)^7 \Bigr) \, \, =  \, \, \,  \, \,
x^7 \,\,  \,+5208\, x^8 \, \, \, +16877196\, x^9 \,\, \, +43972589024\, x^{10}
 \nonumber \\
  \hspace{-0.95in}&& \quad \quad  \,    \quad  \, \quad  \quad   \,
 +101156052918270\,\, x^{11}   \, \,  +215029151897268240\, x^{12}
 \, \,\,\, \, \, \,  + \, \, \, \cdots 
\end{eqnarray}
which is solution of a modular equation
$\, \Gamma_7(x, \, y) \, = \,  \Gamma_7(y, \, x) \, = \, 0\,\, $ that we will
not write here, but can easily be obtained from its
rational parametrization~\cite{SuperMaier}:
\begin{eqnarray}
\label{gamma7}
  \hspace{-0.98in}&&  
x  = \,      {{t} \over {(t^2 \, +13\,t \, +49) \cdot \, (t^2 \, +5\,t  \,+1)^3}},  
\quad
y  = \,    {{t^7} \over {(t^2 \, +13\,t \, +49) \cdot \, (t^2 \, +245\,t  \,+2401)^3}}.
\nonumber 
\end{eqnarray}                             
This series (\ref{mod7}) is solution of an order-eight linear differential operator
$\, L_8 \, = \, \, L_1 \oplus \, L_7$, which is the direct sum of an
order-one  linear differential operator $\, L_1$
with a rational function solution of the form
\begin{eqnarray}
\label{gamma7L1}
\hspace{-0.98in}&&  \quad \quad \quad 
x \cdot \, {{p_6(x) } \over {
    (11356800389480448000000\, x^2 \, +34848505552896000\, x \, +1 )^3 }}, 
\end{eqnarray}
where $\, p_6(x)$ is a polynomial of degree six, 
and an order-seven irreducible
linear differential operator $\, L_7$.

\vskip .1cm 

$\bullet$  Let us also consider
\begin{eqnarray}
  \label{mod8}
  \hspace{-0.98in}&&  \, \, \,   
  \tilde{X}\Bigl(\tilde{Q} (x)^{8} \Bigr) \, \, =  \, \,  \,\, \,\,   \,
 x^8 \, \,\,  +5952\, x^9 \,\,  +21502368\, x^{10} \, \, +61310179840\, x^{11}
 \, \, \,\,\,  + \, \, \, \cdots 
\end{eqnarray} 
which is solution of a modular equation
$\, \Gamma_{8}(x, \, y) \, = \,  \Gamma_{8}(y, \, x) \, = \, 0\, $ that we will
not write here, but can easily be obtained from its
rational parametrization~\cite{SuperMaier}:
\begin{eqnarray}
\label{gamma8}
  \hspace{-0.98in}&&      \quad     \quad  \quad  \quad  \quad     \quad     \quad      
 x \, \,  = \,  \, \,  \, \,
     {{t \cdot \, (t+4)^2 \cdot \, (t+8)} \over {
     (t^4 \, +16\, t^3 \, +80\, t^2 \, +128\,t  \,+16)^3}},
\nonumber \\
   \hspace{-0.98in}&&  \quad     \quad   \quad  \quad   \quad  \quad    \quad             
y  \, \, = \, \, \,\,  \,
     {{t^{8} \cdot \, (t+4) \cdot \, (t+8)^2 } \over {
           (t^4 \, +256\, t^3 \, +5120\, t^2 \, +32768\,t  \,+65536)^3}}.
\end{eqnarray}
The polynomial, associated with the modular equation
$\, \Gamma_{8}(x, \, y) \, = \, 0$,  is of degree $\, 12$ in $\, y$ (resp. in $\, x$).
This series (\ref{mod8}) is solution of an order-twelve linear differential operator
$\, L_{12} \, = \, \, L_1 \oplus \, L_{11}$, which is the
direct sum of an order-one  linear differential operator $\, L_1$
with a rational function solution of the form 
\begin{eqnarray}
\label{gamma8rat}
\hspace{-0.98in}&&      \quad  \quad   \quad   \quad
\quad  \quad   \quad  \quad   \quad  \quad   \quad  \quad \quad\quad \quad \quad
x \cdot \, {{  p_{11}(x)} \over {p_4(x)^3 }}, 
\end{eqnarray}
where $\, p_{11}(x)$ is a polynomial of degree eleven, and where $\, p_4(x)$ reads
\begin{eqnarray}
\label{p4}
\hspace{-0.98in}&&   \quad
1080060886113159937649308593750000\, x^4
\,\, -826335556188178615474500000000\, x^3
\nonumber \\
\hspace{-0.98in}&&  \quad  \quad
\, -15705521635909735050750000\, x^2 \,\, \, +8041801037378436000\, x \,\,\,  -1, 
\end{eqnarray}
and an order-eleven  linear differential  operator $\, L_{11}$.
The other {\em algebraic } solution series of $\, \Gamma_{8}(x, \, y) \, = \, 0 \, $ are
the compositional inverse of series (\ref{mod8}), namely 
\begin{eqnarray}
\label{gamma8other2}
  \hspace{-0.98in}&&  \quad     \quad \quad
 \omega \cdot \, x^{1/8} \, \
 -744 \cdot \,   \omega^2 \cdot \, x^{2/8} \, \, \,  +356652  \omega^3 \cdot \, x^{3/8} \, \,\,
  -140361152 \cdot \,   \omega^4 \cdot \, x^{4/8}
 \nonumber \\
 \hspace{-0.98in}&&  \quad   \,\,  \quad \quad \, \,\,\,
  +49336682190  \cdot \,   \omega^5 \cdot \, x^{5/8} \, \,\,
  -16114625669088 \cdot \,   \omega^6 \cdot \, x^{6/8} \, \,\,
  \, \, + \, \, \, \cdots 
\end{eqnarray}
where $\, \omega^8 = \, 1$, 
together with 
\begin{eqnarray}
\label{gamma8other}
  \hspace{-0.98in}&&  \quad  \quad   \, \,   \, 
  -x^2 \,\,\, -1488\, x^3 \,\,\, -2055120\, x^4
      \,\, \, -2864378368\, x^5 \,\,\, -4071821465856\, x^6
  \nonumber \\
 \hspace{-0.98in}&&  \quad     \quad \quad \quad
  \, -5900566305239040\, x^7 \,\, \,-8695398352685449216\, x^8
  \,\, \,\, \, + \, \, \, \cdots 
\end{eqnarray}
which is\footnote[5]{In contrast, note that  $\, y_2(y_1(x)) \, = \, \, y_2(x)$.}
$\, y_1(y_2(x))$
(where $\, y_1$ is (\ref{modularequasol14c}) and $\, y_2$ is (\ref{modularequasol1})),
and
\begin{eqnarray}
\label{gamma8other2}
  \hspace{-0.98in}&&  \quad  \quad \, \, \, \quad \, \, 
 \omega \cdot \, x^{1/2} \, \, \,
  -744 \cdot \,   \omega^2 \cdot \, x^{2/2} \, \,   + 356280 \cdot \,  \omega^3 \cdot \, x^{3/2}
\, \,\,\, -139807616  \cdot \,  \omega^4 \cdot \, x^{4/2}
\nonumber \\
\hspace{-0.98in}&&  \quad   \quad \quad \quad \quad  \quad \quad \, \,
  \, \, + 48938964576  \cdot \,  \omega^5  \cdot \, x^{5/2} \, \, \,
 \, \, + \, \, \, \cdots                  
\end{eqnarray} 
where $\, \omega^2 = \, -1$.
The number of algebraic solution series of  $\, \Gamma_{8}(x, \, y) \, = \, 0$, or $\, L_{12} $,
is $\, 1 \, +1 \, +8 \, +2\, = \, 12$.  

\vskip .1cm 

$\bullet$ The (algebraic) series
\begin{eqnarray}
\label{mod11}
  \hspace{-0.98in}&&  \,
 \tilde{X}\Bigl(\tilde{Q} (x)^{11} \Bigr) \, \, =  \, \,  \,
 x^{11} \,  \,+8184 \, x^{12} \,  \,+38699100\, x^{13} \, \, +138966918112\, x^{14} \,      
    \, \, + \, \, \, \cdots 
\end{eqnarray}
 is solution of a modular equation
$\, \Gamma_{11}(x, \, y) \, = \,  \Gamma_{11}(y, \, x) \, = \, 0\, $ that we will
 not write here. The polynomial, associated with the modular equation
$\, \Gamma_{11}(x, \, y) \, = \, 0$,  is of degree $\, 12$ in $\, y$ (resp. in $\, x$).
 In contrast with the previous examples, this  modular equation
is a {\em genus-one} curve\footnote[1]{See Appendix I in~\cite{ChristolUna},
  which is  the unabridged arXiv version of~\cite{Christol}.}. We, thus,
{\em do not have a rational parametrization} of the modular equation~\cite{SuperMaier}.
This series is solution of  an order-twelve linear differential operator
$\, L_{12} \, = \, \, L_1 \oplus \, L_{11}$, which is the direct sum of
an order-one linear differential operator $\, L_1$
with a rational function solution
\begin{eqnarray}
\label{mod11ratsol}
\hspace{-0.98in}&&  \, \, \,
\quad \quad \quad \quad  \quad \quad \quad \quad \quad\quad \quad \quad \quad\quad \quad \quad
x \cdot \, {{p_{11} } \over { p_4^3}}, 
\end{eqnarray}
where $\, p_{11}$ is polynomial of degree eleven,
and where $\, p_4$ reads
\begin{eqnarray}
\label{mod11ratsol}
\hspace{-0.98in}&&  \, \, \, \, \, \, \,
\quad \quad \quad \quad  \quad 
1577314437358442913340940353536000000000000\, x^4
\nonumber \\
 \hspace{-0.98in}&&  \, \, \, \quad  \quad \quad \quad  \quad \quad 
 \, -496864268553728774541064273920000000000 \, x^3
\nonumber  \\
 \hspace{-0.98in}&&   \quad  \quad  \quad \quad  \quad  \quad  \quad  \quad 
 +45688143672322270430861721600000000\, x^2
 \nonumber \\
 \hspace{-0.98in}&&  \, \, \,\quad \quad  \quad \quad \quad  \quad  \quad  \quad  \quad\,
   +98823634118413525094400000 \, \, x \, \, \, +1,
\end{eqnarray}
and an order-eleven
linear differential  operator $\, L_{11}$.

\vskip .1cm

$\bullet$ The (algebraic) series
\begin{eqnarray}
\label{mod12}
  \hspace{-0.98in}&&  \,  
 \tilde{X}\Bigl(\tilde{Q} (x)^{12} \Bigr)  \, =  \, \,  \,  
 x^{12} \,  \, \, +8928 \, x^{13} \, \, +45538416 \, x^{14} \, \, + 174773255424 \, x^{15} \, \,     
     \, + \, \, \, \cdots 
\end{eqnarray}
is solution of a modular equation
$\, \Gamma_{12}(x, \, y) \, = \,  \Gamma_{12}(y, \, x) \, = \, 0$, that we will
not write here, but can easily be obtained from its
rational parametrization~\cite{SuperMaier}:
\begin{eqnarray}
\label{mod12rat}
  \hspace{-0.98in}&&  \, 
 x   = \, 
 t \cdot \,
 {{(t+2)^3\, (t+3)^4\, (t+4)^3\, (t+6)} \over { (t^2 +6\,t +6)^3 \cdot \, p_{6}^3}},
\quad 
y  = \,
t^{12} \cdot \,
{{(t+2)\, (t+3)^3\, (t+4)^4\, (t+6)^3 } \over { (t^2 +12\,t +24)^3 \cdot \,   q_{6}^3}},
 \nonumber 
\end{eqnarray}
where:
\begin{eqnarray}
\label{mod12ratpol}
  \hspace{-0.98in}&&  \,  \,  \quad   \quad 
 p_{6} \, \, = \, \, \,\,
 t^6 \,\, \, +18\,t^5\,\, \,+126\, t^4 \,\,\, +432\, t^3 \,\,\,
 +732\, t^2 \,\,\, +504\, t \,\, +24, 
 \\
  \hspace{-0.98in}&&  \,  \,  \quad   \quad 
q_{6} \, \, = \, \, \,\,
t^6 \,\, +252\, t^5 \,\, +4392\, t^4 \,\, +31104\, t^3 \,\, +108864\, t^2
\,\, +186624\, t \,\, \, +124416.
  \nonumber 
\end{eqnarray}
The polynomial, associated with the modular equation
$\, \Gamma_{12}(x, \, y) \, = \, 0$,  is of degree $\, 24$ in $\, y$ (resp. in $\, x$).
The  (algebraic) series solutions of the  modular equation
$\, \Gamma_{12}(x, \, y) \, = \, 0\, $ are solutions of an order-$24$
linear differential operator $\, L_{24} \, = \, \, L_1 \oplus \,   L_{23}$,
which is the direct sum of an order-one linear differential
operator $\, L_1$,
with a rational function solution
\begin{eqnarray}
\label{mod12ratpolsol}
\hspace{-0.98in}&&  \,  \,  \, \,   \quad  \quad    \quad   \quad   \quad   \quad  \quad   \quad
x \cdot \, {{ p_{23}} \over { (6549518250000\,x^2\, -2835810000\,x \, +1)^3 \cdot \, p_6^3 }} 
\end{eqnarray}
where $\, p_{23}$ is a polynomial of degree 23, and where $\, p_6$ is the polynomial
\begin{eqnarray}
\label{mod12ratpolsolp6}
  \hspace{-0.98in}&&  \,  \,  \,   \,  \,  \,  \quad   \quad   \quad  \quad  
  42889619864187195342544128412237640625000000000000\, x^6
  \nonumber \\
 \hspace{-0.98in}&&  \,  \,  \quad  \quad  \quad   \quad  \quad   
 \, +3869372376492639837782614434923625000000000000\, x^5
  \nonumber \\
 \hspace{-0.98in}&&  \,  \,  \quad  \quad   \quad   \quad   \quad  \quad  
 +34904627315764077727184412247908187500000000\, x^4
 \nonumber \\
 \hspace{-0.98in}&&  \,  \,  \quad \, \quad   \quad  \quad  \quad \quad      \quad  
 +1007059405271040783775694468925000000000\, x^3
  \nonumber \\
 \hspace{-0.98in}&&  \,  \,  \quad \, \quad   \quad  \quad   \quad \quad   \quad   \quad    
 \, +280179539493990596285512318134750000\, x^2
 \nonumber \\
 \hspace{-0.98in}&&  \,  \,  \quad \quad \, \quad   \quad  \quad    \quad   \quad  \quad   \quad  
 \, -22804995243537595825782822000\, x \,\, \,  +1, 
\end{eqnarray}
and an order-$23$ linear
differential operator  $\, L_{23}$.
The other (algebraic) series are, respectively,
the compositional inverse of series (\ref{mod12}), namely
\begin{eqnarray}
\label{mod12other2}
  \hspace{-0.98in}&& \,   \quad \quad  \,
    \omega \cdot \, x^{1/12}  \,\, \,\, -744 \cdot \, \omega^2 \cdot \, x^{2/12}
 \,\, \, \, +356652 \cdot \, \omega^3 \cdot \, x^{3/12}
\,\, \, -140361152 \cdot \, \omega^4 \cdot \, x^{4/12}
\nonumber \\
 \hspace{-0.98in}&&  \,\,   \quad \quad \quad \quad 
   \, +49336682190 \cdot \, \omega^5 \cdot \, x^{5/12}
\,\,\,  -16114625669088 \cdot \, \omega^6 \cdot \, x^{6/12}
\nonumber \\
  \hspace{-0.98in}&&  \,\, \,  \quad  \quad \quad \quad \quad \quad 
+4999042477430456 \cdot \, \omega^7 \cdot \, x^{7/12}
\, \,\, \,\, + \, \, \, \cdots 
\end{eqnarray}                  
where $\, \omega^{12} \, =\, 1$, together with
\begin{eqnarray}
\label{mod12other1}
  \hspace{-0.98in}&&  
-x^3 \, -2232\, x^4 \, -3911868\, x^5 \, -6380015304\, x^6 \, -10139549171670\, x^7 \,
  \, \, + \, \, \, \cdots 
\end{eqnarray}
which is nothing but $\, y_3(y_1(x)) \,= \, \,  y_1(y_3(x))$ (with $\, y_3$
given by (\ref{modularequasol13}), and
$\, y_1$ given by (\ref{modularequasol14c})),
and 
\begin{eqnarray}
\label{mod12other3}
  \hspace{-0.98in}&&  \,\quad \quad \quad \quad 
 \omega \cdot \, x^{1/3} \,\, \,  \,-744 \cdot \, \omega ^2 \cdot \, x^{2/3}
  \,  \,  \,\, -356652 \cdot \, x \, \, \,
\,  +140361400 \cdot \, \omega \cdot \, x^{4/3}
\nonumber \\
  \hspace{-0.98in}&&  \,\quad \quad \quad \quad \quad  \quad \,
-49337051214 \cdot \, \omega ^2\cdot \, x^{5/3} \,\,\,
 -16114891018176 \cdot \, x^{6/3}
\nonumber \\
\hspace{-0.98in}&&  \,\quad \quad \quad \quad \quad \quad   \quad   \quad  \,
+4999181715881876 \cdot \, \omega \cdot \, x^{7/3}
  \, \,\, \,\, + \, \, \, \cdots 
\end{eqnarray}                  
where $\, \omega^{3} \, =\, 1$, 
\begin{eqnarray}
\label{mod12other4}
  \hspace{-0.98in}&&  \,\,\, \,  \, 
\omega \cdot \, x^{4/3} \,\, \, +992 \cdot \, \omega \cdot \, x^{7/3} \,\,\,
 -744 \cdot \, \omega^2 \cdot \, x^{8/3}\,\,\,  +1123568 \cdot \, \omega \cdot \, x^{10/3}
\,\, \, \, + \, \, \, \cdots                   
\end{eqnarray}
where $\, \omega^{3} \, =\, -1$, and
\begin{eqnarray}
\label{mod12other5}
  \hspace{-0.98in}&&  \,\, \,\,\, \,  \, 
 \omega \cdot \, x^{3/4}\,\,\, \, -744 \cdot \, \omega ^2 \cdot \, x^{6/4} \,\,\,
 +558 \cdot \, \omega \cdot \, x^{7/4} \, \, \, +356652 \cdot \, \omega ^3 \cdot \, x^{9/4}       
  \, \, \,\, + \, \, \, \cdots                   
\end{eqnarray}            
where $\, \omega^{4} \, =\, 1$. 
This gives $\, 1 \, +1 \, + 12 \, + 3\, +3 \, +4 \, = \, \, 24 \, \, $
algebraic series, solutions of $\, \Gamma_{12} \, = \, 0 \,  \, $ and $\, L_{24}$.

{\bf Remark \ref{modularcurve}.6:}  Recalling the algebraic series $\, y_3$,
given by (\ref{modularequasol13}),
and the algebraic series $\, y_1$ given by (\ref{modularequasol14c}),
one can see that the algebraic series (\ref{mod12other1}) is {\em nothing but}:
\begin{eqnarray}
\label{relationy3Y1}
  \hspace{-0.98in}&&  \,\,  \quad \, \, \, \,
 y_3(y_1(x)) \, \, = \, \, \,  y_1(y_3(x))
 \, \, = \, \, \, \,  \,  -x^3 \,\, \, \, -2232\, x^4 \,\, \, -3911868\, x^5
 \,\,   \, \,  \, + \, \, \, \cdots 
\end{eqnarray} 

\vskip .1cm

$\bullet$ The (algebraic) series
\begin{eqnarray}
\label{mod16}
  \hspace{-0.98in}&&  \, \, \,  \,   \quad \quad   
 \tilde{X}\Bigl(\tilde{Q} (x)^{16} \Bigr)  \, =  \,\, \,  \,\, 
 x^{16} \,\,\,  \,+ 11904 \, x^{17} \, \, + 78431040 \, x^{18} \,\, \, + 378584548352 \, x^{19}
 \nonumber \\
  \hspace{-0.98in}&&  \, \quad \,  \,  \,  \quad \quad \,\, \quad \, \quad \quad \quad 
 +1496557573544352\, x^{20} \, \,\, \,\, + \, \, \, \cdots 
\end{eqnarray}
is solution of a {\em modular equation}
$\, \Gamma_{16}(x, \, y) \, = \,  \Gamma_{16}(y, \, x) \, = \, 0$, that we will
not write here, but can easily be obtained from its rational parametrization~\cite{SuperMaier}:
\begin{eqnarray}
\label{mod16rat}
  \hspace{-0.98in}&&  \,  \,  \,   \,  \,  
 x \,  \,   = \, \, \,
 t \cdot \,  {{ (t+2)^4 \, (t+4) \, (t^2 +4\,t +8)} \over { p_{8}^3}},
\quad    \,  \,  
y  \,  \, = \,\, \,
  t^{16} \cdot \,  {{  (t+2) \, (t+4)^4 \, (t^2 +4\,t +8)} \over {  q_{8}^3}},
 \nonumber 
\end{eqnarray}
where:
\begin{eqnarray}
\label{mod25ratp10}
  \hspace{-0.98in}&&  \, \, \,  \,  \quad  
 p_{8} \, \, = \, \, \, \,
 t^8 \, +16\, t^7 \, +112\,t^6 \, +448\, t^5 \, +1104\, t^4
  \, +1664\, t^3 \, +1408\, t^2 \, +512\, t \, +16, 
                     \nonumber \\
\hspace{-0.98in}&&  \,  \, \, \,  \quad  
  q_{8} \, \, = \, \, \, \,
  t^8 \,\, +256\, t^7 \, \,+5632\, t^6 \,\, +53248\, t^5 \,\, +282624\, t^4
                   \,\, +917504\, t^3 \,\, +1835008\, t^2
 \nonumber \\
\hspace{-0.98in}&&  \,  \,  \quad  \quad  \quad  \quad    \quad  \quad   \quad  
 \, +2097152\, t \,\, +1048576.
\end{eqnarray}                    
The polynomial in the modular equation
$\, \Gamma_{16}(x, \, y)  \, = \, 0\, $ is of degree $\, 24$ in $\, y$
(resp. in $\, x$), and, thus, has twenty four algebraic solution series,
corresponding to the  series (\ref{mod16}),
the compositional inverse of series (\ref{mod16}), namely 
\begin{eqnarray}
\label{together2}
  \hspace{-0.98in}&&  \, \quad \quad \quad  \quad  \quad   \,  \, \, \quad  
 \omega \cdot \, x^{1/16} \, \, \, -744 \cdot \, \omega^2 \cdot \,  x^{2/16} \,\, \, \,
  +356652 \cdot \, \omega^3 \cdot \,  x^{3/16}
\nonumber \\
\hspace{-0.98in}&&  \,  \quad  \quad \quad  \quad  \quad  \quad \quad   \,  \quad    \quad  \quad   \, \,
 -140361152  \cdot \, \omega^4 \cdot \,  x^{4/16}
  \, \, \,  \, + \, \, \, \cdots 
\end{eqnarray}
where $\, \omega^{16} \, = \, \, 1$, together with
\begin{eqnarray}
\label{together1}
  \hspace{-0.98in}&&  \,    \quad  \quad   \quad  
  -x^4  \,  \,\, -2976\, x^5 \,  \,\,  -6322896\, x^6 \, \,  \,
   -11838151424\, x^7 \, \, \,  -20872495229904\, x^8
  \nonumber \\
\hspace{-0.98in}&&  \,  \,   \quad    \quad   \quad  \quad   \quad  \quad   \quad  
 \, -35647177059836928\, x^9  \,  \, \, \, + \, \, \, \cdots       
\end{eqnarray}   
and
\begin{eqnarray}
\label{together3}
  \hspace{-0.98in}&&  \, \quad  \quad  \quad  \quad   \quad  \quad  
 \omega \cdot \, x^{1/4} \,  \, \,\, -744 \cdot \, \omega^2 \cdot \,  x^{2/4} \, \,  \,\,
  +356652 \cdot \, \omega^3 \cdot \,  x^{3/4}
\nonumber \\
\hspace{-0.98in}&&  \,  \quad  \quad \quad \quad
\quad  \quad  \quad   \quad  \quad   \quad   \,  \, \,  \,
 -140361152  \cdot \, \omega^4 \cdot \,  x^{4/4}
 \,\, \, \, \,\, + \, \, \, \cdots 
\end{eqnarray}
where $\, \omega^{4} \, = \, \, -1$, and: 
\begin{eqnarray}
\label{together4}
  \hspace{-0.98in}&&  \, \quad  \quad  \quad \quad \quad \quad   \quad  \quad   
\omega \cdot \, x \, \, \,\,  \,  \, -744 \cdot \, \omega \cdot \, (\omega -1) \cdot \, x^2
  \,  \, \,\, \, + \, \, \, \cdots    
\end{eqnarray}
where $\, \omega^{2} \, = \, \, -1$.
The  (algebraic) series solutions of the  modular equation
$\, \Gamma_{16}(x, \, y) \, = \, 0\, $ are solution of an order-$24$
linear differential operator $\, L_{24} \, = \, \, L_1 \oplus \,   L_{23}$,
which is the direct sum of an order-$23$ linear differential operator $ \, L_{23}$, 
and an order-one linear differential
operator $\, L_1$,
with a rational function solution
\begin{eqnarray}
  \label{mod16ratpolsol}
  \quad  \quad  \quad \quad \quad \quad \quad \quad \quad \quad \quad
  x \cdot \, {{p_{23} } \over { p_8^3}}, 
\end{eqnarray}
where $\, p_{23}$ is a polynomial of degree 23, and where $\, p_8$
is the polynomial
\begin{eqnarray}
  \label{mod16ratpolsolp8}
  \hspace{-0.98in}&& 
  15926143836920796849094002857387135460968690480161221686575776100158691406250000\,x^8
  \nonumber \\
 \hspace{-0.98in}&&
 -6042818923606714182438083804301870179528875596947614517314453125000000000000\,x^7
 \nonumber \\
 \hspace{-0.98in}&&   
 +4900698705373764641365354757988280247136785578572898154329101562500000000\,x^6
 \nonumber \\
 \hspace{-0.98in}&&
 +46721890317786185410700227174952944124137546155237676733203125000000000\, x^5
 \nonumber \\
 \hspace{-0.98in}&&  
 +81580198367732340212612911642019252294658707587093110574218750000\, x^4
 \nonumber \\
 \hspace{-0.98in}&&  \,  \quad  
 +736154608709059015006498116049282929703692588255135000000000\, x^3
 \nonumber \\
 \hspace{-0.98in}&&  \,  \quad  \quad 
 +259399171372225204966661002550162965440584749500000\, x^2
 \nonumber \\
 \hspace{-0.98in}&&  \,  \quad  \quad  \quad
 -64670563924749466394147714711210760000\, x \, \, \, \, +1.
\end{eqnarray}

{\bf Remark \ref{modularcurve}.7:}  Recalling the algebraic series $\, y_4$,
given by (\ref{modularequasol14}),
and the algebraic series $\, y_1$ given by (\ref{modularequasol14c}), one can see that the
algebraic series (\ref{together1}) is
{\em nothing}\footnote[1]{In contrast  $\, y_4(x) \, \, = \, \, \,  y_4(y_1(x))$.} {\em but}:
\begin{eqnarray}
\label{relationy1y4}
  \hspace{-0.98in}&&  \,\quad   \quad  \quad  \quad  \quad \, \,
  y_1(y_4(x))
  \, \, = \, \, \, \,  \,   -x^4  \,\, \, \,   -2976\, x^5 \,\,  \,  \,  -6322896\, x^6
  \,  \,\, \,  \, + \, \, \, \cdots 
\end{eqnarray} 
We thus have $\, 1 \, +1 \, +16 \, +4  +\, 2 \, = \, \, 24\, $
{\em algebraic solutions} of the {\em  modular equation}
$\, \Gamma_{16}(x, \, y) \,  \, = \, 0$,  and also solutions of $\, L_{24}$.

\vskip .1cm
\vskip .1cm

$\bullet$ The (algebraic) series
\begin{eqnarray}
\label{mod18}
  \hspace{-0.98in}&&  \, \quad  \quad \quad  \quad   
 \tilde{X}\Bigl(\tilde{Q} (x)^{18} \Bigr)  \, =  \,\, \,\,  
 x^{18} \,\,  \, + 13392 \, x^{19} \,\, \, + 98198568 \, x^{20} \,\, \, + 522607392000 \, x^{21}
 \nonumber \\
  \hspace{-0.98in}&&  \, \quad \quad \quad \quad  \quad    \,\, \quad \, \quad \quad \quad 
 + 2259156547520244 \, x^{22}
\, \,\,  \, + \, \, \, \cdots 
\end{eqnarray}
is solution of a modular equation
$\, \Gamma_{18}(x, \, y) \, = \,  \Gamma_{18}(y, \, x) \, = \, 0$, that we will
not write here, but can easily be obtained from its rational parametrization~\cite{SuperMaier}:
\begin{eqnarray}
\label{mod18rat}
  \hspace{-0.98in}&& \quad   \quad  \quad \quad  \,  \,  \,   \,  \,  
 x \,  \,   = \, \, \,\, \,
 t \cdot \,  {{ (t+2)^9 \cdot \, (t+3)^2 \cdot \, (t^2 +3\, t +3)^2 \, (t^2 +6\, t+12)} \over {
   (t^3 +6\,t^2 \,+12\,t +6)^3 \cdot \, p_{9}^3}},
\nonumber \\
   \hspace{-0.98in}&&   \quad  \quad  \quad  \quad \,  \,  \,   \,  \,                     
y  \,  \, = \,\, \,\, \,
 t^{18} \cdot \,  {{ t+2)^2 \cdot \, (t+3)^9 \cdot \, (t^2 +3\,t +3)\, (t^2 +6\,t +12)^2} \over {
  (t^3 +12\,t^2 +36\,t +36)^3 \cdot \,  q_{9}^3}},
\end{eqnarray}
where:
\begin{eqnarray}
\label{mod18ratp9}
  \hspace{-0.98in}&&  \,  \, \, \, \quad \quad   \quad 
 p_{9} \, \, = \, \, \, \,
   t^9  \,\, +18\, t^8 \,\, +144\, t^7 \, \,+666\, t^6 \,\, +1944\, t^5 \,\,+3672\, t^4
 \nonumber \\
\hspace{-0.98in}&&  \,  \,  \quad   \quad \quad \quad \quad \quad \quad \quad 
 \, +4404\, t^3 \, \, +3096\, t^2 \, \, +1008\, t \,\,  +24, 
 \nonumber \\
\hspace{-0.98in}&&  \,  \, \, \, \quad   \quad  \quad 
q_{9} \, \, = \, \, \, \,
t^9\,\,   +252\, t^8 \,\,  +4644\, t^7 \, \,  +39636\, t^6 \, \, +198288\, t^5
\,\,  +629856\, t^4 \, 
\nonumber \\
  \hspace{-0.98in}&&  \,  \,  \quad   \quad  \quad  \quad  \quad    \quad  \quad   \quad
 +1294704\, t^3 \, \,  +1679616\, t^2 \,\,  +1259712\, t \,\,   +419904.
  \end{eqnarray}                    
The polynomial in the modular equation
$\, \Gamma_{18}(x, \, y)  \, = \, 0\, $ is of degree $\, 36 \, $ in $\, y$
(resp. in $\, x$), and thus
has thirty-six algebraic solution series, corresponding to the  series (\ref{mod18}),
together with the compositional inverse of series (\ref{mod18}), namely 
\begin{eqnarray}
\label{18together1}
  \hspace{-0.98in}&&  \, \quad \quad  \quad \quad  \quad   \quad    \quad  
 \omega \cdot \, x^{1/18} \, \, \, \, -744 \cdot \, \omega^2 \cdot \,  x^{2/18}  \, \, \,\, 
 +356652 \cdot \, \omega^3 \cdot \,  x^{3/18}
 \nonumber \\
  \hspace{-0.98in}&&  \,  \quad  \quad  \quad \quad \quad  \quad   \quad    \quad  \quad   \, \,
 -140361152  \cdot \, \omega^4 \cdot \,  x^{4/18}
  \, \, \, \, \, + \, \, \, \cdots 
\end{eqnarray}
where $\, \omega^{18} \, = \, \, 1$, together with
\begin{eqnarray}
\label{18together2}
  \hspace{-0.98in}&&  \, \quad  \quad  \quad \quad  \quad   \quad  
\omega \cdot \, x^2 \, \,  \,\,  +1488 \cdot \, \omega \cdot \, x^3 \,\,  \, \,
 +(744 \, +2055120 \cdot \, \omega) \cdot \, x^4
  \nonumber \\
\hspace{-0.98in}&&  \,  \, \quad \quad  \quad \quad    \quad   \quad  \quad   \quad  \quad   \quad  
+(2214144  \,  + 2864378368 \cdot \, \omega) \cdot \, x^5
\, \, \,\, \, + \, \, \, \cdots       
\end{eqnarray}
where $\, 1 \, +\omega \, +\omega^2 \, = \, 0$, and
\begin{eqnarray}
\label{18together3}
  \hspace{-0.98in}&&  \, \quad  \quad  \quad    \quad    \quad  \quad  
 \omega \cdot \, x^{1/2} \, \, \, \, -744 \cdot \, \omega^2 \cdot \,  x^{2/2} \, \, \,\, 
  +(356652 \cdot \, \omega^3\, +372 \cdot \, \omega)  \cdot \,  x^{3/2}
\nonumber \\
 \hspace{-0.98in}&&  \,  \quad  \quad \quad   \quad \quad  \quad  \quad   \quad   \, \,
 +(139807616 \, \omega^2 \, +140361152) \cdot \,  x^{4/2}
 \, \,\,   \,\, \, + \, \, \, \cdots 
\end{eqnarray}
where $\, \omega^{4} + \omega^2 \, +1\, = \, \, 0$,  and
\begin{eqnarray}
\label{18together4}
  \hspace{-0.98in}&&  \, \quad  \quad \quad    \quad   \quad  \quad   
 \omega \cdot \, x^{2/9}  \, \, \,  \, \, -744 \cdot \, \omega^2  \cdot \, x^{4/9} \,\, \, \,
 +356652 \cdot \, \omega^3  \cdot \, x^{6/9} \,
  \,   \, \, + \, \, \, \cdots    
\end{eqnarray}
where $\, \omega^{9} \, = \, \, 1$,
 and: 
\begin{eqnarray}
\label{18together5}
  \hspace{-0.98in}&&  \, \quad  \quad \quad   \quad  \quad   
 \omega \cdot \, x^{9/2}  \, \, \,  \, \, +3348 \cdot \, \omega  \cdot \, x^{11/2} \,\, \,\, 
 + 7735986 \cdot \, \omega  \cdot \, x^{13/2} \, \,
   \,  \, \, + \, \, \, \cdots    
\end{eqnarray}
where $\, \omega^{2} \, = \, \, 1$.
We thus have $\, 1\, +18 \, +2 \, +4 \, +9 \, +2 \, = \, \, 36 \, $ {\em algebraic} series
of the modular equation $\, \Gamma_{18}(x, \, y)  \, = \, 0$.
These algebraic series are solutions of an order-$36$ linear differential operator $\, L_{36}$
which is the {\em direct-sum} of an order-$35$ linear differential operator $\, L_{35}$, 
and
an order-one linear differential operator
with a rational function solution
\begin{eqnarray}
\label{18together5rat}
\hspace{-0.98in}&& \quad  \quad  \quad \quad \quad  \quad
\quad  \quad  \quad  \quad \quad  \quad  \quad  \quad 
 x \cdot \, {{p_{35}} \over { p_3^3 \cdot p_9^3}},
\end{eqnarray}
where $\, p_{35}$ is a polynomial of degree 35, where $\, p_3$ reads
\begin{eqnarray}
  \label{18together5ratp3}
  \hspace{-0.98in}&& \, \, \, \, \, 
  1879994705688000000000\,x^3 \, \, -224179462188000000\,x^2 \,\,  +151013228706000\, x \, \,-1,
  \nonumber 
\end{eqnarray}
and $\, p_9$
reads:
\begin{eqnarray}
  \label{18together5ratp9}
  \hspace{-0.98in}&&
  141600617083186841426749541059379178266125496444877735060776646144000000000000000000000000000\,x^9
  \nonumber \\
  \hspace{-0.98in}&&
  -91940358193098820927255075706021981712433442298247865135275206912000000000000000000000000\, x^8
  \nonumber \\
  \hspace{-0.98in}&&
  +23575127643124642999337421097401673608067186617214973940390237696000000000000000000000\, x^7
 \nonumber \\
  \hspace{-0.98in}&&
  +472396958753110140888731003718496465436906620981212557050651392000000000000000000\, x^6
 \nonumber \\
  \hspace{-0.98in}&&
  +9765441515113592938914449083136683600667888100995340182803776000000000000000\, x^5
  \nonumber \\
  \hspace{-0.98in}&&
  +14038814920070381530487981789718895908787729404706990128672000000000000\, x^4
  \nonumber \\
  \hspace{-0.98in}&&+6390980152781882840426709358572754975540778747694537696000000000\, x^3
  \nonumber \\
  \hspace{-0.98in}&&
  +7559858588621896366536922746878187128255472000000\, x^2
  \nonumber \\
  \hspace{-0.98in}&&
  +3443855962300764146093216928806375326182000\,\, x \, \,\, -1.
\end{eqnarray}

\vskip .1cm

\section{Beyond pullbacked $\, _2F_1$ hypergeometric functions: a selected Heun function.}
\label{beyond}

Let us show that the results displayed on the classical modular curves
and their associated  modular forms with pullbacked
$\, _2F_1$ hypergeometric functions, also work on Shimura curves~\cite{Shimura}
and their associated automorphic forms~\cite{Ford} with a Heun function {\em which
cannot be reduced to  pullbacked} $\, _2F_1$  {\em hypergeometric functions}.

Recalling\footnote[1]{See also subsection 2.3 of~\cite{GlobNilp}.} Krammer's
counterexample to Dwork's conjecture~\cite{Moeller,Dettweiler,Dwork}, let us
consider the Heun function
$\, Heun(81, \, 1/2, \, 1/6, \, 1/3, \, 1/2, \, 1/2; \, x)$ which is solution of an order-two
linear differential operator $\, L_2$ which is {\em  globally nilpotent}~\cite{GlobNilp}. The
series expansion
of this Heun function is {\em not globally bounded}~\cite{Christol,ChristolUna}.
Let us introduce the following  function $\, F(x)$:
\begin{eqnarray}
\label{FHeun}
\hspace{-0.98in}&&    \,  \,\, \,
F(x) \, \, = \, \, \,
x^{1/2} \cdot \, \Bigl(1 -{{x}\over {81}} \Bigr)^{1/2} \cdot \, (1-x)^{1/2} \cdot \,
HeunG\Bigl(81,{{1} \over {2}}, \, {{1} \over {6}}, \, {{1} \over {3}},
  \, {{1} \over {2}}, \, {{1} \over {2}}; \, \, \, x\Bigr)^2,
\end{eqnarray}
or, more simply, the following (non globally bounded) series:
\begin{eqnarray}
\label{FHeunsquare}
\hspace{-0.98in}&&    \,  \,\, \,
81 \cdot \, F(x)^2 \, \, = \, \, \, x \cdot \, (81-x) \cdot \, (1-x) \cdot \,
HeunG\Bigl(81,{{1} \over {2}}, \, {{1} \over {6}}, \, {{1} \over {3}},
  \, {{1} \over {2}}, \, {{1} \over {2}}; \, \, \, x\Bigr)^4
\nonumber \\
\hspace{-0.98in}&&  \quad  \,  \,
\,  = \,  \, 81\, x \, -78\, x^2 \,  -{{137} \over {81}} \, x^3\,  -{{3892} \over{6561}} \, x^4 \,
-{{44495} \over {177147}} \, x^5 \, -{{1900594} \over {14348907}} \, x^6 \, \, \, + \, \, \, \cdots 
\end{eqnarray}
Let us consider the Schwarzian equation associated with the order-two
linear differential operator $\, L_2$. The corresponding function
$\, W(x)$ reads (see subsection (\ref{Schwarzian})): 
\begin{eqnarray}
\label{WHeun}
\hspace{-0.98in}&&   \quad  \quad  \quad   \,  \,\, \,  \,\, \,
W(x) \, \, = \, \, \,
- {{35\,x^4 \, -3680\, x^3 \, +244242\, x^2 \, -244944\, x \, +177147
  } \over { 72 \cdot \, x^2 \cdot \, (x-1)^2 \cdot \, (x-81)^2}}.  
\end{eqnarray}
One can actually verify that $\, W(x)$, given by  the {\em rational function} (\ref{WHeun}),
can actually also be written  {\em in terms of the Heun function} (\ref{FHeun}):
\begin{eqnarray}
\label{relatWHeun}
\hspace{-0.98in}&&   \,  \quad \quad  \quad 
W(x) \, \, = \, \, \,
{{F''(x)} \over {F(x)}} \, \, -{{1} \over {2}} \cdot \, \Bigl({{F'(x)} \over {F(x)}}\Bigr)^2
\, \, = \, \, \,
\Bigl({{F'(x)} \over {F(x)}}\Bigr)' \, +{{1} \over {2}} \cdot \, \Bigl({{F'(x)} \over {F(x)}}\Bigr)^2. 
\end{eqnarray}
Note that introducing a ``nome''
\begin{eqnarray}
\label{DD-finiteexampleagain}
  \hspace{-0.95in}&& \quad  \,  \, \quad \quad \quad
 {{  Q(x)' }  \over {  Q(x)  }} \, \, = \,  \, \, {{ 1}  \over { F(x)}}
 \quad \quad \quad  \,  \,  \hbox{or:} \quad \quad \, \,  \,  \quad
   Q(x) \, \, = \, \, \,  \exp\Bigl( \int^x \, {{dx} \over {F(x)}}  \Bigr),   
\end{eqnarray}
relation (\ref{relatWHeun}) is nothing but relation (\ref{Harnad214}), namely:
\begin{eqnarray}
\label{Harnad214app}
\hspace{-0.95in}&& \quad  \quad  \quad  \quad  \quad 
  W(x) \, \,  \, =  \, \, \, \, \, - \, \{Q(x), \, x \} \, \, \, \, 
 - {{1} \over {2  \cdot \, Q(x)^2 }} 
 \cdot  \Bigl({{ d Q(x)} \over {d x}} \Bigr)^2.
\end{eqnarray}
A one-parameter series $\, y(a, \, x)$ is actually
solution\footnote[1]{Note that there are no solutions of the form
  $\, a \cdot \, x^n \, + \, \, \cdots\, $ with $\, n \, \ge \, 2$, since $\, W(x)$ is not of the form
  $\, -1/2/x^2 \, + \, \cdots$. The series expansion of $\, W(x)$ reads $\, -3/8/x^2 \, + \, \cdots$}
of the Schwarzian equation
\begin{eqnarray}
\label{condition1n1again}
\hspace{-0.95in}&& \quad \quad  \quad \quad  \quad \quad  
 W(x)  \, \, \,  \, -W(y(x)) \cdot  \, y'(x)^2 \, \, \,  \,+ \,  \{ y(x), \, x\} 
\, \,\, \, = \,\, \, \,  \, 0, 
\end{eqnarray}
with $\, W(x)$ given by (\ref{WHeun}):
\begin{eqnarray}
\label{SchwarzHeun}
\hspace{-0.98in}&&   \quad  
y(a, \, x) \, = \,\, a \cdot \, x \, \, -{{26} \over {81}} \cdot \, a \cdot \, (a-1) \cdot \, x^2 \, \,
+{{1} \over {6561}} \cdot \, a \cdot \, (a-1) \cdot \, (243\,a \,-1109) \cdot \, x^3
\nonumber \\
\hspace{-0.98in}&&   \quad  \quad  \quad  \quad  \, \, 
-{{2} \over {3720087}} \cdot \, a \cdot \, (a-1) \cdot \, (4013\,a^2\,-62326\,a\, +201028)\cdot \, x^4
\, \,\, + \, \, \, \cdots 
\end{eqnarray}
One can easily verify the functional equation:
\begin{eqnarray}
\label{FuncHeun}
\hspace{-0.98in}&&   \quad \quad  \quad \quad  \quad \quad \quad 
a \cdot \,   F\Bigl( y(a, \, x)\Bigr)^2
\, \, = \, \, \, \, F(x)^2 \cdot \, {{d y(a, \, x) } \over {dx}}^2,  
\end{eqnarray}
or:
\begin{eqnarray}
\label{FuncHeun2}
\hspace{-0.98in}&&  \quad  \quad \quad \quad   \quad \quad \quad 
a^{1/2} \cdot \,   F\Bigl( y(a, \, x)\Bigr)
\, \, = \, \, \, \, F(x) \cdot \, {{d y(a, \, x) } \over {dx}}.
\end{eqnarray}
One can also verify the following compositional formula:
\begin{eqnarray}
\label{composition2Heun}
\hspace{-0.95in}&& \,  \, \quad  \quad \quad  \quad  \quad  \quad \quad \quad \quad \quad
 y\Bigl(a, y(a', \, x)\Bigr) \, \, = \, \,  \, y(a\, a', \, \, x).
\end{eqnarray}
Let us introduce the two $\, a \, \rightarrow \, \, 0$ and  $\, a \, \rightarrow \, \, \infty$
limits of the one-parameter series (\ref{SchwarzHeun}):
\begin{eqnarray}
\label{compositiona0bisHeun}
  \hspace{-0.95in}&& \quad  \quad \quad \quad 
\tilde{Q}(x) \, \, = \, \, \,
  \lim_{a \rightarrow \, 0}  {{ y(a, \, \, x) } \over {a}} \, \, = \, \, \, \, \,\, 
  x \, \, \, \, +{{26} \over {81}} \, x^2 \,\, \, \,  + {{1109} \over {86561}} \, x^3
  \, \, \, +{{402056} \over {3720087}}\, x^4 
 \nonumber  \\
  \hspace{-0.95in}&& \,  \,  \quad  \quad \quad  \quad \quad \quad \quad 
  + {{2565526} \over {33480783}}\, x^5  \,\, \,  + {{471402140 } \over { 8135830269}}\, x^6 
  \,\,  \, \,  + \,\, \, \cdots
\end{eqnarray}
and:
\begin{eqnarray}
\label{compositionainftybisHeun}
  \hspace{-0.95in}&& \, \, \, \quad  \quad \quad \quad 
\tilde{X}(x) \, \, = \, \, \,
\lim_{a \rightarrow \, \infty}  \, y\Bigl(a, \, \, {{x} \over {a}}\Bigr)  \, \, = \, \, \, \, \,
x \,  \,  \, \, -{{26} \over {81}} \, x^2 \, \, \,  \,  +{{1} \over {27}} \, x^3
\, \, \,  \,  -{{8026} \over {3720087}}\, x^4\, 
 \nonumber  \\ 
\hspace{-0.95in}&& \, \quad  \quad  \quad \quad \quad \quad   \quad  \, 
  \, +{{38603} \over {301327047}}\, x^5 \, \,\,  \, - {{3200 } \over {301327047 }}\, x^6 
 \, \,\,\, \,  + \,\, \, \cdots 
\end{eqnarray}
One  verifies that the one-parameter series (\ref{SchwarzHeun}) {\em is actually of the form}:
\begin{eqnarray}
\label{oftheformHeun}
 \hspace{-0.95in}&&  \quad  \quad \quad  \quad \quad \quad   \quad \quad \quad \quad \quad 
y(a, x) \, \, = \, \, \,    \tilde{X}\Bigl( a \cdot \, \tilde{Q}(x) \Bigr). 
\end{eqnarray}
From the exact decomposition (\ref{oftheformHeun}),
together with the fact that  the one-parameter series (\ref{SchwarzHeun})
is such that $\, y(1, \, x) \, = \, \, 1$, one deduces immediately that the series (\ref{compositionainftybisHeun})
is actually the {\em compositional inverse} of the series (\ref{compositiona0bisHeun}).

\vskip .1cm

\section{Very simple  polynomial examples for $\, F(x)$.}
\label{miscellancorresp}

Let us display some very simple examples for $\, F(x)$, and the corresponding
one-parameter functions $\, y(a, \, x)$, solutions of the Schwarzian
equation (\ref{wherecond}).

$\bullet$ For $\, F(x)\, = \, x \cdot \, (1\, +p\, x)$,
one has $\, W(x) \, = \, \, -1/2/x^2/(1+p\, x)^2$,  and
a one-parameter function $\, y(a, \, x)$, solution of the Schwarzian
equation (\ref{wherecond}) with that $\, W(x)$ reads: 
\begin{eqnarray}
\label{family_a1}
 \hspace{-0.98in}&& \,  \, \quad \quad \quad \quad \quad \quad \quad \quad \quad
  y(a, \, x)  \, \, = \, \, \,  \,
  {{a \cdot \, x } \over {  (1\, \, +p\, x) \, \, -a \cdot \, p \cdot \, x}}. 
\end{eqnarray}
It is straightforward to see that $\, y(a, \, x)$ can be written
$\, y(a, \, x)  = \,   \tilde{X}(a \cdot \, \tilde{Q}(x))$, where:
\begin{eqnarray}
\label{family_a1XQ}
\hspace{-0.98in}&&
 \quad \quad \quad \quad \quad
\tilde{X}(x)  \, \, = \, \, \, {{x } \over { 1 \, \, - p \cdot \, x}}
\quad \quad  \, \, \hbox{and:} \quad \quad \quad 
\tilde{Q}(x)  \, \, = \, \, \, {{x } \over { 1 \, \, + p \cdot \, x}}.
\end{eqnarray}

$\bullet$ For $\, F(x)\, = \, x\, +p$,  one has $\, W(x) \, = \, \,  -1/(x +p)^2/2$, and
a one-parameter function $\, y(a, \, x)$, solution of
$\, F(y(a, \, x)) = \, F(x) \cdot \, {{\partial (a, \, x)} \over {\partial x}}$ and of the Schwarzian
equation (\ref{wherecond}) with that $\, W(x)$ reads ($p$ is fixed): 
\begin{eqnarray}
  \label{family_a3}
 \quad \quad \quad  \quad
  y(a, \, x)  \, \, = \, \, \,    a \cdot \,(x \, +p) \, \, \, \, -p.
\end{eqnarray}

$\bullet$ For $\, F(x)\, = \, x^2$,  one has\footnote[1]{More generally one has
  $\, W(x)= \, 0$ for  $\, F(x)\, = \, q \cdot \, (x +p)^2$. For $\, W(x)=0$
  the solution of the Schwarzian equation(\ref{wherecond}) 
  is a three-parameter solution:
  $ y(x)  \, \, = \, \, \,  \,   {{a \, x \, +b} \over {c \, x \, +d}}$.}
$\, W(x) \, = \, \,  0$, and
a one-parameter function $\, y(a, \, x)$, solution of
$\, F(y(a, \, x)) = \, F(x) \cdot \, {{\partial (a, \, x)} \over {\partial x}}$ and of
the Schwarzian
equation (\ref{wherecond}) with  $\, W(x)\, = \, \,  0\, $ reads: 
\begin{eqnarray}
\label{family_a4}
\quad  \quad
  y(a, \, x)  \, \, = \, \, \, \,   {{x} \over { 1 \, \,\,  \, - \, \ln(a) \cdot \, x}}.
\end{eqnarray}

 $\bullet$ For $\, F(x)\, = \, p$, one has $\, W(x) \, = \, \,  0$,
the one-parameter function $\, y(a, \, x)$, solution of
$\, F(y(a, \, x)) = \, F(x) \cdot \, {{\partial (a, \, x)} \over {\partial x}}$ and of the Schwarzian
equation (\ref{wherecond}) with  $\, W(x) \, = \, \,  0$ reads ($p$ is fixed):
 \begin{eqnarray}
 \label{family_a5}
 \quad  \quad
 y(a, \, x)  \, \, = \, \, \,   x \,\, \, \,  +p \cdot \, \ln(a).
 \end{eqnarray}

 $\bullet$ For $\, F(x)\, = \, p \cdot \, x$,  one has\footnote[5]{More generally when
  $\, W(x)=\, -1/2/x^2$
  the solution of the Scwharzian equation (\ref{wherecond}) with that $\, W(x)$,
  is a three-parameter solution:
  $\,  \ln\Bigl(y(x)\Bigr)  \, \, = \, \, \,  \, $
  $  {{a \, \ln(x) \, +b} \over {c \, \ln(x) \, +d}}$. } $\, W(x) \, = \, \,  -1/2/x^2$,
 and the one-parameter function
 $\, y(a, \, x)$, solution of $\, F(y(a, \, x)) = \, F(x) \cdot \, {{\partial (a, \, x)} \over {\partial x}}$
 and of the Schwarzian
equation (\ref{wherecond}) with that $\, W(x)$ reads ($p$ is fixed):
 \begin{eqnarray}
   \label{family_a6}
 \quad  \quad 
  y(a, \, x)  \, \, = \, \, \,  \,  a^p \cdot \,  x.
\end{eqnarray}

 All these one-parameter functions (\ref{family_a1}),  (\ref{family_a3}),
 (\ref{family_a4}),  (\ref{family_a5}),  (\ref{family_a6})   verify the  composition rule:
\begin{eqnarray}
\label{composition2app}
\hspace{-0.95in}&& \,  \, \quad   \quad  \quad  \quad  \quad \quad \quad \quad \quad
 y\Bigl(a, y(a', \, x)\Bigr) \, \, = \, \,  \, y(a\, a', \, \, x).
\end{eqnarray}
All these one-parameter
 functions (\ref{family_a1}),  (\ref{family_a3}),
 (\ref{family_a4}),  (\ref{family_a5}),  (\ref{family_a6})
 verify:
\begin{eqnarray}
  \label{family_a6also}
\hspace{-0.95in}&& \,  \, \quad  \quad \quad  \quad \, 
  F \Bigl(y(a, \, x) \Bigr) \, \, = \, \, \,  a \cdot \,  {{\partial y(a, \, x) } \over { \partial a}}
\, \, = \, \, \, F(x) \cdot  {{\partial y(a, \, x) } \over { \partial x}}.
\end{eqnarray}

\vskip .1cm


\vskip .5cm

\vskip .5cm

\vskip .5cm

\vskip .5cm

\vskip .5cm



\begin{thebibliography}{99}

\bibitem{Noetherian} S. Boukraa, S. Hassani and J-M.Maillard,
  {\em Noetherian mappings}, Physica {\bf D 185}, (2003) pp. 3-44

\bibitem{FactoBirat}  S. Boukraa and J-M.Maillard, 
 {\em Factorization properties of birational mappings}, Physica {\bf A 220} (1995) 403-470.
  
\bibitem{YBE} M.P. Bellon, J-M. Maillard and C. Viallet,
  {\em Infinite discrete symmetry group for the Yang-Baxter equations, Vertex models},
  Physics Letters {\bf B 260}, (1991) pp. 87-100.

\bibitem{BeMaVi92} M.P. Bellon, J-M. Maillard and C-M. Viallet,
{\em Quasi-Integrability of the sixteen vertex model}, 
Phys.Lett. {\bf B 281}, (1992), pp. 315--319 
 
\bibitem{Fatou} P. Fatou, Sur les \'equations fonctionnelles, 
Journal of Symbolic Comutation {\bf 94}, (2019), pp.90-104.

\bibitem{Ritt} J. F. Ritt, On the iteration of rational function, 
Trans. Amer. Math. Soc. {\bf 21} (1920), pp. 348-356.

\bibitem{Fatou2} P. Fatou, Sur les fonctions qui admettent 
plusieurs th\'eor\`emes de multiplication, 
C. R. acad. Sci. Paris, S\'er. I Math  {\bf 173}, (1921), pp. 571-573.

\bibitem{Eremenko} A. E. Eremenko, On some functional equations connected
with iteration of rational functions, 
Leningrad Math. J. Vol. 1, (1990), pp. 905-919.

\bibitem{Barry} B. M. McCoy and J-M. Maillard,  
{\em The anisotropic Ising correlations as elliptic integrals:  duality and differential equations}, 
J. Phys. {\bf A 49}, Number 43, (2016) 434004 (24pp), Special Issue in honour of A. J. Guttmann,  
 arXiv:1606.08796v4 [math-ph]

\bibitem{Ford} L. Ford, {\em Automorphic Functions}, (1929) AMS Chelsea Publishing. 
 
\bibitem{Stiller}  P. F. Stiller,
 {\em Classical Automorphic Forms and Hypergeometric Functions},
 Journ. of Number Theory, {\bf 28}, no. 2,  219-232, (1988). 

\bibitem{Schwarzian2}  J. McKay and A. Sebbar, {\em Fuchsian groups, automorphic functions 
and Schwarzians},  Math. Ann. {\bf 318}, (2000) pp.255-275.

\bibitem{Schwarzian}  J. McKay and A. Sebbar, {\em Fuchsian groups, Schwarzians, 
and theta functions}, C. R. Acad. Sci. Paris, {\bf 327}, S\'erie I, (1998) pp.343-348.

\bibitem{Koblitz} N. Koblitz, {\em Introduction to Elliptic Curves and Modular Forms},
  Graduate Texts in Mathematics, Second Edition, 1993, Springer 

\bibitem{1-2-3} J. H. Bruinier, G. van der geer, G. Harder and D. Zagier, {\em The 1-2-3 of Modular Forms},
  Universitext, 2008, Springer 

\bibitem{Berndt} G.E. Andrews, B.C. Berndt, {\em Ramanujan's Lost Notebook}, 2012, Part I, Part III, Part V, Springer 
  
\bibitem{ModularForms} Henri Cohen and Fredrik Stromberg, {\em Modular Forms: A Classical Approach},
  Graduate Studies in Mathematics, {\bf 179}, American Mathematical Society 

\bibitem{Hilbert} Paul B. Garrett, {\em  Holomorphic Hilbert Modular Forms},
  Wadsworth and Brooks/Cole Advanced Books and Software, Pacific Grove, CA, 1990

\bibitem{SuperMaier} R. Maier, {\em On rationally parametrized modular equations}, 
J. Ramanujan Math. Soc (2009), pp. 1-73,
 (2006), arXiv:0611041v3 [math.NT]

\bibitem{McMullen} C. McMullen, {\em Amenability, Poincar´e series and quasiconformal maps},
  Invent. Math. {\bf 97} (1989), 95–127.

\bibitem{Fricke} R. Fricke, F. Klein, {\em Vorlesungen über die Theorie der automorphen Funktionen},
  1–2 , Teubner (1926)

\bibitem{Janos} J. Kollár, {\em Shafarevich maps and automorphic forms}, (1995), M. B. Porter Lectures, Princeton University Press

\bibitem{Dalzell} D.P. Dalzell, {\em Theory of the theta-Fuchsian functions}, London Math. Soc. pp. 539-558

\bibitem{Farkas}
Farkas, Hershel M. (2008). {\em Theta functions in complex analysis and number theory},
In Alladi, Krishnaswami (ed.). Surveys in Number Theory. Developments in Mathematics. {\bf 17}, Springer-Verlag.
pp. 57–87. 

\bibitem{Schoeneberg} B. Schoeneberg, (1974). {\em IX. Theta series". Elliptic modular functions},
Die Grundlehren der mathematischen Wissenschaften, {\bf 203}, Springer-Verlag. pp. 203–226.


\bibitem{Migdal}
http://www.physics.fsu.edu/courses/Spring05/phy6938-02/decimation.pdf

\bibitem{Fisher} M.E. Fisher, 
{\em Renormalization group theory: Its basis
 and formulation in statistical physics}, 
Reviews of Modern Physics, {\bf 70}, No. 2,  (1998) pp. 653--681

\bibitem{Renorm} A. Bostan, S. Boukraa, S. Hassani, J-M. Maillard, J-A. Weil, N. Zenine and N. Abarenkova,
{\em Renormalization, isogenies and rational symmetries of differential equations},
  Advances in Mathematical Physics, Hindawi Pub. Volume 2010,  ID 941560,  44 pages

\bibitem{DDFinite} Antonio Jiménez-Pastor, Veronika Pillwein,
  {\em A computable extension for D-finite functions: DD-finite functions},
  Journal of Symbolic Computation {\bf  94},  2019, pp. 90-104

 \bibitem{Schwarz1} Y. Abdelaziz and J-M. Maillard,
{\em Modular forms, Schwarzian conditions, and symmetries of differential equations in physics},
(2017) J.Phys. {\bf A 50}: Math. Theor. 215203

\bibitem{Schwarz2} Y. Abdelaziz and J-M. Maillard,
{\em Schwarzian conditions for linear differential operators with selected differential Galois groups},
(2017) J.Phys. {\bf A 50}: Math. Theor. 465201 (33p). 

\bibitem{Heegner} S. Boukraa, S. Hassani, J-M. Maillard and  N. Zenine,
	{\em Singularities of $\, n$-fold integrals of the Ising class 
          and the theory of elliptic curves},
	\newblock J. Phys. {\bf A 40}: Math. Theor (2007) 11713-11748
	\newblock http://arxiv.org/pdf/math-ph/0706.3367
 
\bibitem{bo-ha-ma-ze-07b}
S. Boukraa, S. Hassani, J.-M. Maillard and N. Zenine,
{\em Landau singularities and singularities of holonomic integrals of the Ising class},
J. Phys. A: Math. Theor. \textbf{40} (2007) 2583--2614
and  arXiv:math-ph/0701016v2 

\bibitem{Canada} B. C. Berndt and H. H. Chan, 
{\em Ramanujan and the Modular $j$-Invariant},
 Canad. Math. Bull. {\bf 42} (1999) 427-440

\bibitem{Andrews} G.E. Andrews and B.C. Berndt, Chapter 17 pp. 373-393, 
in {\em Ramanujan's Lost Notebook}, Part I, 52005) Springer

\bibitem{Atkin} H. H. Chan and M.-L. Lang, {\em Ramanujan's modular equations
    and Atkin-Lehner involutions}, Israel Journal of Mathematics, 
  {\bf 103}, (1998) pp. 1--16.

\bibitem{Hermite} C. Hermite, {\em Sur la th\'eorie des \'equations modulaires}, 
 Comptes Rendus Acad. Sci. Paris {\bf 49}, 16-24, 110-118, and 141-144, 1859 
Oeuvres compl\`etes, Tome II. Paris: Hermann, p. 61, 1912. 

\bibitem{Hanna} M. Hanna, {\em The Modular Equations}, Proc. London Math. Soc. {\bf 28}, 46-52, 1928.

\bibitem{Morain} F. Morain, {\em Calcul du nombre de points sur une courbe elliptique 
dans un corps fini: aspects algorithmiques}, Journal de Th\'eorie des Nombres de Bordeaux, 
tome 7, (1995) pp.255-282 and https://eudml.org/doc/247643

\bibitem{Weisstein} Weisstein, Eric W. "Modular Equation." From MathWorld - A Wolfram Web Resource.
 http://mathworld.wolfram.com/ModularEquation.html

\bibitem{IsingCalabi} A. Bostan, S. Boukraa, 
     S. Hassani,  M. van Hoeij, J-M. Maillard, 
         J-A. Weil, N. J. Zenine, 
	{\em The Ising model: from elliptic curves to
         modular forms and Calabi-Yau equations},
	\newblock J. Phys. {\bf A 44}: Math. Theor.  (2011) (43 pp) 045204
        \newblock I0P Select, 
	\newblock and arXiv: 1007.69804 v1  [math-ph]
        \newblock and hal-00684883, version 1

\bibitem{IsingCalabi2} M. Assis, S. Boukraa, S. Hassani, 
       M. van Hoeij, J-M. Maillard,  B.M. McCoy
       {\em Diagonal Ising susceptibility: elliptic integrals, 
       modular forms and Calabi-Yau equations}, 
       \newblock J. Phys. {\bf A 45}: Math. Theor. (2012) 075205, [32 pages].
       \newblock IOP Select paper and Highlights of 2012
       \newblock arXiv:arXiv:1110.1705v2 [math-ph]   

\bibitem{Christol} A. Bostan, S. Boukraa, G. Christol, S. Hassani, J-M. Maillard, 
	{\em Ising $\, n$-fold integrals 
          as diagonal of rational functions and integrality of series expansions: 
        integrality versus modularity}, (2012) J. Phys. A: Math. Theor. {\bf 46} 185202

\bibitem{ChristolUna} A. Bostan, S. Boukraa, G. Christol, S. Hassani, J-M. Maillard, 
	{\em Ising $\, n$-fold integrals 
          as diagonal of rational functions and integrality of series expansions: 
          integrality versus modularity (unabridged version)},
           (2012)  arXiv:1211.6031v1 [math-ph] 

\bibitem{Jacobi} A.M. Legendre, Trait\'e des fonctions elliptiques, vol. {\bf 3}, Paris, 1825-1828.

\bibitem{Nova}  C.G.J. Jacobi, Fundamenta Nova Theoriae Functionum Ellipticarum K\"onigsberg, 1829.


\bibitem{What} V. Ovsienko and S. Tabachnikov, 
 {\em What is ... the schwarzian Derivative ?}, (2009), 
Notice of the AMS, {\bf 56}, pp. 34-36

\bibitem{DD} A. Jiménez-Pastor and V. Pillwein,
  {\em A computable extension for D-finite functions: DD-finite functions}
Bull. Soc. Math. France {\bf 47}, (1919), pp. 161-271.

\bibitem{DD1} A. Jiménez-Pastor, {\em  DD-finite functions in Sage},
  S\'eminaire Lotharingien de Combinatoire {\bf 82B}
  (2019) Article 101, 8pp. 

\bibitem{Candelas} P. Candelas, X. de la Ossa, P. Green and L. Parkes,
 {\em A pair of Calabi-Yau manifolds as an exactly
soluble superconformal theory}, Nucl. Phys. {\bf B359}, (1991), pp. 21-74.

\bibitem{Doran}
C. F. Doran,
 {\em Picard-Fuchs Uniformization and Modularity of the Mirror Maps},
 Comm. Math. Phys. {\bf 212}, pp. 625-647, (2000). 

\bibitem{Doran2}
C. F. Doran,
 {\em Picard-Fuchs Uniformization: Modularity of the Mirror Map
 and Mirror-Moonshine},  CRM Proc. Lecture Notes, {\bf 24}, Amer. Math. Soc.
pp. 257-281, Providence
and arXiv:math/9812162v1,  (1998). 

\bibitem{LianYau}
B.H. Lian and S-T. Yau,
 {\em Mirror Maps, Modular Relations and Hypergeometric Series II},
 Nuclear Phys. {\bf B 46}, Proceedings Suppl. Issues 1-3, (1996) pp.248-262
 and arXiv: hepth/950753v1 (1995)

\bibitem{Kratten}
C. Krattenthaler and T. Rivoal,
 {\em On the Integrality of the Taylor Coefficients of Mirror Maps},
 Communications in Number Theory (2009), Volume: {\bf 3},
 http://www-fourier.ujf-grenoble.fr/~rivoal
 and arXiv:0709.1432v3[math.NT]

\bibitem{Selected}  S Boukraa and J-M Maillard, {\em Selected non-holonomic functions
 in lattice statistical mechanics and enumerative combinatorics}, 
2016, J. Phys. {\bf A 49}: Math. Theor (29 pages) 074001
and arXiv:1510.04651v1 [math-ph] 

\bibitem{IsTheFull} A. J. Guttmann, I. Jensen, J-M. Maillard, J. Pantone,
{\em Is the full susceptibility of the square-lattice Ising model a differentially 
algebraic function ?}, (2016) J. Phys. {\bf A 49}: Math. Theor.  504002 (36 pages)
Special Issue in honour of A. J. Guttmann, 
and arXiv:1607.04168v2 [math-ph]

\bibitem{Replicable}  J. McKay and A. Sebbar, {\em Replicable functions: an introduction}, Springer

\bibitem{Replicable2} A. El Basraoui and J. McKay,
  {\em The Schwarzian equation for completely replicable functions}, 
LMS J. Comput. Math. {\bf 20} (1) (2017) 30-52 

\bibitem{Replicable3} B. Heim  and A. Murasea, {\em Completely replicable functions} and symmetries, 
Abhandlungen aus dem Mathematischen Seminar der Universit\"at Hamburg, (2019).

\bibitem{Replicable4} D. Ford, J. McKay and S. Norton, {More on replcable functions},
  Communications in Algebra {\bf 22}, (1994) pp. 5175-5193 (2007)

\bibitem{Replicable5}  A. El Basraoui, {\em Modular functions and replicable functions}, (2004) 

\bibitem{Replicable6} D. Alexander, C. Cummins, J. McKay and C. Simons,
  {\em Completely replicable functions}, Groups,
combinatorics and geometry (Durham, 1990), London Mathematical Society Lecture Note Series {\bf 165}
(Cambridge University Press, Cambridge, 1992) 87–98.

\bibitem{Replicable7} B Heim,   {\em Completely replicable functions and symmetries}
    Volume 89, pages 169–177, (2019) Springer 

\bibitem{Replicable8} Chang Heon Kim and Ja Kyung Koo, {\em Super-replicable functions N(j1,N)
and periodically vanishing property}, J. Korean Math. Soc. 2007; 44(2): 343-371

\bibitem{Casale} G. Casale, {\em Enveloppe Galoisienne d'une application rationnelle 
de $\mathbb{P}_1$}, Publicacions Matem\`atiques, Vol. {\bf 50}, No. 1 (2006), pp. 191-202
Published by: Universitat Aut\`onoma de Barcelona,  
arXiv[math/0503424]

\bibitem{Casale2} G. Casale, {\em An introduction to Malgrange pseudogroup},
	SMF - Séminaires et Congr\`es {\bf 23} (2011)

\bibitem{Casale3} G. Casale, {\em El grupoide de Galois de una transformación racional},
	VIII Escuela Doctoral intercontinental de Matem\`aticas PUCP-UVa 2015
CIMPA Research school "Transformation Groups and Dynamical Systems"

\bibitem{Casale4} G. Casale and Julien Roques, {\em Dynamic of rational 
 mappings and difference Galois theory}, Int. Math. Res. Notices 2008 (2008)

\bibitem{Casale5} G. Casale
{\em Sur le groupo\"ide de Galois d'un feuilletage},
Th\`ese de doctorat effectu\'ee sous la direction d'Emmanuel Paul et Jean-Pierre Ramis,
soutenue le 09/07/2004

\bibitem{Casale6} G. Casale, {\em D-enveloppe d'un diff\'eomorphisme de $(C, \, 0)$}, 
Annales de la Facult\'e des Sciences de Toulouse, Math\'ematiques, 
Tome XIII, (2004) pp. 515-538.

\bibitem{Casale7} G. Casale, {\em Morales-Ramis Theorems via Malgrange pseudogroup},
 Tome 59, (2009), pp. 2593-2610., 
Annales de l'institut Fourier, 

\bibitem{Malgrange} B. Malgrange, {\em On nonlinear differential Galois Theory}, 
Ann. of Math. 23B:2 (2002), pp. 219-226. 

\bibitem{Hindawi} A. Bostan, S. Boukraa, S. Hassani, J-M. Maillard, 
J-A. Weil, N. Zenine, and N. Abarenkova, 
{\em  Renormalization, isogenies and rational symmetries and differential equations}, 
Advances in Mathematical Physics,
Volume {\bf 2010} (2010), Article ID 941560, 44 pages, 
https://www.hindawi.com/journals/amp/2010/941560/

\bibitem{buium} A. Buium, {\em Geometry of differential polynomial functions III: moduli spaces}, 
Amer. J. Math. 117 (1995) 1-73.

\bibitem{Shimura} Y. Abdelaziz, S. Boukraa, C Koutschan and J-M. Maillard,
{\em Heun functions and diagonals of rational functions},
(2020) J.Phys. {\bf A 53}: Math. Theor. 075206  (24p). 

\bibitem{Goro} G. Shimura, {\em Correspondances modulaires et les fonctions 
$\, \zeta$ de courbes alg\'ebriques}, Journal of the Math. Soc. of Japan, 
  {\bf 10}, (1958), pp. 1-27.
  
\bibitem{Shimura_Una} Y. Abdelaziz, S. Boukraa, C Koutschan and J-M. Maillard,
{\em Heun functions and diagonals of rational functions (unabridged version)},
arXiv:1910.10761 [math-ph] (2019)  (64p). 

\bibitem{Moeller} I. Bouw and M. M\"oeller,
  {\em Differential equations associated with nonarithmetic Fuchsian groups},
arXiv:0710.5277v1 [math.AG] 2007 

\bibitem{Dettweiler} M. Dettweiler and S. Reiter, {\em On globally nilpotent differential equations},
Journal of Differential Equations {\bf 248},  2010, pp. 2736-2745

\bibitem{Dwork} B. Dwork, {\em Differential operators with nilpotent $\, p$-curvature},
American Journal of Mathematics {\bf 112} 1990 pp. 749-786

\bibitem{KrattenthalerRivoal}
C. Krattenthaler and T. Rivoal,
 {\em Analytic properties of 
   of Mirror Maps}, J. Aust. Math. Soc. {\bf 92} (2012) pp.195-235
 
\bibitem{Chrisconj}  Y. Abdelaziz, C. Koutschan and J-M. Maillard, {\em On Christol’s conjecture},
  2020 J. Phys. A: Math. Theor. 53 205201

\bibitem{GlobNilp} A. Bostan, S. Boukraa, S. Hassani, J.-M. Maillard, J.-A. Weil, and N. Zenine,
  {\em Globally nilpotent diﬀerential operators and the square Ising model},
J. Phys. A: Math. Theor. {\bf 42}   (2009) 125206
(50pp) and arXiv:0812.4931

\bibitem{extension}  S. Boukraa et J-M. Maillard,
  {\em The lambda extension of the Ising correlation functions $C(M,N)$}, 
2023,  J. Phys. A: Math. Theor. {\bf 56} 085201 and 
arXiv:2209.07434v2

\bibitem{lambdaPainl} S. Boukraa, C. Cosgrove, J-M. Maillard, and B. M. McCoy,
{\em Factorization of Ising correlations
$\, C(M, \, N)$ for $\nu =\, -k$ and $\, M + N$ odd,
$\, M \le \, N$,  $\, T < T_c$ and their lambda extensions}
  2022
J. Phys. A: Math. Theor. {\bf 55} 405204

\bibitem{Belyi3} M. van Hoeij, R. Vidunas,  
{\em Belyi functions for hyperbolic hypergeometric-to-Heun transformations}, 
Journal of Algebra (2015), Volume: {\bf 441}, pp. 609-659, 
arXiv:1212.3803v3[math.AG] 

\bibitem{Belyi} J. Sijsling and J. Voight, {\em On computing Belyi maps}, 
Publications math\'ematiques de Besan\c{c}con (2014), 
Issue: 1, pp.73-171,  arXiv:1311.2529v3[math.NT] 

\bibitem{Belyi2} R. Vidunas, A. V. Kitaev,  {\em Computation of highly ramified coverings}, 
 Math. Comp. (2009), Volume: {\bf 78}, pp.2371-2395,
arXiv:0705.3134[math.AG]

\bibitem{Belyi4} L. S. Khadjavi and V. Scharaschkin,  {\em Belyi maps and Elliptic Curves}, 
http://myweb.lmu.edu/lkhadjavi/belyielliptic.pdf 

\bibitem{Belyi5} D. Masoero,  {\em Painlev\'e I, Coverings of the Sphere and Belyi Functions},
 Constr Approx (2014), Volume: {\bf 39}, Issue: 1, pp.43-74, arXiv:1207.4361v2[math.ph] 

\bibitem{maier-05} R. Maier,
{\em On reducing the Heun equation to the hypergeometric equation},
J. Differential Equations {\bf 213} (2005), no. 1, pp. 171-203.

\bibitem{Automat} A.J. Guttmann and J-M. Maillard, {\em Automata and the susceptibility 
of the square lattice Ising model modulo powers of primes}, (2015)
 J. Phys. {\bf A 48}: Math. Theor (22 pages) 47001, Special Issue dedicated to R.J. Baxter
and arXiv:1507.02872v2 [math-ph] 
   
  
\bibitem{bo-gu-ha-je-ma-ni-ze-08}
S. Boukraa, A.J. Guttmann, S. Hassani, I. Jensen, J.-M. Maillard, B. Nickel
 and N. Zenine,
 {\em Experimental mathematics on the magnetic susceptibility of
 the square lattice Ising model},
 J. Phys. A: Math. Theor. {\bf 41} (2008) 455202 (51pp)
 and  arXiv:0808.0763


\end{thebibliography}
\end{document}